\documentclass[11pt]{article}
\usepackage{feynmp,epsfig}
\begin{document}\setlength{\unitlength}{1mm}
\begin{fmffile}{azrevfig}

\def\bef{\begin{figure}}
\def\eef{\end{figure}}
\newcommand{\ans}{ansatz}
\newcommand{\be}[1]{\begin{equation}\label{#1}}
\newcommand{\beq}{\begin{equation}}
\newcommand{\ee}{\end{equation}}
\newcommand{\beqn}[1]{\begin{eqnarray}\label{#1}}
\newcommand{\eeqn}{\end{eqnarray}}
\newcommand{\bd}{\begin{displaymath}}
\newcommand{\ed}{\end{displaymath}}
\newcommand{\mat}[4]{\left(\begin{array}{cc}{#1}&{#2}\\{#3}&{#4}
\end{array}\right)}
\newcommand{\matr}[9]{\left(\begin{array}{ccc}{#1}&{#2}&{#3}\\
{#4}&{#5}&{#6}\\{#7}&{#8}&{#9}\end{array}\right)}
\newcommand{\matrr}[6]{\left(\begin{array}{cc}{#1}&{#2}\\
{#3}&{#4}\\{#5}&{#6}\end{array}\right)}
\newcommand{\cvb}[3]{#1^{#2}_{#3}}
\def\lsim{\raise0.3ex\hbox{$\;<$\kern-0.75em\raise-1.1ex
\hbox{$\sim\;$}}}
\def\gsim{\raise0.3ex\hbox{$\;>$\kern-0.75em\raise-1.1ex
\hbox{$\sim\;$}}}
\def\abs#1{\left| #1\right|}
\def\simlt{\mathrel{\lower2.5pt\vbox{\lineskip=0pt\baselineskip=0pt
           \hbox{$<$}\hbox{$\sim$}}}}
\def\simgt{\mathrel{\lower2.5pt\vbox{\lineskip=0pt\baselineskip=0pt
           \hbox{$>$}\hbox{$\sim$}}}}
\def\unity{{\hbox{1\kern-.8mm l}}}
\def\epr{E^\prime}
\newcommand{\al}{\alpha}
\def\16p{16\pi^2}
\newcommand{\eps}{\varepsilon}
\def\ep{\epsilon}
\def\ga{\gamma}
\def\Ga{\Gamma}
\def\om{\omega}
\def\OM{\Omega}
\def\la{\lambda}
\def\La{\Lambda}
\def\al{\alpha}
\newcommand{\ov}{\overline}
\renewcommand{\to}{\rightarrow}
\renewcommand{\vec}[1]{\mbox{\boldmath$#1$}}
\def\tm{{\widetilde{m}}}
\def\mcirc{{\stackrel{o}{m}}}
\def\dem{\delta m^2} 
\def\sint{\sin^2 2\theta} 
\def\tant{\tan 2\theta} 
\def\tanL{\tan 2\theta^L}
\def\tanR{\tan 2\theta^R}
\newcommand{\tanb}{\tan\beta}
\def\brf{{\mathbf f}}
\def\bbf{\bar{\bf f}}
\def\bF{{\bf F}}
\def\bbF{\bar{\bf F}}
\def\bFp{{\bf F^\prime}}
\def\bbFp{\bar{\bf F^\prime}}
\def\bY{{\mathbf Y}}
\def\by{{\mathbf y}}
\def\bX{{\mathbf X}}
\def\bS{{\mathbf S}}
\def\bM{{\mathbf M}}
\def\bA{{\mathbf A}}
\def\bB{{\mathbf B}}
\def\bG{{\mathbf G}}
\def\bI{{\mathbf I}}
\def\bb{{\mathbf b}}
\def\bh{{\mathbf h}}
\def\bg{{\mathbf g}}
\def\bla{{\mathbf \la}}
\def\bmu{\mathbf m }
\def\bunity{{\mathbf 1}}
\def\cA{{\cal A}}
\def\cB{{\cal B}}
\def\cC{{\cal C}}
\def\cD{{\cal D}}
\def\cF{{\cal F}}
\def\cG{{\cal G}}
\def\cH{{\cal H}}
\def\cI{{\cal I}}
\def\cL{{\cal L}}
\def\cM{{\cal M}}
\def\cO{{\cal O}}
\def\cQ{{\cal Q}}
\def\cR{{\cal R}}
\def\cS{{\cal S}}
\def\cT{{\cal T}}
\def\dfrac#1#2{{\displaystyle\frac{#1}{#2}}}
\newcommand{\tphi}{\tilde{\phi}}
\newcommand{\Tphi}{\tilde{\Phi}}
\def\Bsi{{\bar{\Psi}}}
\newcommand{\bx}{\bar{\rm X}} 
\newcommand{\wx}{{\rm X}} 
\newcommand{\bv}{\bar{\rm V}} 
\newcommand{\wv}{{\rm V}} 
\newcommand{\tl}{\tilde{l}} 
\newcommand{\tq}{\tilde{q}}
\newcommand{\tu}{\tilde{u}}
\newcommand{\td}{\tilde{d}}
\newcommand{\tuc}{\tilde{u}_c} 
\newcommand{\tdc}{\tilde{d}_c} 
\newcommand{\tec}{\tilde{e}_c} 
\newcommand{\TQ}{\tilde{Q}} 
\newcommand{\TU}{\tilde{U}}
\newcommand{\TE}{\tilde{E}} 
\newcommand{\TUC}{\tilde{U}_c} 
\newcommand{\TEC}{\tilde{E}_c} 
\newcommand{\TQC}{\tilde{Q}_c} 
%

\begin{titlepage}

\begin{flushright}
DFAQ-99/TH/06 \\ 
DFPD-99/TH/29 \\ 
March 2000
\end{flushright}

\vspace{2.0cm}

\begin{center}

{\Large \bf 
 Predictive Grand Unified Textures\\
\vspace{0.3cm}
 for Quark and Neutrino Masses and Mixings} \\

\vspace{0.7cm}

{\large \bf Zurab Berezhiani${}^{a,b,}$\footnote{
E-mail address: berezhiani@fe.infn.it }
and Anna Rossi${}^{c,}$\footnote{
E-mail address: arossi@pd.infn.it }
}
\vspace{5mm}

{\it ${}^a$ Dipartimento di Fisica, 
Universit\`a di L'Aquila,  
I-67010 Coppito, AQ, and \\
INFN, Laboratori Nazionali del Gran Sasso, I-67010 Assergi, AQ, Italy}

{\it ${}^b$ The Andronikashvili Institute of Physics, 
Georgian Academy of Sciences, \\ 
380077 Tbilisi, Georgia} 

{\it  ${}^c$ Dipartimento di Fisica, 
Universit\`a di Padova and\\ 
INFN Sezione di Padova, I-35131 Padova, Italy. 
}
\end{center}

\vspace{10mm}

\begin{abstract}
\noindent

We propose new textures for the fermion Yukawa matrices 
which are  generalizations of the so-called Stech ansatz. 
We discuss how these textures can be realized in 
supersymmetric  grand unified models with horizontal 
symmetry $SU(3)_H$ among the fermion generations.  
In this framework the mass and mixing hierarchy of fermions 
(including neutrinos) can emerge in a natural way. 
We emphasize the central role played by the $SU(3)_H$ 
adjoint Higgs field which reduces $SU(3)_H$  
to $U(2)_H$ at the GUT scale. 
A complete $SO(10)\times SU(3)_H$ model is presented
in which the desired Yukawa textures can be obtained 
by symmetry reasons.  
The phenomenological implications of these textures 
are thoroughly investigated.  
Among various realistic possibilities for the Clebsch 
factors between the quark and lepton entries, 
we find three different solutions  which provide 
excellent fits of the quark masses and CKM mixing angles. 
Interestingly, all these solutions predict the correct 
amount of CP violation via the CKM mechanism, and, 
in addition, lead to an appealing pattern of the neutrino 
masses and mixing angles.  
In particular, they all predict  nearly maximal 23 mixing 
and small 12 mixing in the lepton sector, respectively 
in the range needed for the explanation of the atmospheric and 
solar neutrino anomaly.

\end{abstract}
\end{titlepage}


\section{Introduction}

In the last years the interest for the issue of neutrino mass 
generation has been renewed, especially by the SuperKamiokande 
data on atmospheric neutrinos \cite{ANP}.     
These data can be explained by $\nu_\mu-\nu_\tau$ oscillation 
within the following  parameter range (at 99 $\%$ CL):
\beqn{AN}
&&
\delta m^2_{\rm atm}= (1- 9)\times 10^{-3}~ {\rm eV}^2,
\nonumber \\ 
&&
\sin^2 2\theta_{\rm atm}>0.8
\eeqn     
(the best-fit values are 
$\dem_{\rm atm}=3\times 10^{-3}~\mbox{eV}^2$ and  
$\sint_{\rm atm}=0.99$),  
while the explanation through the $\nu_\mu -\nu_e$ 
oscillation is strongly disfavoured \cite{Lisi}.  
On the other hand, the solar neutrino data \cite{SNP}  
can be interpreted in terms  of $\nu_e$ oscillation 
into $\nu_\mu$, $\nu_\tau$, or their mixture 
within the following parameter range \cite{BKS,BKS2}:
\beqn{SN}
&&
\delta m^2_{\rm sol}= (3-10)\times 10^{-6}~{\rm eV}^2,
\nonumber \\ 
&& 
\sin^2 2\theta_{\rm sol}=(0.1-1.4)\cdot 10^{-2} 
\eeqn     
(the best-fit values are 
$\dem_{\rm sol}=5\times 10^{-6}~\mbox{eV}^2$ 
and  $\sint_{\rm sol}=7\times 10^{-3}$), the so called 
small-mixing angle MSW solution \cite{MSW}.\footnote{
We concentrate mainly on this solution, whose parameter 
range naturally emerges in the models presented below.  
One has to remark, however, that the solar neutrino data 
are still ambiguous and cannot discriminate other possible 
solutions. In particular, the recent analysis show 
that the large-mixing angle MSW and long wavelength ``just-so'' 
oscillation solutions  have comparable 
statistical significance \cite{Lisi,BKS2}. } 
Therefore, 
the experimental data provide a strong evidence 
that like  charged fermions, also neutrinos 
have masses and weak mixing. 
Namely, barring the less natural possibility that the 
neutrino mass eigenstates $\nu_{1,2,3}$ are strongly 
degenerate and  assuming that $m_3 >m_2> m_1$, 
the $\delta m^2$ ranges in eqs.  (\ref{AN}) and (\ref{SN}) 
directly translate into   
\beqn{nu-spectr} 
&&
m_3 \simeq (\dem_{\rm atm})^{1/2} =  
(3 - 10)\cdot 10^{-2} ~ {\rm eV}, \nonumber \\
&&
m_2 \simeq (\dem_{\rm sol})^{1/2} =  
(1.7 - 3.3)\cdot 10^{-3} ~ {\rm eV} ,
\eeqn 
which indicates that the neutrino mass ratio $m_3/m_2$ 
is similar to that of charged fermions, 
$m_\tau/m_\mu$ or $m_b/m_s$.  
However, the neutrino mixing pattern inferred from 
(\ref{AN}) and (\ref{SN}) 
is drastically different from the well-established 
structure of the Cabibbo-Kobayashi-Maskawa (CKM) mixing
of quarks.  

The explanation of the fermion flavour structure  
is one of the most challenging problems of particle physics,  
and the neutrino case constitutes a part of this issue. 
In the standard model the charged fermion masses emerge 
from the Yukawa terms:\footnote{
In the following we mainly 
deal with the supersymmetric model, though the general features 
of our discussion are equally valid for the non-supersymmetric 
case.
} 
\be{Yuk}
u^c_i \bY_u^{ij} q_j H_u  
+ d^c_i \bY_d^{ij} q_j H_d + e^c_i \bY_e^{ij}l_j H_d, 
\ee
where $q_i=(u,d)_i$, $u^c_i$, $d^c_i$ and 
$l_i=(\nu,e)_i$, $e^c_i$ are the quark and lepton fields 
of three families ($i=1,2,3$), 
and $H_{u,d}$ are the Higgs doublets, with the 
vacuum expectation values (VEVs)  
$v_{u,d}$ determining the electroweak scale, 
$(v_u^2 +v_d^2)^{1/2} =v_W=174$ GeV. 
There is no renormalizable term that gives rise to the 
neutrino masses. However, the latter can get 
Majorana masses from the lepton-number violating 
higher order operator   
\be{Yuk-nu}
\frac{1}{M_L}\, l_i \bY_\nu^{ij} l_j H_u^2 ~, 
~~~~~  \bY_\nu = \bY_\nu^T
\ee
which is cutoff by some large scale $M_L$, e.g. the grand 
unification or Planck scale \cite{BEG}.\footnote{
Any known mechanism for the neutrino masses reduces to 
such an effective operator. E.g., in the `see-saw' scheme 
\cite{seesaw} it is obtained after integrating out 
heavy-neutral fermions with  Majorana masses $\sim M_L$. 
}     
Hence, the charged fermion masses are $\sim v_W$ and 
the neutrino masses are $\sim v_W^2/M_L$ 
which makes it easy to understand why the latter are so light.  

The quark mixing is originated from the misalignment 
between the Yukawa matrices $\bY_u$ and $\bY_d$. 
Analogously, the neutrino mixing is due to misalignment  
between $\bY_e$ and $\bY_\nu$. In particular, one can 
choose a basis where $\bY_u$ and $\bY_\nu$ are diagonal,  
$\bY_u= {\rm Diag}(Y_u,Y_c,Y_t)$ and 
$\bY_\nu= {\rm Diag}(Y_1,Y_2,Y_3)$. In this basis, the quark 
and lepton mixing angles are determined respectively 
by the form of $\bY_d$ and $\bY_e$ 
which in general remain non-diagonal. The latter 
are to be diagonalized by the bi-unitary transformations: 
\beqn{unit} 
&&
U'^T_{d}\bY_d U_{d}=
\bY_d^{D}=\mbox{Diag}(Y_d,\,Y_s,\,Y_b) 
\nonumber \\ 
&&
U'^T_{e}\bY_e U_{e}=
\bY_e^{D}=\mbox{Diag}(Y_e,\,Y_\mu,\,Y_\tau) ,
\eeqn
with the unitary matrices $U_{d,e}$ rotating the 
left handed (LH) states and with $U'_{d,e}$ rotating 
the right handed (RH) ones. 
Thus, in this basis the CKM matrix of weak mixing 
between the physical quark states $(u,c,t)$ and $(d,s,b)$
is $V_q=U_d$, while the matrix $V_l=U_{e}^\dagger$   
determines the properties of the  weak current between the 
charged leptons $(e,\mu,\tau)$ and the neutrinos. 
Namely, $V_l$ relates the neutrino flavour eigenstates 
$(\nu_e,\nu_\mu,\nu_\tau)$ to the mass eigenstates 
$(\nu_1,\nu_2,\nu_3)$, and describes   
the neutrino oscillation phenomena. 

The unitary matrices 
\be{CKM}
V_q = \matr{V_{ud}}{V_{us}}{V_{ub}} {V_{cd}}{V_{cs}}{V_{cb}} 
{V_{td}}{V_{ts}}{V_{tb}} , ~~~~~
V_l= \matr{V_{e1}}{V_{e2}}{V_{e3}}
{V_{\mu1}}{V_{\mu2}}{V_{\mu3}} 
{V_{\tau1}}{V_{\tau2}}{V_{\tau3}} 
\ee
are usually   parameterised as \cite{maiani}: 
\be{MNS}  
V_{q,l}= \matr{c_{12}c_{13}}{s_{12}c_{13}}{s_{13}e^{-i\delta}}
{-s_{12}c_{23}-c_{12}s_{23}s_{13}e^{i\delta}}
{c_{12}c_{23}-s_{12}s_{23}s_{13}e^{i\delta}} {s_{23}c_{13}} 
{s_{12}s_{23}-c_{12}c_{23}s_{13}e^{i\delta}}
{-c_{12}s_{23}-s_{12}c_{23}s_{13}e^{i\delta}} {c_{23}c_{13}}_{q,l}  
\ee 
where $s_{ij}$ ($c_{ij}$)  stand for the sines (cosines) 
of the three mixing angles $\theta_{12}$, $\theta_{23}$ and 
$\theta_{13}$ while $\delta$  is the CP violating phase. 
(In the case of Majorana 
neutrinos, the lepton mixing matrix contains two 
additional phases $\delta_{1,2}$,  factorised out as  
$V_l\cdot{\rm Diag}(1,e^{i\delta_1},e^{i\delta_2})$, 
which, however, are not relevant for neutrino oscillations.)
In the following, we distinguish the 
angles and phases in $V_q$ and $V_l$ by the superscripts 
`$q$' and `$l$', respectively.    

The explanation of the fermion mass and mixing pattern   
is beyond the capacities of the standard model. 
The Yukawa matrices $\bY_{u,d,e}$, 
are arbitrary complex $3\times 3$ matrices, 
not constrained by any symmetry property. 
Concerning the neutrinos, apart from the Yukawa factors 
$\bY_\nu$ in (\ref{Yuk-nu}), also the lepton-number violation 
scale $M_L$ is an arbitrary parameter.   
The magnitude of the latter can be inferred from 
the mass value $m_3$ in (\ref{nu-spectr}). 
Namely, from $m_3\sim Y_3 v_W^2/M_L$ we obtain 
\be{Y3}
Y_3^{-1} M_L \sim 10^{15} ~ {\rm GeV},  
\ee 
which is rather close to the GUT scale $M_G\sim 10^{16}$ GeV. 
In particular, the normalization of the operator (\ref{Yuk-nu}) 
as $M_L=M_G$ implies $Y_3\sim 10$. 
Let us remark that there is no contradiction 
if the effective coupling constant $Y_3$ is large. 
E.g. in the context of the see-saw mechanism 
\cite{seesaw} $Y_3$ is determined by the  ratio  
$Y_{\rm Dirac}^2/Y_{\rm Majorana}$ which can easily happen 
to be   
$\sim 10$ even if the Yukawa constants 
$Y_{\rm Dirac}$ and $Y_{\rm Majorana}$ are order 1.

One aspect of the flavour problem is related to the  
inter-family mass/Yukawa hierarchy among fermions \cite{ICTP}.  
For example, the top Yukawa constant is $Y_t\sim 1$,  
while the first-generation constants $Y_{e,u,d}$ are much 
smaller,  $\sim 10^{-5}$. 
Concerning the neutrinos,  they also seem to indicate 
a mass hierarchy similar to that of the charged fermions. 
In particular, eqs. (\ref{nu-spectr}) show that 
the ratio ${Y_3}/{Y_2}$ lies in the range $10-60$. 
There are no data restricting the ratio 
$Y_2/Y_1$, but it  may also  be large.
For the sake of comparison, the mass ratios between 
different families are summarized in the Table \ref{tab1}.   

On the other hand, the mixing structure is very different 
between  quarks and leptons (see Table \ref{tab2}). 
The most striking feature is that the atmospheric neutrino 
data favour the nearly maximal 23 mixing of neutrinos, 
while the 23 mixing of quarks is very small.  
On the contrary, the MSW solution implies a very small 
12 mixing angle for neutrinos as compared to the
sizeable 12 angle of the Cabibbo mixing for quarks.  

\begin{table}
 \begin{center}
 \begin{tabular}{||l||l||}
        \hline \hline
$~~~~2^{nd}/1^{st}$ & ~~~~~~ $3^{rd}/2^{nd}$  \\  
        \hline \hline
${Y_c}/{Y_u} = 250-1000$ & 
${Y_t}/{Y_c} =230-330_{(\mu=M_W)}$, $320-1000_{(\mu=M_G)}$ \\ 
    \hline
${Y_s}/{Y_d} =17-25$ & 
${Y_b}/{Y_s} =30-85_{(\mu=M_W)}$,  $33-120_{(\mu=M_G)}$ \\  
    \hline
${Y_\mu}/{Y_e} = 207$ & 
${Y_\tau}/{Y_\mu}=17$ \\  
    \hline
${Y_2}/{Y_1} >1 $ & 
${Y_3}/{Y_2} =10 - 60$ \\
        \hline \hline
 \end{tabular}
\caption[]{\small Quark and lepton mass ratios between 
second/first and third/second families. 
All mass ratios are independent of the renormalization 
scale $\mu$ except $Y_t/Y_c$  and $Y_b/Y_s$ 
(in the case of reasonably small $\tan\beta\equiv v_u/v_d$).  
For  definiteness, the latter ratios are evaluated 
at the electroweak scale, $\mu=M_W$, 
while at the GUT scale, $\mu=M_G\simeq 10^{16}$ GeV, 
they are scaled respectively 
as $B_t^{-3}$ and $B_t^{-1}$, with the factor $B_t=B(Y_t)$ 
varying from 0.9 to 0.7 for the top Yukawa constant 
at the GUT scale in the interval $Y^G_t= 0.5-3$ 
(see details in section 4). }
 \label{tab1}
 \end{center}
 \end{table} 


It is widely believed that the fermion flavour 
structure can be understood within the context of the 
grand unified theories with horizontal symmetries.  
The latter acts between the fermion families and thus 
can help to constrain the form of the Yukawa matrices 
$\bY_{u,d,e,\nu}$.  
In particular, it can dictate 
the hierarchical and closely aligned structures 
of $\bY_u$ and $\bY_d$ and thus simultaneously 
explain the large mass splitting among 
different families 
as well as the smallness of the quark mixing angles.
However, the origin of the large neutrino mixing 
is not {\it a priori} clear.
Thus, the crucial point that must be explained by any 
realistic flavour theory is the dramatic difference between 
the quark and lepton mixing angles.\footnote{
In last years many efforts have been applied to derive 
the neutrino mixing pattern from particular textures 
of the neutrino mass matrix \cite{all}. 
Differently from some point of view adopted in \cite{all},  
we  believe that a satisfactory scheme can be obtained in 
the context of a well-defined and complete flavour theory 
including the charged fermions. 
Such attempts have been taken e.g. in refs. \cite{GUT} 
where the neutrino mixing pattern was discussed 
in the context of grand unified and horizontal symmetries.}

\begin{table}
 \begin{center}
 \begin{tabular}{||c|c|c||}
        \hline \hline
         & $\mbox{\it q(uark)}$ &$\mbox{\it l(epton)}$  \\
        \hline
        $\theta_{12}$ & $12.7\pm0.1$ & $2.4\pm 1.0$  \\
        \hline
        $\theta_{23}$ & $2.3\pm0.1$ & $45\pm13$  \\
        \hline
        $\theta_{13}$ & $0.21\pm0.03$ & $<15$\\  
        \hline \hline
 \end{tabular}
 \caption[]{\small Quark and leptonic mixing angles (in $^\circ$). 
The values of $\theta^l_{23}$ and $\theta^l_{12}$ 
correspond to the 99\% CL regions of the AN and SN  
data fitting, eqs. (\ref{AN}) and (\ref{SN}),   
while the upper bound for $\theta_{13}^l$ 
comes from the combination of CHOOZ \cite{CHOOZ} and atmospheric neutrino 
data \cite{Lisi}. 
The values of $\theta^q_{12}$, $\theta^q_{23}$ and 
$\theta^q_{13}$ reflect the present data on 
$|V_{us}|$, $|V_{cb}|$ and $|V_{ub}/V_{cb}|$ \cite{maiani}. 
The last two values depend on the renormalization scale 
and for $\mu=M_G$ they should be 
scaled by a factor $B_t$, i.e. become smaller by $(10-30) \%$ . 
}
 \label{tab2}
 \end{center}
 \end{table} 

In a previous paper \cite{BR} 
we have outlined that such a complementary pattern of 
quark and lepton mixings can naturally emerge   
in the context of  grand unified theories. 
In the  $SU(5)$ model the quark and lepton states 
of each family are combined in the multiplets  
$\bar{5}_i=(d^c, l)_i$,  $10_i =(u^c, e^c, q)_i$,   
and the Higgs doublets $H_u$ and $H_d$  fit respectively 
into the representations $H\sim 5$ and $\bar H\sim \bar5$.  
The terms responsible for the fermion masses read as: 
\be{Yuk-su5}
10_i \bG_u^{ij}10_j H + 
\bar5_i \bG^{ij} 10_j \bar{H} + 
\frac{1}{M_L}\bar5_i\bG_\nu^{ij}\bar5_j H^2 ,  
\ee
where the Yukawa constant matrices $\bG_{u}$ and 
$\bG_{\nu}$ are symmetric due to $SU(5)$ symmetry reasons  
while the form of $\bG$ is not constrained. 
After the $SU(5)$ symmetry breaking 
these terms reduce to the standard model terms  
(\ref{Yuk}) and (\ref{Yuk-nu}),  with  
\beqn{relat}
&&
\bY_u=\bG_u, ~~~~ \bY_\nu=\bG_\nu , \nonumber \\ 
&&
\bY_d=\bG , ~~~~~~ \bY_e=\bG^T .  
\eeqn 
Hence, in the basis where the matrices 
$\bG_u$ and $\bG_\nu$ are  diagonal, 
the rotation angles of the LH leptons coincide 
with that of the RH quarks: $U_e=U'_d$, and vice versa, 
$U_d=U'_e$. (One has to remark, 
that in grand unification the rotation angles of RH 
states also have  a physical sense 
as  they define the fermion mixing in 
B- and L-violating currents.) 
Hence, if $G^{23}\sim G^{33}$ while $G^{32}\ll G^{33}$,  
the large 23 neutrino mixing will occur together with  
the small 23 mixing of quarks \cite{GUT,BR}.    
In other words, the phenomenology requires that  
the matrix $G$ should have a strongly asymmetric form.  

In particular, a very interesting situation emerges 
in the case of  Fritzsch-like texture \cite{Fritzsch} 
with vanishing $G^{22}$ and $G^{11}$ entries. 
In addition, the entry $G^{33}$ can be taken as 
$SU(5)$ singlet thus maintaining 
the $b-\tau$ Yukawa unification, 
while the  off-diagonal entries $G^{23}$ etc. 
should contain some $SU(5)$ Clebsches to avoid the 
wrong relations $Y_{d,s}=Y_{e,\mu}$ of the minimal $SU(5)$.
This means that the masses 
of the third generation emerge from the Yukawa couplings 
to the Higgs 5-plet, while the lighter generation masses 
are contributed also by 45-plet.  
In fact, it is not  necessary to introduce 
the elementary Higgs 45-plet. 
Instead, the off-diagonal entries $\bG^{ij}$  can be regarded 
as operators dependent on the adjoint Higgs $\Phi\sim 24$ 
of $SU(5)$, $\bG^{ij}=\bG^{ij}(\Phi)$.\footnote{ This 
should be  understood as expansion series  $\bG^{ij}(\Phi)= 
G^{(0)ij} + G^{(1)ij}\frac{\Phi}{M} + ...$, 
where $M$ is some cutoff mass larger than the 
GUT scale (e.g. the string scale). 
This means that the off-diagonal terms   
emerge from the effective higher-order operators 
$\frac{\Phi}{M}\bar{H}\bar5_i G^{(1)ij}10_j$ etc., just 
on the same footing as the last term in (\ref{Yuk-su5}). 
Since in general the operator $\Phi\cdot \bar{H}$ is represented 
by the tensor product $24\times \bar5=\bar5 + \ov{45}$,  
it can distinguish the corresponding 
entries in the matrices $\bY_e$ and $\bY_d$ and hence avoid 
the lepton-quark degeneracy of the minimal $SU(5)$ theory.
} 
In this case the relation $\bY_d=\bY_e^T$ is not exact 
anymore, but it fulfills with the precision of these  
Clebsch factors.  Nevertheless, this property ensures 
that between the 23 mixing angles of quarks and leptons
the following relation is fulfilled 
with a good accuracy \cite{BR}: 
\be{rule23} 
\tan\theta_{23}^q \tan\theta_{23}^l \simeq 
\left(\frac{m_\mu m_s}{m_\tau m_b}\right)^{1/2}. 
\ee
This product rule indeed works remarkably well. 
It demonstrates a `see-saw' correspondence between the 
lepton and quark mixing angles and tells us that whenever 
the neutrino mixing is large, $\tan\theta_{23}^l \sim 1$,   
the quark mixing angle comes out small and 
in the correct range, $\tan\theta_{23}^q \sim 0.04$.   

Namely, in ref. \cite{BR} we have considered the following 
Yukawa pattern: 
\beqn{Fred}
&&
\bY_u = \matr{Y_u}{0}{0} {0}{Y_c}{0} {0}{0}{Y_t}, ~~~~~
\bY_d = \matr{0}{A_d}{0} {A'_d}{0}{B_d} {0}{B'_d}{D} = 
\matr{0}{A_d}{0} {\frac{1}{a}A_d}{0}{B_d} {0}{\frac{1}{b}B_d}{D},
\nonumber \\ 
&&
\bY_\nu = \matr{Y_1}{0}{0} {0}{Y_2}{0} {0}{0}{Y_3}, ~~~~~
\bY_e^T = \matr{0}{A_e}{0} {A'_e}{0}{B_e} {0}{B'_e}{D} = 
\matr{0}{A_e}{0} {\frac{1}{a}A_e}{0}{B_e} 
{0}{\frac{1}{b}B_e}{D}, 
\eeqn 
where 
the parameters $b=B_e/B'_e=B_d/B'_d$ and 
$a=A_e/A'_e=A_d/A'_d$ reflect the possible asymmetry 
between the off-diagonal entries in $\bY_{d,e}$.  
In addition, the Clebsches $A_e/A_d = k_A$ and 
$B_e/B_d = k_B$ are allowed to be different from 1 
and their values can be extracted  from the quark and 
lepton masses.

Indeed, the texture predicts the following 
relations between the Yukawa eigenvalues \cite{BR}: 
\be{bsd}
\frac{Y_b} {Y_{\tau} } = Z, ~~~~~ 
\frac{Y_s-Y_d}{Y_\mu-Y_e}= \frac{1}{k_B^2 Z}, ~~~~~ 
\frac{Y_dY_s}{Y_eY_\mu}= \frac{1}{k_A^2 Z}, 
\ee
and 
\be{s/d} 
\frac{m_s}{m_d} + \frac{m_d}{m_s} -2 =  
\frac{k_A^2}{Z k_B^4} 
\left(\frac{m_\mu}{m_e} + \frac{m_e}{m_\mu} -2 \right) , 
\ee
where 
\be{btau}
Z =\sqrt{1-(b+b^{-1})(1-k_B^{-2})\frac{m_\mu-m_e}{m_\tau}} ~.
\ee
In particular, the phenomenologically correct picture 
for the fermion masses emerges if the asymmetry parameter 
$b$ is large enough, $b\sim 10$, and when  
$k_A\simeq 1$ and $k_B\simeq 2$ \cite{BR}. 
In this case the relation  (\ref{s/d}) leads to the correct 
prediction for the strange/down quark mass ratio.  

On the other hand, for the 23 mixing angles in (\ref{CKM}) 
we obtain the relations: 
\be{mix23} 
\tan\theta_{23}^l \simeq b^{1/2} \sqrt{{m_\mu\over m_\tau}}, ~~~~  
\tan\theta_{23}^q \simeq b^{-1/2}\sqrt{{m_s\over m_b}},   
\ee
from which  the product rule (\ref{rule23}) is immediate.
These relations also point to a large asymmetry 
factor $b$.\footnote{
The symmetric case $b=1$, i.e. $\bY_{d,e}$ having the 
familiar Fritzsch texture, is obviously excluded,   
since the 23 mixing of quarks is too big, 
$\theta^q_{23} = (6-10)^\circ$, while it is too small 
for leptons, $\theta^l_{23} = 13.7^\circ$  
(cfr. the experimental values in   Table \ref{tab2}). 
}  
In particular, for $b\sim 10$ the neutrino mixing becomes 
nearly maximal, $\theta_{23}^l \sim 45^\circ$, whereas  
the quark mixing angle gets small, 
$\theta_{23}^q \sim (2-3)^\circ$.

In the general case, for arbitrary $a$, 
the product rule similar to that in (\ref{mix23}) can be 
obtained also for 12 mixing angles. 
However, from  phenomenological grounds one can 
deduce that no significant asymmetry should occur 
in the 12 sector of the matrices (\ref{Fred}).  
Namely, the successful relation for the Cabibbo angle: 
\be{mix12} 
\tan\theta^q_{12} \simeq \sqrt{\frac{m_d}{m_s}} ,
\ee
indicates that $A_d\simeq A'_d$, so that  $a=1$ is appropriate. 
Interestingly, in this case  
we obtain for the leptonic 12 mixing \cite{BR}:  
\be{mix12_l}
\tan\theta^l_{12} 
\simeq   \sqrt{\cos\theta_{23}^l \frac{m_e}{m_\mu}} .
\ee
Given that $\theta_{23}^l\sim 45^\circ$, 
this implies
$\theta_{12}^l \sim 3^\circ$, which is in the range  
relevant to  the MSW solution of the SN problem 
(cfr. Table \ref{tab2}). 
 
Therefore, for $a=1$ the ansatz (\ref{Fred}) depends 
on six parameters, so it is highly predictive. 
Among these, three Yukawa entries $A_e,B_e,D$ 
can be expressed in terms of lepton couplings 
$Y_{e,\mu,\tau}$ and $b$.  
Thus 9 physical quantities, 
the Yukawa eigenvalues $Y_{d,s,b}$     
and the mixing angles in $V_q$ and $V_l$, 
are left as functions of three Clebsch factors, 
$k_B,k_A$ and $b$.
Hence, at the GUT scale 
these are connected by six relations like (\ref{bsd}), 
(\ref{mix23}), etc.   
As discussed above, in the first approximation 
these relations well reproduce the observed pattern 
of the quark and lepton mixing angles for $b\sim 10$ 
and $k_B\simeq 2$, $k_A\simeq 1$. However, a 
closer inspection shows that for precise Clebsch 
values $k_B=2$ and $k_A=1$, which could be 
motivated in the GUT context, and $a=1$ which 
could follow from the $U(2)_H$ horizontal symmetry, 
the predictions for $|V_{us}|$ and $|V_{cb}|$ 
 substantially deviate from the corresponding experimental 
values which presently are known with a very 
good accuracy. 
The case summarized above implies a real CKM matrix since  
$\bY_u$ is taken diagonal and so all phases in $\bY_d$ 
can be absorbed by the field re-definitions. 
Nevertheless, it would be interesting to address 
the CP issue in a scenario in which $\bY_u$ is still 
kept diagonal.

In the present paper we suggest a new grand unified 
ansatz for the fermion masses, which is an extension 
of the pattern (\ref{Fred}) of ref. \cite{BR}. 
Namely, we consider the case when the 
matrices $\bG_u$ and $\bG_\nu$ are both diagonal 
and related as $\bG_\nu \propto \bb'^{-1}\cdot\bG_u$,  
with $\bb'$ being a diagonal matrix 
$\bb'={\rm Diag}(1,1,b')$.  
Regarding $\bG$, we assume that it has the diagonal part 
$\propto \bG_u$ and the off-diagonal one 
$\propto \bb^{-1}\cdot\bA(\Phi)$, 
where  $\bA$ is  an antisymmetric matrix with 
$\Phi$-field dependent entries inducing different 
Clebsch factors for the down quarks and charged leptons, and 
$\bb={\rm Diag}(1,1,b)$.   
In other words, we consider the Yukawa textures:
\beqn{Stech}
& \bY_u = \bY_u^D, 
& \bY_d = \rho\bY_u + \bb^{-1}\bA_d , \nonumber \\ 
& \bY_\nu= \eta \bb'^{-1}\bY_u ,  
& \bY_e^T = \rho\bY_u  + \bb^{-1}\bA_e ,  
\eeqn 
where $\rho$ and $\eta$ are proportionality 
coefficients. 
For $b=1$ this pattern resembles the so-called 
{\it Stech} texture proposed long time ago 
by Stech \cite{Stech} and independently 
by Chkareuli and one of the authors \cite{su3H},   
where the off-diagonal entries of $\bY_{d}$ were assumed 
to be antisymmetric, $\bY_d = \rho \bY_u + \bA_d$.  
This case, however, is completely excluded by 
experimental data. 
As for our texture, the matrices $\bb^{-1}\bA_{d,e}$  
remain antisymmetric only in the 12 block, 
while the other entries are skew due to the 
factor $b\neq 1$. 
In the explicit form the Yukawa matrices read as:    
\beqn{Stech-exp}
& 
\bY_u = \matr{Y_u}{0}{0} {0}{Y_c}{0} {0}{0}{Y_t}, 
& 
\bY_d = \matr{\frac{Y_u}{Y_t}D}{A_d}{C_d} 
{-A_d}{\frac{Y_c}{Y_t}D}{B_d}
{-\frac1b C_d}{-\frac1b B_d}{D},  
\nonumber \\ 
& 
\bY_\nu = \eta \matr{Y_u}{0}{0} {0}{Y_c}{0} 
{0}{0}{\frac{1}{b'}Y_t}, ~~ 
&
\bY_e^T = \matr{\frac{Y_u}{Y_t}D}{A_e}{C_e} 
{-A_e}{\frac{Y_c}{Y_t}D}{B_e}
{-\frac1b C_e}{-\frac1b B_e}{D}  .
\eeqn 
Clearly this texture 
represents an extension of the 
Fritzsch-like `zero-texture' (\ref{Fred}) 
considered in \cite{BR}.  
In particular, for vanishing 13 and 31 entries, 
$C_{d,e}=0$, this texture essentially reduces to the latter  
as the diagonal 11 and 22 entries 
in $\bY_{d,e}$ (\ref{Stech}) are quite small,   
since the  ratios $Y_u/Y_t$ and $Y_c/Y_t$ are 
much smaller than $1/3$ and $2/3$ Yukawa ratios in the 
down quark and charged leptons. 
In practice, as we show below, one can safely take the  
limit $Y_u/Y_t \to 0$ and ignore the 11 entries in 
$\bY_{d,e}$. 
As for the 22 entry, though it is small, it can provide 
significant corrections (in particular, to $V_{cb}$).  
In principle, it also contains the relative 
(un-removable) CP violating phase which however 
cannot provide a sufficient amount of CP violation.  
This can be remedied by non-vanishing  $C_{e,d}$. 

Moreover, 
we show that for large values of $b$ ($\sim 10$), 
the texture (\ref{Stech-exp}) provides 
a very appealing description of the quark and 
lepton mass and mixing structures.   
Interestingly, the case $b'\sim b\sim 10$ also provides 
nice relations between neutrino and up-quark masses, 
$Y_2/Y_3 = b'Y_c/Y_t$ (cfr. Table 1).  

Hence, the complete pattern (\ref{Stech-exp}) 
is first  motivated  on the phenomenological  
point of view as it can fit the physical observables with 
excellent precision.  
From a theoretical point of view, we show that these 
textures can be justified  in the framework of grand 
unification 
by invoking also the concept of horizontal symmetry 
among three fermion families \cite{John,su3H}. 

The paper is organized as follows.
In Sect. 2 we discuss some theoretical tools  
for building-up the textures (\ref{Stech}) 
in a  realistic and predictive manner 
in the framework of supersymmetric $SU(5)$ model 
with $SU(3)_H$ horizontal  symmetry.   
The discussion in Sect. 3 is meant to provide an existence 
proof of those realizations in the framework of 
$SO(10)\times SU(3)_H$ model.
Three particular ans\"atze are singled out, 
characterized by different assignments of the 
Clebsch factors between the quark and lepton entries, 
and in Sect. 4 their phenomenological analysis is presented 
and discussed in all details. 
Finally, we summarise our findings and conclude.

\section{Theoretical framework: GUT and horizontal symmetry}

In this section we would like to sketch the general ideas 
needed to obtain the  Yukawa textures (\ref{Stech}) 
in the $SU(5)$ model with the $SU(3)_H$ horizontal symmetry 
\cite{John,su3H}. 
In the course of the presentation, it will 
become apparent that the discussion has to take a less 
general character as for example some additional symmetries 
should be invoked. This aspect will be faced in Sect. 2.2. 
In the discussion  $SU(5)$ is taken as 
a prototype theory featuring the basic properties 
of grand unification. In a more general context 
one could think of GUTs based on larger groups 
(e.g. $SU(N)$ with $N \geq 6$) 
which contain $SU(5)$ and perhaps also unify 
both $SU(5)$ and $SU(3)_H$. 
A particular model based on $SO(10)\times SU(3)_H$ 
will be presented in the next section.

\subsection{General aspects}

We consider the supersymmetric $SU(5)$ model 
with the horizontal symmetry $SU(3)_H$ where three 
fermion families are unified in chiral superfields: 
\be{f}
\bar5_i=(d^c,l)_i \sim (\bar5,3), ~~~~~~
10_i = (u^c,q,e^c)_i \sim (10,3), 
\ee 
($i=1,2,3$ is $SU(3)_H$ index),  
while the Higgs superfields $5,\bar5$ 
are singlets of $SU(3)_H$:
\be{H}
H\sim (5,1), ~~~~~~ \bar{H}\sim (\bar5,1).
\ee
We also assume that the theory is invariant under R-parity. 
In other words, we impose the matter parity $Z_2$ 
under which the matter superfields change the sign 
while the Higgs ones are invariant.\footnote{  
The $SU(3)_H$ symmetry itself does not not prevent the 
$R$-parity violating couplings 
$\eps^{ijk}10_i \bar{5}_j \bar{5}_k$ 
from where the $B-L$ violating terms 
$u^c d^c d^c,~ q d^c l,~ e^c l l$ arise. 
For a general discussion on the relation between 
$R$ parity and horizontal symmetries, see \cite{ZE}. 
In particular,  in case of the horizontal symmetry 
$SU(4)_H$ the R-parity could emerge as an 
accidental symmetry. One has to remark, however, 
that in many cases considered below the 
matter parity automatically follows from the 
discrete symmetries imposed on the model.}

Since the fermion bilinears transform as 
$3\times 3 = \bar{3} + 6$, their ``standard'' Yukawa 
couplings to the Higgses are forbidden by the horizontal 
symmetry.\footnote{
The following discussion can  equally apply in both the    
cases of global and local horizontal symmetry, 
though it would be more appealing  
to regard $SU(3)_H$ as gauge symmetry, like $SU(5)$.
} 
Hence, the fermion masses can be induced only by 
higher order operators involving a set of ``horizontal'' 
Higgs superfields $X^{ij}$ in two-index representations 
of $SU(3)_H$: 
symmetric $X_s^{ij}\equiv S^{\{ij\}}\sim (1,\bar6)$ and 
antisymmetric $X_a^{ij}\equiv A^{[ij]} = 
\eps^{ijk} A_k\sim (1,3)$:\footnote{
The theory may also contain  conjugated Higgses 
$\bar{X}_{ij}$ in representations $\bar{S}\sim (1,6)$ 
and $\bar{A}\sim (1,\bar3)$.  
These usually are needed for writing non-trivial 
superpotential terms in order to generate the 
horizontal VEVs (see Appendix A). 
The conjugated fields, however, do not couple the 
fermion   superfields (\ref{f}) and thus do not 
contribute to the quark and lepton masses.}  
\be{hoo}
\cO_u =\frac{S^{ij}}{M} 10_i 10_j H ,~~~~~
\cO= \frac{S^{ij}+A^{ij}}{M} \bar{5}_i 10_j \bar{H} , ~~~~~
\cO_\nu  = \frac{S^{ij}}{M^2} \bar{5}_i \bar{5}_j H^2
\ee
where $M$ is some large scale (the flavour scale).
Needless to say, 
by $SU(5)$ symmetry reasons, the  
antisymmetric Higgses $A$ can participate only in $\cO$. 
Therefore, in the low-energy limit the operators (\ref{hoo}) 
reduce to the Yukawa couplings:   
\be{hoo2}
\bY_u \sim \frac{\langle S\rangle}{M} , ~~~~~
\bY_d,\bY_e^T \sim 
\frac{\langle S\rangle + \langle A \rangle}{M} , ~~~~~
\frac{\bY_\nu}{M_L} \sim  \frac{\langle S\rangle}{M^2} . 
\ee
In this way, the fermion mass hierarchy can be 
naturally linked to the hierarchy of the horizontal-symmetry 
breaking scales \cite{su3H,PLB85}. 
For more details on the horizontal VEV structures, see 
e.g. \cite{BDJC}. 

In particular, let us assume that the horizontal Higgses 
include one sextet $S$ with a VEV taken diagonal:
\be{S} 
\langle S^{ij} \rangle = 
\matr{{\cal S}_1}{0}{0} {0}{{\cal S}_2}{0} {0}{0}{{\cal S}_3} , 
~~~~~ \cS_3 \gg \cS_2 \gg \cS_1 ,  
\ee
and a set of triplets $A_n^{ij} \equiv \eps^{ijk} A_{nk}$,  
$n=1,2,3$, 
which in general have the VEVs towards all three components:
\be{A}
\sum_n \langle A_n^{ij} \rangle = 
\matr{0}{\cA_3}{\cA_2} {-\cA_3}{0}{\cA_1} {-\cA_2}{-\cA_1}{0} , 
~~~~~ \cA_1 > \cA_2 > \cA_3 .  
\ee
(see Appendix A). Then from (\ref{hoo2}) we see that 
$Y_t\sim 1$ implies $\cS_3\sim M$ 
which indeed can naturally arise from the Higgs sector 
as we show in Appendix A.  
Similarly one can  expect that also $Y_{b,\tau}\sim 1$ 
which  would require large $\tan\beta$ regime. 
However, as we shall see below, 
in realistic schemes also  small 
$\tan\beta$  can be naturally accommodated. 
In addition, the operator $\cO_\nu$ translates into 
$Y_3/M_L \sim\cS_3/M^2\sim Y_t/M$. 
Therefore,  the flavour scale $M$ is of the order of  
the B-L violating scale $M_L\sim10^{15}$ GeV (cfr. (\ref{Y3})). 
The closeness of the flavour scale with 
the GUT scale $M_G \sim 10^{16}$ GeV suggests that 
they may have a common origin.
(The mismatch between the estimate of $M$ and $M_G$  
can be due to some spread of 
the coupling constants in the theory.) 
  
It is well known that the effective operators (\ref{hoo})  
may arise entirely from a renormalizable theory, 
after integrating out some heavy degrees of freedom 
\cite{FN}, and thereby the flavour scale can be  
related to the mass scale of the latter.  
It is also plausible to think that this mass scale 
is set by large (order GUT scale) VEV of some Higgs 
$\Omega$, $\langle \Omega\rangle \sim M $. 

In particular, one can introduce vector-like  states 
in the $10$- and $\ov{10}$-representations of $SU(5)$:  
\be{T}
T^i = (U^c,Q,E^c)^i \sim (10,\bar{3})~, ~~~~~~  
\ov{T}_i = (U,Q^c,E)_i \sim (\ov{10},{3}) ,  
\ee 
and the $SU(5)$-singlet fields
\be{N}
N^i \sim (1,\bar3), ~~~~~~ \ov{N}_i   \sim (1,3), 
\ee 
and consider the following Yukawa couplings 
in the superpotential: 
\beqn{ren}
&& 
W_T= 10HT + \bar{5}\bar{H}T + \ov{T} X 10 + T \Omega \ov{T}
\nonumber \\ 
&&
W_N= \bar5 HN + X_s\ov{N}^2 + N \Omega \ov{N}
\eeqn
(family indices are suppressed, and order 1 coupling 
constants are understood in each term). 
Clearly, in $W_T$ the horizontal Higgs $X$ can be  
both of the type $X_s=S$ or $X_a=A$, 
whereas in $W_N$ only the symmetric 
Higgses $X_s$ can participate.  
Since  $T,\ov{T}$ contain heavy states with 
electric charges $2/3$, $-1/3$ and $-1$, their 
exchange can induce the operators 
$\cO_u$ and $\cO$ relevant for the masses of all 
charged fermions: up quarks, down quarks and leptons.  
Analogously, the operator  $\cO_\nu$ for  
the neutrino masses emerges via the exchange 
of the singlet states $N,\ov{N}$. 
The relevant diagrams are shown in Fig. \ref{fig-T}.

Therefore, after integrating out the heavy states,
we obtain the operators (\ref{hoo}) in the form 
\beqn{hoo1}
&&
\cO_u= 10 (H \bM_T^{-1}X + X^T\bM_T^{-1}H) 10 = 
2\cdot 10 H (\bM_T^{-1}S) 10 + 
10 H (\bM_T^{-1}A - A\bM_T^{-1}) 10 , \nonumber \\ 
&&
\cO= \bar5 (\bar{H} \bM_T^{-1}X) 10 =   
\bar5 \bar{H} (\bM_T^{-1}S + \bM_T^{-1}A) 10 ,  
\nonumber \\ 
&&
\cO_\nu= 
\bar{5} H \bM_N^{-1} S \bM_N^{-1}  H \bar{5} 
\eeqn  
where $\bM_T,\bM_N$ are  $3\times 3$ mass matrices of 
the heavy states, induced from their couplings to 
the Higgs $\Omega$. 
Therefore, if $\Omega $ is a gauge singlet,  
then $\bM_T,\bM_N$  will be degenerate in family space, 
$\bM_{T,N} \sim M\cdot\bunity$,  where 
$\bunity = {\rm Diag}(1,1,1)$ is the unit matrix of $SU(3)_H$.  
In this case, if $X$  are $SU(5)$ singlets, only the symmetric 
fields $X_s=X_s^T$ can contribute $\cO_u$ in (\ref{hoo1}),  
while the contributions of the antisymmetric ones 
$X_a=-X_a^T$ cancel out. On the other hand, 
both $X_s$ and $X_a$ can contribute to $\cO$. 
Therefore, the operators (\ref{hoo1}) reduce to 
the Yukawa couplings of the form:
\be{hoo3}
\bY_u \sim \bM_T^{-1} \langle S \rangle , ~~~~~
\bY_d,\bY_e^T \sim \bM_T^{-1} \langle S \rangle + 
\bM_T^{-1} \langle A \rangle , ~~~~~
M_L^{-1}\bY_\nu \sim  \bM_N^{-2} \langle S\rangle .
\ee
More explicitly, these can be written as:
\be{stech-q} 
\bY_u\propto \bS, ~~~~ \bY_\nu \propto \bS, ~~~~ 
\bY_d = \bY_e^T \propto \bS + \bA , 
\ee 
where $\bS=M^{-1} \langle S \rangle$ and 
$\bA=M^{-1} \langle A \rangle$ are respectively  
the symmetric and antisymmetric matrices reflecting the form 
of the VEVs (\ref{S}) and (\ref{A}). 
(We omit the dimensionless factors 
which keep track of the coupling constants 
in (\ref{ren}) and are not specified at the moment).

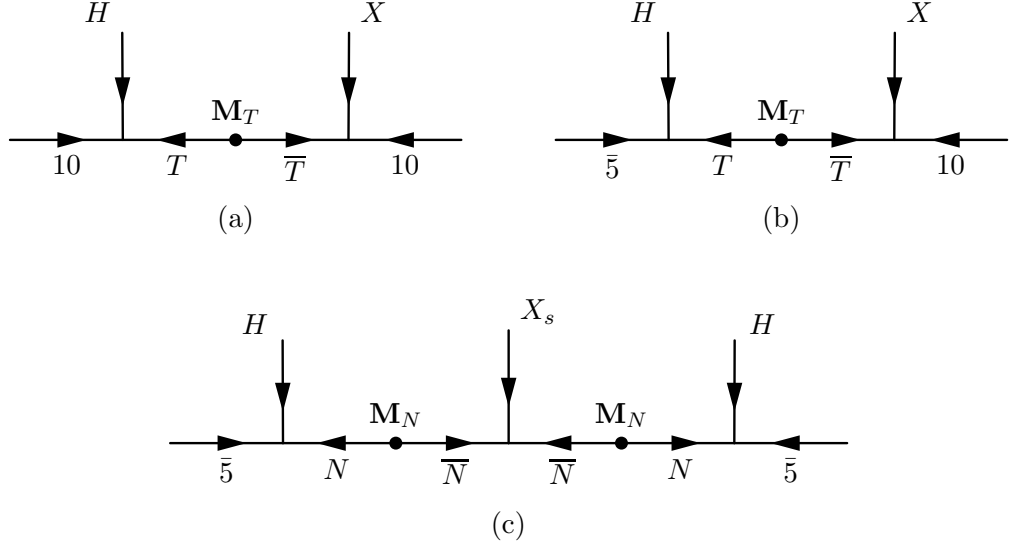
\begin{figure}
\begin{center}
%
\begin{fmfgraph*}(60,30)
\fmfleft{t1}
\fmfright{t2}
\fmftop{x1,H,S,X,x2} \fmfbottom{a}
\fmf{fermion,label=$10$}{t1,v1}
\fmf{fermion,label=$T$,label.side=left}{v2,v1}
\fmf{fermion,label=$\ov{T}$}{v2,v3}
\fmf{fermion,label=$10$,label.side=left}{t2,v3}
\fmfv{label=${\bf M}_T$,label.angle=90}{v2}
\fmfv{label=(a),label.dist=7,label.angle=90}{a}
\fmffreeze
\fmf{fermion}{H,v1}  \fmf{fermion}{X,v3}
\fmfdot{v2}  
\fmflabel{$H$}{H}
\fmflabel{$X$}{X}
\end{fmfgraph*}
\hspace{1cm}
\begin{fmfgraph*}(60,30)
\fmfleft{t1}
\fmfright{t2}
\fmftop{x1,H,S,X,x2}
\fmfbottom{b}
\fmf{fermion,label=$\bar{5}$}{t1,v1}
\fmf{fermion,label=$T$,label.side=left}{v2,v1}
\fmf{fermion,label=$\ov{T}$}{v2,v3}
\fmf{fermion,label=$10$,label.side=left}{t2,v3}
\fmfv{label=${\bf M}_T$,label.angle=90}{v2}
\fmfv{label=(b),label.dist=7,label.angle=90}{b}
\fmffreeze
\fmf{fermion}{H,v1}
\fmf{fermion}{X,v3}
\fmfdot{v2}  
\fmflabel{$\bar{H}$}{H}
\fmflabel{$X$}{X}
\end{fmfgraph*}\\
\vspace{1.cm}
\begin{fmfgraph*}(90,30)
\fmfleft{t1}
\fmfright{t2}
\fmftop{x1,H,S,X,S1,H1,x2}
\fmfbottom{b}
\fmf{fermion,label=$\bar{5}$}{t1,v1}
\fmf{fermion,label=$N$,label.side=left}{v2,v1}
\fmf{fermion,label=$\ov{N}$}{v2,v3}
\fmf{fermion,label=$\ov{N}$,label.side=left}{v4,v3}
\fmf{fermion,label=$N$}{v4,v5}
\fmf{fermion,label=$\bar{5}$,label.side=left}{t2,v5}
\fmfv{label=${\bf M}_N$,label.angle=90}{v2}
\fmfv{label=${\bf M}_N$,label.angle=90}{v4}
\fmfv{label=(c),label.dist=6,label.angle=90}{b}
\fmffreeze
\fmf{fermion}{H,v1}
\fmf{fermion}{X,v3}
\fmf{fermion}{H1,v5}
\fmfdot{v2,v4}  
\fmflabel{$H$}{H}
\fmflabel{$H$}{H1}
\fmfv{label=$X_s$,label.dist=5,label.angle=30}{X}
\end{fmfgraph*}
\caption{\small The heavy fermion exchanges 
giving rise to the effective higher order operators 
for the fermion masses: (a) for up quarks, 
(b) for down quarks and charged leptons,  
(c) for  neutrinos.
\label{fig-T}}
\end{center}
\end{figure}

It should be immediately noted, however, that the texture 
(\ref{stech-q}) is far from being realistic for many 
reasons: 

\noindent
(A) 
$\bY_d \propto \bY_u + \bA$  
with antisymmetric $\bA$ leads to very big 23 mixing, 
$s_{23}^q\simeq (Y_s/Y_b)^{1/2} \sim 0.1 - 0.2$, 
much above the experimental value of $V_{cb}$.

\noindent
(B) the relation $\bY_\nu \propto \bY_u$ implies  
$Y_1:Y_2:Y_3 = Y_u:Y_c:Y_t$ and thus 
too  big a hierarchy between the neutrino mass 
eigenvalues $m_3$ and $m_2$  (cfr. Table \ref{tab1}).

\noindent
(C) finally, $\bY_d=\bY_e^T$ features the minimal $SU(5)$ 
degeneracy of the Yukawa couplings, $Y_{d,s,b}= Y_{e,\mu,\tau}$, 
whose drawbacks have been already recalled.

The latter problem can be easily cured by making use  
of the adjoint superfield $\Phi$ which 
breaks $SU(5)$ down to $SU(3)\times SU(2)\times U(1)$ 
by the  VEV  proportional to the 
hypercharge generator:  
$\langle\Phi\rangle \propto \lambda_y
\equiv {\rm Diag}(1/3,1/3,1/3,-1/2,-1/2)$.  
Thus if   
the matrix $A$ in (\ref{hoo2}) effectively contains 
$\Phi$,  $A=A(\Phi)$, effective Clebsch factors emerge 
discriminating quarks from leptons.
Therefore, at least some of the antisymmetric Higgses 
$A$ can be taken in the mixed representation $A\sim (24,3)$.\footnote{ 
Alternatively, one can take  all the  
antisymmetric Higgses as $SU(5)$ singlets, 
$A\sim (1,3)$, but assume that some of them 
act in combination with $\Phi \sim (24,1)$, 
to end up with  the effective operator 
$\frac{A \Phi}{M^2}\bar{5}\cdot10 \bar{H}$ (cfr. (\ref{hoo}))   
so that the tensor product $\bar{H}\cdot \Phi\cdot A$ 
effectively contains both $(\bar5,3)$ and 
$(\ov{45},3)$ representations. Such an attitude 
will be taken in constructing the $SO(10)$ model in Sect. 3.}  
The tensor product is  
$\bar5\times 24 = \bar5 + \ov{45} + 70$, where only the first two 
terms are relevant for the fermion bilinear 
$\bar5 \times 10 = 5 + 45$. 
Then the unwanted
relations $Y_{d,s}=Y_{e,\mu}$ 
can be avoided while maintaining the successful relation 
$Y_b \simeq Y_\tau$. 

The first two problems can be solved in a similar way, by 
making use of the adjoint Higgs of $SU(3)_H$.
The masses of the heavy fermions 
transform as $3\times \bar{3} = 1+8$ representations. 
Hence one can assume that the set of Higgses $\Omega$ 
contains, besides the singlet $I\sim (1,1)$,  also 
the superfield $\Sigma \sim (1,8)$ with the VEV 
$\langle\Sigma\rangle$ pointing towards the $\lambda_8$ 
generator:\footnote{
In the following of the paper, the symbol $\Omega$ stands 
for a set of Higgs fields in real  
representations of the gauge groups. When needed the content 
of $\Omega$ will be specified as singlet    
$I \sim (1,1)$,  octet $\Sigma \sim (1,8)$ or 24-plet 
$\Phi \sim (24,1)$.}  
\be{VEV-Sigma}
\langle \Sigma \rangle \sim M \cdot 
\matr{1}{0}{0} {0}{1}{0} {0}{0}{-2}~. 
\ee
This VEV breaks 
$SU(3)_H$ down to $SU(2)_H\times U(1)_H$ and thus 
discriminates the third 
fermion family from the others \cite{AB}. 
Hence, the heavy-fermion mass matrices  $\bM_{T,N}$  
maintain only the $SU(2)_H$ invariant form  
$\bM = M\cdot {\rm Diag}(1,1,b)$, 
with $b$ in general different from 1.  
In this way, from (\ref{hoo3}) we can obtain 
the desired  Yukawa pattern (\ref{Stech}).\footnote{ 
In group-theoretical language this can be rephrased as follows. 
The tensor product  $A \cdot \Sigma$  reads as 
$3\times 8 = 3 +\bar6 + 15$ where both 3 and $\bar6$ 
can match the fermion bilinears $ 3\times 3$. 
In this way, the off-diagonal entries in $\bY_{e,d}$ are 
not anymore antisymmetric -- only the 12 sector keeps 
on the antisymmetry owing to the residual 
$SU(2)_H$ symmetry. } 

Notice however, that for achieving our goal, 
we have implemented   the adjoint structures 
($\Sigma$ into $M$ and $\Phi$ into $A$) 
in eq. (\ref{hoo3}) which is not the right place. 
Indeed, (\ref{hoo3}) follows from (\ref{hoo1}) 
if and only if neither $M$ nor $A$ 
contain adjoint representations of $SU(5)$ or 
$SU(3)_H$. For example, if $A$ contains the  
$SU(5)$ adjoint, then the tensor product 
$A\times H \sim (24,3)\times (5,1)$ effectively contains the  
(45,3) representation which would 
induce off-diagonal  antisymmetric contribution also 
to $\bY_u$. On the other hand, 
if $A$'s are $SU(5)$ singlet, but $\bM_T$ contains the 
$SU(3)_H$ octet, then $(1,3) \times (1,8) \times (5,1)$ 
effectively contains $(5,\bar{6})$ which would still give rise 
to symmetric off-diagonal entries in $\bY_u$. 
Therefore, in either case 
the antisymmetric VEVs $\langle A\rangle$ 
contribute the first term in (\ref{hoo}) and so induce 
 off-diagonal entries in $\bY_u$ ruining in such a way 
the desired  pattern.
Tackling this issue requires more  theoretical ingredients 
and tools which will be discussed below.

\subsection{Some specific realizations}

Thus, we have seen that in order to arrive 
at the desired textures (\ref{Stech}), 
both the $SU(3)_H$ singlet $I\sim (1,1)$ and octet 
$\Sigma \sim (1,8)$ are needed for 
generating the heavy fermion masses. 
In addition, the diagrams contributing via the 
$S$ and $A$ type Higgses are to be ``differentiated'' 
by some additional symmetry so that $S$ will contribute 
to all the matrices $\bY_u$, 
$\bY_\nu$ and $\bY_{d,e}$, while $A$'s only to $\bY_{d,e}$. 
Hence, we  have to assume that besides the 
local $SU(5)\times SU(3)_H$ symmetry, the theory 
is invariant under some set $\cG$ of abelian symmetries 
which may contain e.g. non-anomalous or pseudo-anomalous 
gauge $U(1)$ factors, continuous or discrete R-symmetry, 
discrete symmetries like $Z_2$, $Z_3$, etc.  

First, on the basis of the considerations outlined in 
the previous section, we put forward the needed couplings 
in the superpotential. Second, those will be  motivated  
by some additional symmetry of $\cG$.

Now about the first step. We assume that the flavour scale 
is determined by the VEVs of the singlet $I\sim (1,1)$ 
and the octet $\Sigma \sim (1,8)$, respectively 
$\langle I\rangle = M$ and 
$\langle \Sigma \rangle = xM\cdot {\rm Diag}(1,1,-2)$, 
$x\sim 1$. Then the  combination $I+\Sigma$  provides 
order $M$ masses to the heavy fermions, splitting 
the third generation from the first two and maintaining 
the  latter degenerate. As in the previous subsection, we take 
the horizontal VEVs in the form (\ref{S}) and (\ref{A}) and 
define the dimensionless VEV matrices as  
$\bS \equiv M^{-1}\langle S\rangle$ and 
$\bA_n \equiv M^{-1}\langle A_n\rangle$.

$\bullet$ 
We further assume that the superpotential $W_T$ (\ref{ren}) 
related to heavy 10-plets (\ref{T}) involves some set 
of horizontal Higgses $X$, which in fact consists of 
the sextet $S \sim (1,\bar6)$.  
Therefore, $W_T$  becomes:   
\be{ren-T} 
W_T= f 10HT + g\bar{5}\bar{H}T + \la_T\ov{T} X 10 
+  T (\alpha_T I+\beta_T\Sigma) \ov{T} ,  
\ee 
where we explicitly indicate the order 1 coupling constants.   
So the diagonal entries in $\bY_u$ and $\bY_{d,e}$ are  
induced respectively via the diagrams (a,b) of Fig. \ref{fig-T}.   
In order to respect the $b-\tau$ Yukawa unification, 
the 24-plet $\Phi\sim (24,1)$ should not couple to $T\ov{T}$.

$\bullet$ 
The couplings in $W_N$ are left as in (\ref{ren}) 
and they induce 
$\bY_\nu$ via the diagram (c) in Fig. \ref{fig-T}:  
\be{ren-N} 
W_N= h\bar5 HN +  \la_N S\ov{N}^2 + 
N (\alpha_N I+\beta_N \Sigma)\ov{N} .  
\ee 
Thus, $\bY_\nu$ is diagonal in the VEV basis (\ref{S}). 

$\bullet$ 
Let us consider another set of horizontal Higgses $X'$, 
which contains antisymmetric fields $A_n$ in the 
representations $(1,3)$ and $(24,3)$,  
and assume that they contribute only to $\bY_{d,e}$ 
via the exchange of other heavy fermions.  
The simplest possibility is to introduce additional 
states  in 5- and $\bar5$-representations of $SU(5)$: 
\be{F}
\ov{F}^i = (D^c,L)^i \sim (\bar{5},\bar{3}) , ~~~~~~  
{F}_i = (D,L^c)_i  \sim (5,3)~,
\ee 
having the following couplings in the superpotential:  
\be{ren-F} 
W_F= \la_F \bar5 X' F + g'\bar{F} \bar{H}'10 +  
 F (\alpha_F I+\beta_F \Sigma) \ov{F} ,  
\ee
where $\bar{H}'$ is another Higgs $\bar5$-plet 
(or $\ov{45}$-plet) of $SU(5)$. 
The corresponding diagram is shown in Fig. \ref{fig-F}(a). 
In the following, such a scenario will be called 
$F$-scheme. 

Alternatively, instead of the $F +\bar{F}$ 
one can involve  an additional pair of 10-plets 
\be{T'}
T'^i = (U'^c,Q',E'^c)^i \sim (10,\bar{3})~, ~~~~~~  
\ov{T}'_i = (U',Q'^c,E')_i \sim (\ov{10},{3}) ,  
\ee
with the superpotential terms: 
\be{ren-T1}
W_{T'} = g' \bar5 \bar{H}'T' + \la'_T \ov{T}' X' 10 
+ T' (\alpha'_T I+ \beta'_T \Sigma) \ov{T}' .
\ee 
The corresponding exchange is shown in Fig. \ref{fig-F}(b).
This scenario will be referred to as the $T'$-scheme. 

In either scheme, the standard Higgs doublet $H_d$ 
should be regarded a superposition of the doublet 
components in $\bar{H}$  and $\bar{H}'$:  
$H_d = c_\omega \bar{H}'_2 - s_\omega\bar{H}_2$, 
which is rendered light after arranging the 
doublet-triplet splitting, 
while the orthogonal combination 
$s_\omega \bar{H}'_2 + c_\omega\bar{H}_2$  
is left heavy, with mass $\sim M_G$    
(here $\omega$ is a mixing angle and 
$c_\omega(s_\omega) =\cos\omega(\sin\omega)$). 
On the other hand, we assume that 
$H_u$ comes entirely from $H$, while the  field 
$H'$, though  present, is just a spectator 
in the Yukawa sector. On the contrary, $H_d$ 
is contained in both $\bar{H}$ and $\bar{H}'$ 
with weights $s_\omega$ and $c_\omega$ respectively.  

Once the fields $A_n=(X_a)_n$ contain the mixed representation 
$(24,3)$ and/or $\bar{H}'$ is 45-plet, the diagrams 
in Fig. \ref{fig-F} induce off-diagonal entries in $\bY_d$ 
and $\bY_e^T$ with different Clebsch factors.


\begin{figure}
\begin{center}
%
\begin{fmfgraph*}(60,30)
\fmfleft{t1}
\fmfright{t2}
\fmftop{x1,H,S,X,x2}
\fmfbottom{b}  
\fmf{fermion,label=$\bar{5}$}{t1,v1}
\fmf{fermion,label=$F$,label.side=left}{v2,v1}
\fmf{fermion,label=$\bar{F}$}{v2,v3}
\fmf{fermion,label=$10$,label.side=left}{t2,v3}
\fmfv{label=$\bM_F$,label.angle=90}{v2}
\fmfv{label=(a),label.dist=5,label.angle=90}{b}
\fmffreeze  
\fmf{fermion}{H,v1}
\fmf{fermion}{X,v3} 
\fmfdot{v2}
\fmflabel{$X'$}{H}
\fmflabel{$\bar{H}'$}{X}
\end{fmfgraph*}
\hspace{1cm}
\begin{fmfgraph*}(60,30)
\fmfleft{t1}
\fmfright{t2}
\fmftop{x1,H,S,X,x2}
\fmfbottom{b}
\fmf{fermion,label=$\bar{5}$}{t1,v1}
\fmf{fermion,label=$T'$,label.side=left}{v2,v1}
\fmf{fermion,label=$\ov{T}'$}{v2,v3}
\fmf{fermion,label=$10$,label.side=left}{t2,v3}
\fmfv{label=$\bM'_T$,label.angle=90}{v2}
\fmfv{label=(b),label.dist=5,label.angle=90}{b}
\fmffreeze
\fmf{fermion}{H,v1}
\fmf{fermion}{X,v3}
\fmfdot{v2}  
\fmflabel{$\bar{H}'$}{H}
\fmflabel{$X'$}{X}
\end{fmfgraph*}
%
\caption{\small The diagrams inducing the off-diagonal 
contributions to $\bY_{d,e}$: 
(a) exchange of 5-plets ($F$-scheme),  
(b) exchange of additional 10-plets ($T'$-scheme). 
\label{fig-F}}
\end{center}
\end{figure}
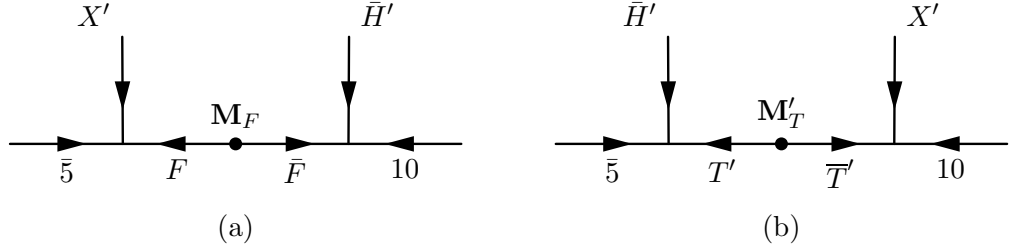


Let us consider in some more details e.g. the $F$-scheme. 
After substituting the VEVs of 
the horizontal Higgses,  the couplings  
(\ref{ren-T}), (\ref{ren-N}) and (\ref{ren-F}) 
give rise to the following  field-dependent mass matrices for 
the up quark and neutrino states:
\be{MlU}
\begin{array}{ccc}
 & {\begin{array}{ccc} \,q ~& \,\,\,\,\;Q &\,\,\,\, \,\,~U 
\end{array}}\\ \vspace{2mm}
\begin{array}{c}
u^c \\ Q^c \\ U^c   \end{array}\!\!\!\!\!&{\left(\begin{array}{ccc}
0 & \brf H_u & \bmu^T_u \\ 
\bmu_u & \bM_T & 0 \\ 
\brf H_u & 0 & \bM_T  \end{array}\right)} 
\end{array} \! , ~~~~~~~~
\begin{array}{ccc}
 & {\begin{array}{ccc}\nu ~& \,\,\,\,\;N &\,\,\,\, \,\,~N^c  
\end{array}}\\ \vspace{2mm}
\begin{array}{c}
\nu \\ N \\ N^c\end{array}\!\!\!\!\!&{\left(\begin{array}{ccc}
0 & \bh H_u & 0   \\
 \bh H_u & 0 & \bM_N \\ 
0 &  \bM_N& \bmu_\nu  \end{array}\right)} 
\end{array} \! , 
\ee 
and for the down quark and charged lepton states: 
\be{Ml}
\begin{array}{ccc}
 & {\begin{array}{ccc}\,q ~& \,\,\,\,\;~~Q &\,\,\,\, \,~D
\end{array}}\\ \vspace{2mm}
\begin{array}{c}
d^c \\ Q^c \\ D^c\end{array}\!\!\!\!\!&{\left(\begin{array}{ccc}
0 & \bg \bar{H}_2 & \bmu_d \\
\bmu_u & \bM_T & 0 \\ 
\bg' \eta_d \bar{H}'_2 & 0 & \bM_F  \end{array}\right)}
\end{array} \! , ~~~~~
\begin{array}{ccc}
& {\begin{array}{ccc}l
~& \,\,\,\,\;E &\,\,\,\, \,\,~L  \end{array}}\\ \vspace{2mm}
\begin{array}{c}
e^c \\ E^c \\ L^c\end{array}\!\!\!\!\!&{\left(\begin{array}{ccc}
0 & \bmu^T_u &\bg' \eta_e \bar{H}'_2\\
\bg \bar{H}_2 & \bM_T & 0 \\
\bmu^T_l & 0 & \bM_F  \end{array}\right)}
\end{array} \! , 
\ee
where $\eta_{d,e}$ are the Clebsch factors dependent on the 
$SU(5)$ content of $\bar{H}'$. Namely, $\eta_e=\eta_d$ 
if $\bar{H}'$ is $\bar5$-plet and $\eta_e=-3\eta_d$ if 
$\bar{H}'$ is $\ov{45}$-plet.  
In the above, 
each entry is a $3\times 3$ matrix in the $SU(3)_H$ space.    
In particular $\brf= f\bunity$, $\bh =h\bunity$, etc. 
are flavour-blind ($SU(3)_H$ degenerate)  matrices. 

As long as the heavy fermion mass matrices are induced 
by $I$ and $\Sigma$, they will be diagonal and have    
$SU(2)_H$ invariant shape $\bM_\Psi = a_\Psi M \bb_\psi$  
($\Psi={T,F,N}$), where  
\be{shape}
\bb_\Psi =  
\matr{1}{0}{0} {0}{1}{0} {0}{0}{b_\Psi} , ~~~~  
a_\psi = \alpha_\Psi + x\beta_\Psi , ~~~~ 
b_\psi = \frac{\alpha_\Psi - 2x\beta_\Psi}
{\alpha_\Psi + x\beta_\Psi}~ .  
\ee
These will give rise to the deformation factors 
$\bb$ and $\bb'$ in (\ref{Stech}) whose relations with the 
factors $\bb_{\psi}$ are easy to catch -- at any rate 
they will be shown  later.

Thus, the main information on the fermion flavour pattern 
is contained in the matrices $\bmu_{u,d,l,\nu}$.
The matrices $\bmu_{u,\nu}$ 
are proportional to the VEV 
$\langle S\rangle = {\rm Diag}(\cS_1,\cS_2,\cS_3)$,  
$\bmu_u = \la_T M\bS$ and $\bmu_\nu=\la_N M\bS$.   
As for the matrices $\bmu_{d,e}$, they are antisymmetric 
and emerge  from the VEVs of the triplets $A_n$ 
sketched in  eq.  (\ref{A}).  
As far as the latter set contain $(1,3)$ and $(24,3)$ 
representations, all those  entries in general are 
non-trivial $5\times 5$ matrices,   
$\tilde{\cA}_k = \cA_k(\unity + \zeta_k \lambda_y)$, 
where $\unity$ and $\lambda_y$  are the unit and hypercharge 
matrices in $SU(5)$ space, respectively 
and $\zeta_k$ measures the relative weight of the 
singlet and 24-plet VEVs at each component ($k=1,2,3$). 
Thus, the corresponding entries in 
$\bmu_{d}$ and $\bmu_l^T$ differ by  
Clebsch factors $C^{d,e}_k = 1+\zeta_k y_{d,l}$, 
where $y_d=1/3$ and $y_l=-1/2$ are 
the hypercharges of the $d^c$ and $l$ fragments of 
$\bar5$-plets.  

Therefore, after integrating out the heavy states, 
we obtain an effective low energy theory with the 
following Yukawa matrices:
\beqn{Y-all}
&&
\bY_u = f(\bM_T^{-1}\bmu_u +\bmu_u^T \bM_T^{-1}) 
\nonumber \\
&&
\bY_d = g s_\omega \bM_T^{-1} \bmu_u + 
g' c_\omega \eta_d \bmu_d \bM_F^{-1}  
\nonumber \\  
&&
\bY_e^T = g s_\omega \bM_T^{-1} \bmu_u + 
g' c_\omega \eta_e \bmu_l \bM_F^{-1}    
\eeqn
and 
\be{Ynu}
\frac{\bY_\nu}{M_L} = h^2 \bM_N^{-2}\bmu_\nu . 
\ee
Therefore, $\bY_u$ and $\bY_\nu$ get contribution 
only from $S$: 
\be{u,nu}
\bY_u = \frac{2f\la_T}{a_T} \bb_T^{-1}\bS, ~~~~ 
\frac{\bY_\nu}{M_L} = \frac{h^2 \la_N}{a_N^2 M} \bb_N^{-2}\bS~,  
\ee  
and thus both have diagonal forms related as:  
\be{Yu}
\frac{\bY_\nu}{M_L} = \eta\bb'^{-1} \frac{\bY_u}{M},  
\ee
where $\bb'={\rm Diag}(1,1,b')$~, and  
\be{eta}
b'=\frac{b_N^2}{b_T}, ~~~~  
\eta = \frac{h^2\la_N a_T}{2f \la_T a_N^2}~.
\ee

As for $\bY_{e,d}$, 
they have  a diagonal  contribution $g s_\omega/(2f)\bY_u$   
from the first term in (\ref{Y-all}), and  off-diagonal 
entries $ \bA_n \bb_F^{-1}$ from the  
the second term.
Therefore, the full matrices $\bY_{d,e}$  
can be rewritten as  
\be{Yuk-de}
\bY_d = \rho \bY_u + \bb^{-1}\bA_d, ~~~~~~
\bY_e^T = \rho \bY_u + \bb^{-1}\bA_e , 
\ee
where $\rho = \frac{g}{2f}s_\omega$ and 
$\bb={\rm Diag}(1,1,b)$, with $b =b_F^{-1}$, 
while the other dimensionless coefficients, such as  
the Clebsch factors related to the representations of 
$\bar{H}'$ and $A$, are absorbed into the definition of 
the entries of  the antisymmetric matrices 
\be{Ade}
\bA_{d,e}=
\matr{0}{A_3^{d,e}}{A_2^{d,e} }  
{-A_3^{d,e}}{0}{A_1^{d,e} }  
{-A_2^{d,e} }{-A_1^{d,e} }{0} ,  
\ee
so that the relative Clebsch factors are defined as  
$k_i = A^e_i/A^d_i = \frac{\eta_e C^e_i}{\eta_d C^d_i}$.  
\noindent 
Putting everything together, we can write the Yukawa matrices as: 
\beqn{Stech-F}
& 
\bY_u = \matr{Y_u}{0}{0} {0}{Y_c}{0} {0}{0}{Y_t}, 
& 
\bY_d = \matr{\rho Y_u}{A^d_3 }{A^d_2 } 
{-A^d_3}{\rho Y_c}{A^d_1}
{-\frac1b A^d_2}{-\frac1b A^d_1}{\rho Y_t},  
\nonumber \\ 
& 
\bY_\nu=\eta\matr{Y_u}{0}{0}{0}{Y_c}{0}{0}{0}{\frac{1}{b'}Y_t}, 
&
\bY_e^T = \matr{\rho Y_u}{A^e_3}{A^e_2} 
{-A^e_3}{\rho Y_c}{A^e_1}
{-\frac1b A^e_2}{-\frac1b A^e_1}{\rho Y_t} . 
\eeqn

The following comments are in order. 

$\bullet$ First of all, 
$Y_t \sim 1$ implies that $\cS_3 \sim b_T M$, 
while the relations 
$Y_u : Y_c : Y_t \sim \eps^2 : \eps : b_T^{-1}$, 
well fits the upper quark mass hierarchy  
for $\eps \sim 10^{-2}$ and $b_T\sim 0.1-1$. 
As already noted in Sect. 2.1 this situation 
can  naturally emerge from the pattern of the 
horizontal Higgs superpotential (see Appendix A). 

$\bullet$ Since the (3,3) element in $\bY_{d,e}$ is 
also related to largest scale $\cS_3$, 
we have approximate $b-\tau$ Yukawa unification at the 
GUT scale: $Y_\tau\simeq Y_b\simeq \rho Y_t$. 
On the other hand, as in principle 
the mixing angle $\omega$ between the doublets in 
$\bar{H}$ and $\bar{H}'$ can be  small, i.e. 
$\sin\omega < 1$, the case $Y_b/Y_t \ll 1$ is plausible 
even if the constants $f$ and $g$ are both of order 1. 
Therefore, the regime of moderate $\tan\beta$ 
is quite a natural possibility,  in which case 
the renormalization group equations can be substantially 
simplified. In the following we  assume that regime, 
though the case with large $\tan\beta$ should not 
be excluded.  

$\bullet$ The large 23  lepton mixing  implies 
$A^e_1 \sim Y_\tau$. This also seems natural in the context 
of the model as  the largest triplet VEV $\cA_1$ 
should be of the order of the flavour scale scale $M$ 
(see Appendix A). 
On the other hand, the small 23 mixing of quarks 
needs $A^d_1/b < 0.1 Y_b$, i.e. a large asymmetry parameter, 
$b\sim 10$, which in turn implies $b_F\sim 0.1$. 
In this case 
we observe an inverse hierarchy between the first two and 
third family of the heavy $F$ states, i.e. 
$M^F_{1,2} > M^F_3$. 

$\bullet$ 
Concerning the neutrino Yukawa eigenvalues, they obey 
the relations  
\be{nu-pattern}
\frac{Y_1}{Y_2} = \frac{Y_u}{Y_c} \ll 1, ~~~~ 
\frac{Y_2}{Y_3} = b' \frac{Y_c}{Y_t}  
\ee 
which also points to large $b'$ (cfr. Table \ref{tab1}), 
$b'\sim 5-100$. In particular, the relation $b'=b$  
could work, which can occur  if $b_T=b_F$ and $b_N=1$ -- 
that is  when the same singlet-octet combination   
$I+\Sigma$ acts in $W_T$ and $W_F$, 
while in $W_N$ only the singlet $I$ 
is present ($\beta_N=0$ in (\ref{ren-N})).

$\bullet$   
On the other hand, we obtain (cfr. (\ref{Y3})) 
\be{Y-3} 
\frac{M_L}{Y_3} = \eta^{-1} b'\frac{M}{Y_t} 
\sim 10^{15} ~{\rm GeV}~. 
\ee
Since $Y_t\sim 1$ and $b'\sim 10$, 
this translates into $\eta \sim 10^2$ if $M\sim 10^{16}$ GeV, 
i.e. order GUT scale.  It should not be, 
however, very surprising to find $\eta \sim 10^2$ due   
to some spread of the coupling constants in the theory 
(cfr. (\ref{eta})).

Let us 
remark that the set $X$ appearing in eq. (\ref{ren-T}) 
could contain also a triplet Higgs $A_3\sim (1,3)$ 
having a VEV towards the third component, 
$\langle A_3^{ij} \rangle = \eps^{ij3} \cA_3$. 
This would induce antisymmetric 12 entries in 
the matrix $\bmu_u$. This, however, 
would not alter the diagonal form of  $\bY_u$, since in 
the combination $\bM_T^{-1}\bmu_u + \bmu_u^T\bM_T^{-1}$ 
the antisymmetric entries cancel out 
(thanks to the residual $SU(2)_H$ symmetry in $\bM_T$). 
However, in $\bY_{d,e}$ 
this will induce antisymmetric 12 entries with  
Clebsch factors $k_e=k_d$.

The $T'$-scheme can be developed 
along the same lines. The only difference will be that 
in this case we have $b=b_{T'}$, so that there should be 
a direct hierarchy between the first two and 
third family of the heavy $T'$ states, i.e. 
$M^{T'}_{1,2} < M^{T'}_3$. In this case 
$b'=b$ can be obtained if $b_{T'}=b_N=b_T$, 
i.e. when $W_{T,T',N}$ share the  same $I+\Sigma$ combination.   
Alternatively, when $b_{T'}=b_N$ but $b_T=1$
(in $W_T$ contributes only  the  singlet $I$), 
we would obtain 
another interesting possibility $b'=b^2$ which can be 
also compatible with the experimental data. Namely, 
it would point towards smaller 3/2 hierarchy between 
the neutrino mass eigenvalues: $Y_3/Y_2 \sim 10$ 
(cfr. Table \ref{tab1}).


We now come to the second step 
to prove that the superpotentials (\ref{ren-T}), (\ref{ren-N}) 
and (\ref{ren-F}) can be justified by  some extra symmetry $\cG$. 
First of all, we assume that 
$\cG$ contains 
a discrete symmetry $\cR$ under which all superfields 
in the theory as well as the superpotential change  
sign. This will allow only  trilinear terms in 
the superpotential (provided that they are consistent  
also with $SU(5)\times SU(3)_H$ and other symmetries in $\cG$), 
while some singlet $Y$ can have also linear term 
$\Lambda^2 Y$. In this situation all VEVs (including 
the GUT and horizontal ones) can emerge from a single 
mass scale $\Lambda$. 

Second, we  introduce a  
$Z_3$ symmetry acting on the  superfields 
 as $\psi \to \psi\exp(\frac{2\pi}{3}i Q_\psi)$, 
with the following `charges' $Q_\psi$: 
\beqn{Q}
Q=0 : ~~~
&& Y, ~\Phi, ~ \bar{H}'; ~~ T,   
\nonumber \\ 
Q=1 : ~~~
&& I, ~\Sigma, ~ \bar{S}, ~ A_n, ~ H, ~ H'; ~~ \bar5, ~F, 
   ~ \bar{F}, ~ {N},       
\nonumber \\ 
Q=-1 : ~~~ 
&& \bar{I},  ~ S, ~ \bar{A}_n, ~ \bar{H}; 
 ~~ 10, ~~ \ov{T}, ~ \bar{N} .
\eeqn 
where as agreed $\Phi \sim (24,1)$, $\Sigma \sim (1,8)$ 
and $I,\bar{I}$ and $Y$ are some singlets, $\sim (1,1)$,

The most general superpotential invariant under $\cR\times Z_3$ 
is restricted to have the following form: 
\beqn{super-H}
W_{\rm Higgs} = 
&& \Lambda^2 Y + Y^3 + Y\Phi^2 + \Phi^3  
+ Y I\bar{I} + I^3 + \bar{I}^3 + I\Sigma^2 + \Sigma^3 + 
\nonumber \\ 
&& Y (S\bar{S} + A_n\bar{A}_n) + S^3 + \bar{S}^3 + A_1A_2A_3 
+ \bar{A}_1\bar{A}_2\bar{A}_3 + 
\nonumber \\ 
&& H (Y+\Phi) \bar{H} + H'(Y+\Phi) \bar{H} + 
H' I \bar{H}'
\eeqn 
and 
\beqn{super-Y}   
W_{\rm Yuk} = 
&& 10(H+H')T + \bar5\bar{H}T + T(I+\Sigma) \ov{T} + 
\ov{T}S10  + 
\nonumber \\ 
&& \bar5 A F + F(I+\Sigma)\bar{F} + \bar{F} \bar{H}' 10 + 
\nonumber \\ 
&& \bar5 H N + NY\bar{N} + S\bar{N}^2 
\eeqn
respectively in the Higgs and Yukawa sectors.
 
It is worth observing  the following features of the 
Higgs potential. 
It has only one mass scale (in the linear term) 
which determines the physical scales present in the 
theory (GUT, flavour, etc.). 
Notice that among other degenerate minima, this 
superpotential allows a solution  when all singlets, 
adjoints and sextet/triplets get VEVs linked to the scale 
$\Lambda$, modulo unspecified  coupling constants in 
(\ref{super-H}). For example, the superpotential 
terms for $S$ and $\bar{S}$ considered in Appendix A, 
emerge from here with $\mu \sim \langle Y\rangle$. 

There are no couplings between $S$, $A$ and $\Sigma$, 
so the superpotential does not fix the relative orientation 
of their  VEVs. However, that will be fixed by supersymmetry 
breaking terms, along the lines discussed in the Appendix A. 
For example, these can be 
D-terms like 
$\frac{1}{M^2}\int d\theta^4 z\bar{z} 
[\sigma S^\dagger\Sigma^\dagger \Sigma S + 
\lambda_n A_n^\dagger \Sigma^\dagger \Sigma A_n]$ etc., 
where $z=m \theta^2$ stands for supersymmetry breaking spurion. 
Depending on the form and sign of the 
coupling constants, these terms can fix the relative 
VEV orientation of $\Sigma$, $S$ and $A_n$ in the $SU(3)_H$ 
space. In particular, the choice $\lambda_{1,2} >0$ 
favours "orthogonal" VEV directions of $A_{1,2}$ 
and $\Sigma$, while $\sigma <0$ prefers the "parallel"  
VEV directions of $S$ and $\Sigma$.  Thus, in this case 
the largest entry in $S$ is positioned at the same place as 
in $\Sigma$. The same for other fields. 
A typical consequence of the fact that the 
horizontal VEV configuration is dynamically fixed 
by small terms, is the presence of light horizontal 
Pseudo-Goldstone Higgses in the theory, 
reminiscent of the familons, even if the horizontal 
symmetry is local \cite{AB}.  
As for the $SU(5)$ part, the adjoint $\Phi$ gets a VEV 
$\sim \Lambda$ and, as far as the 
superpotential contains also $\Phi^3$ term, 
no light fragments of $\Phi$ are left behind. 

Notice that the term $HI\bar{H}'$ is not present 
in (\ref{super-H}) although it seems to 
be allowed by the symmetry. The point is that if 
$H'$ is a 5-plet, then as it has the same   
charges  as $H$, that term can be rotated away 
by redefinition of $H$ and $H'$. Instead, if $H'$ 
and $\bar{H}'$ are 45-plets, then that term is forbidden by  
the $SU(5)$ symmetry. 
Thus, after substituting the VEVs, the bilinears of 
the doublet and 
triplet fragments in $H,H'$ and $\bar{H},\bar{H}'$ 
have the form:
\be{HHHH}
\mat{\la_1Y +\la_2\Phi}{0}{\la_3Y +\la_4\Phi}{I} ,
\ee
where $\la_1, \la_2, \cdots$  are $\cO(1)$ coupling constants 
previously understood. 
For rendering the Higgses $H_{u,d}$ light, 
the 11 element of (\ref{HHHH}) in the corresponding 
doublet sector has to be fine tuned 
by canceling big contributions between $\la_1Y$ and $\la_2\Phi$. 
(Then the triplet components will remain heavy.) 
Therefore, $H_u$ is contained entirely in $H$ 
whereas $H_d$ emerges as a  superposition 
of the doublet fragments in $\bar{H}$ and $\bar{H}'$, 
$H_d = -s_\omega \bar{H}_2 +c_\omega\bar{H}'_2$, 
with $\tan\omega \sim \frac{\langle I \rangle}
{\langle \la_3 Y + \la_4 \Phi\rangle}$, 
while the other combination of these doublets 
is superheavy.   

The Yukawa superpotential (\ref{super-Y}) contains  
all terms of eqs. (\ref{ren-T}), (\ref{ren-N}) and 
(\ref{ren-F}) which lead to  
the Yukawa pattern (\ref{Yuk-de}). 
Furthermore, some specific features have arisen. 
The heavy-states $\bar{N}$ mass matrix emerges 
only by the VEV of singlet $Y$, so $b_N=1$ and 
thus $b' = b^{-1}_T$. Then, 
if $I+\Sigma$ acts in the same reducible combination 
in $W_T$ and $W_F$, i.e. $b_T=b_F$ 
then $\bb' = \bb$, which is phenomenologically 
welcome (\ref{nu-pattern}). 
For example, this could be the case if 
the reducible combination $I+\Sigma$ 
emerges from an adjoint Higgs of a larger 
group containing $SU(3)_H$, e.g. $SU(N+3)_H$.
Following similar lines of reasoning, 
also the $T'$ scheme can be justified.

Finally, in either schemes, 
the Clebsch coefficients $k_{n} = A^e_n/A^d_n$,  
$n=1,2,3$,  depend on the $SU(5)$ content of the tensor 
products $A_n\cdot \bar{H}'$. 
Needless to say that whatever values of the Clebsch factors  
can occur when the Higgses $A_n$ are in 
$\sim(1+24,3)$ with unspecified weights $\zeta_n$ 
of 1 and 24 component. However, it would be more 
attractive (though it as a matter of taste) 
to find a more ``familiar'' origin of 
the Clebsch factors phenomenologically needed.    
In other words, we assume (on similar 
grounds as e.g. in refs. \cite{Clebsch}) 
that different entries   in the Yukawa matrices 
are differentiated  by some 
``clean'' Clebsch factor  
$k_n$ in which the VEV structure of the original GUT 
 is encoded  through  
a specific representation of the 
intermediate heavy fermions (e.g. $F$ or $T'$ 
in our case). 
Therefore, let us think that at each entry, 
the Clebsches $k_n$  can have some specific  
rational values such as $0,\pm\frac13, \pm\frac23, 
\pm 1,\pm\frac32,\pm 2,\pm 3$, etc. , possibly 
related to various  charges 
(electric, hypercharge, $B-L$ etc.).
 
In particular, in the context of $SU(5)$, 
whenever the product $A_n\cdot \bar{H}'$ 
is in $\bar5$- or $\ov{45}$-channel 
(e.g. simply when $A$ is singlet, $A\sim (1,3)$,  
and the Higgs $\bar{H}'$ is in $\bar5$ or 
$\ov{45}$ representations), we have 
$k_n = 1$ and $k_n =-3$, respectively. 
In  Table \ref{tab3} we report the values for $k_n$ 
attainable for  $A\sim (1,3)$ and $A \sim (24,3)$ 
(or $M \sim 24$), too. 

As in the next we shall consider in detail 
specific Clebsch patterns for the Yukawa matrices, 
characterized by different values of $k_{n}$, 
now we would like to keep the discussion on the 
Clebsch factors more general to show 
how we arrive at those specific choices.

\begin{table} 
 \begin{center}
 \begin{tabular}{||c||c|c|c|c||}
        \hline \hline
$$ & $SU(5) {\rm rep.}$ & $A\sim 1,$ & $A\sim 24,$ & $ A\sim 1$, \\ 
$$ & $$ & $M\sim 1 $ & $M\sim 1 $ & $ M\sim 24 $ \\ 
\hline
$ F-{\rm scheme} $ & $H'\sim \bar 5$ & $ 1$ & $ -3/2$ & $ -2/3 $ \\
$$ & $ H'\sim \ov{45}$ & $ -3$ & $ 9/2$ & $2 $ \\ 
\hline
$T'-{\rm scheme}$ & $H'\sim \bar5$ & $1$ & $6$ & $1/6 $ \\ 
$$ & $ H'\sim \ov{45}$ & $-3$ & $ -18$ & $ -1/2 $ \\
        \hline \hline
 \end{tabular}
\caption[]{\small Possible values for the Clebsch factors 
$k_n=A^e_n/A^d_n$, as they can emerge in the 
$F$- and $T'$-versions  for various assignments of the 
$SU(5)$ representations to the Higgses $\bar{H}'$ and $A$ and 
to the heavy-state masses $M$. 
The case when both $A,M\sim 24$ gives the same 
pattern as $A,M\sim 1$. 
Clearly, in each case the Clebsches obtained 
for $\bar{H}'\sim \bar5$ and $\bar{H}'\sim \ov{45}$ 
differ just by a factor $-3$. }
 \label{tab3}  
 \end{center}
 \end{table}

Indeed, we  could think of all these different ans\"atze  in  
terms of larger vertical and/or horizontal symmetries. 
Namely, in these cases $SU(5)$ may emerge just a subgroup 
of some larger GUT symmetry group. (The $SO(10)$ example 
will be considered in the next section). 
Then the adjoint representation would generally emerge 
accompanied by a singlet partner. Hence, the 
corresponding combination $1+24$ can have a VEV 
with two different eigenvalues 
\be{xxxyy}
\langle \Phi_{1+24} \rangle = 
{\rm Diag}(x,x,x,y,y)
\ee
In the context of $F$-scheme 
(exchange of heavy 5-plets) 
this would produce a Clebsch $k=\frac{y}{x}$. 
In the case of 10-plet exchange ($T'$ scheme) the 
Clebsch value  would be $k=\frac{2y}{x+y}$. 

For example, in the context of the $SU(6)$ model, 
which,   
as a matter of fact, 
provides a natural doublet-triplet 
splitting, the adjoint Higgs (35-plet) of $SU(6)$ 
plays a central role \cite{GIFT}. 
This 35-plet ($35=1+24+5+\bar5$, in terms of $SU(5)$ 
subgroup) has a VEV $\sim {\rm Diag}(1,1,1,1,-2,-2)$. 
Therefore, in the context of $F$-like scheme it 
can provide a Clebsch factor $k=-2$, or, in 
$T'$-like scheme, $k=4$ \cite{GIFT}.   

In fact, one can consider more general cases  
of $SU(N)$ group. Such a theory could contain 
adjoint Higgses with different structures which 
break $SU(N)$ in various possible channels.   
By imposing the traceless condition on 
the $SU(N)$ adjoint VEV, it is straightforward 
to obtain the relation between the two eigenvalues 
$x$ and $y$ in (\ref{xxxyy}) and thus 
the relative Clebsch factors projected out 
in the context of the $F$-like and $T'$-like schemes. 

\vspace{0.5cm}
\noindent
I. $\Phi_1$:  $SU(N) \to SU(5)\times SU(N-5)\times U(1)$,  
this Higgs contains only the $SU(5)$ singlet VEV and 
thus can produce only $k=1$.
 
\vspace{0.4cm}
\noindent
II. $\Phi_{2}$: 
$SU(N) \to SU(N-2)\times SU(2)_w\times U(1)$,   
displaying the weak isospin $SU(2)_w$. Hence  
$\langle \Phi_2\rangle = 
{\rm Diag}\cdot (1,1,..., -\frac{N-2}{2}, -\frac{N-2}{2})$ 
which  implemented in  the $F$-like 
scheme,  generates the Clebsches $k=-\frac{N-2}{2}$, 
i.e. $k=-2,-5/2,-3,-7/2,\cdots$ respectively for 
$N=6,7,8,9,\cdots$. 
In the context of $T'$-like schemes the same VEV 
provides  
$k=\frac{2(N-2)}{N-4}$, i.e. 
$k=4,10/3,3,14/5,\cdots$. 

\vspace{0.4cm}
\noindent 
III. $\Phi_3$: $SU(N)\to SU(3)_c\times SU(N-3)\times U(1)$, 
exhibiting  the colour $SU(3)_c$. 
Therefore, 
$\langle \Phi_3\rangle = 
{\rm Diag}\cdot(1,1,1,-\frac{3}{N-3},\cdots, -\frac{3}{N-3})$.  
Hence, this Higgs in the context of $F$ scheme  
gives $k=-\frac{3}{N-3}$, i.e. $k=-1,-3/4,-3/5,-1/2,\cdots$,  
while in $T'$-scheme $k= \frac{6}{6-N}$, i.e.  
$k=\infty,-6,-3,-2,\cdots$   
($k=\infty$ should be understood as $1/k=0$, i.e. 
$A^d=0$ while $A^e$ is non-zero). 

\vspace{0.4cm}
\noindent 
IV. There can be also adjoints with 
special VEV directions, having vanishing eigenvalues 
towards the $SU(2)_w$ components, say 
$\langle \Phi_0\rangle = 
{\rm Diag}\cdot(1,1,1,0,0,-\frac{3}{N-5},\cdots,-\frac{3}{N-5})$.  
It is natural to use such adjoints for the 
doublet-triplet splitting  \cite{suN}. 
Obviously, this would give $k=0$ in either 
$F$- or $T'$-like schemes. 
Analogously, there can be adjoints with vanishing $SU(3)_c$
components, which would lead to $k=\infty$ in $F$-like 
scheme and $k=2$ in $T'$-like scheme.
  
\vspace{0.4cm}
\noindent
Therefore, in the next, while searching the realistic 
ans\"atze, we shall scan all possible Clebsch values for 
$k_n$ or $1/k_n$ among $0, \pm1, \pm2, \pm3$ plus the  above 
$N$-dependent expressions (for all possible $N$).\footnote{
In the context of $SU(N)$ models containing 
$SU(5)\times SU(3)_H$, ($N \geq 8$),  the large 
value of the horizontal Clebsches $b,b'$ could be 
similarly  obtained 
if also the $I + {\Sigma}\sim (1+8)$-like combinations   
emerge from the  VEV of the "big" adjoint 
breaking $SU(N)$ down to $SU(N-1)\times U(1)$, i.e. 
$b= N-1$. 
} 

It should be remarked that within this wide variety, 
we find  only three acceptable solutions which characterise 
three different ans\"atze. We anticipate them here:
\beqn{CL-A} 
&&{\rm ansatz}~~ {\bf A} 
 ~~~~~~~~~k_{1} = k_2=2 ~,~~~~k_{3} =1~ ,\\
&&{\rm ansatz}~~ {\bf B}  ~~~~~~~~~
k_{1}= k_3 =-3~, ~~k_{2}= 0~
\label{CL-B} \\
&& 
{\rm ansatz}~~ {\bf C}  ~~~~~~~~~k_{1} = -3~, ~~~ 
k_{2} = 0~,~~~~~k_{3} =2~
\label{CL-C}
\eeqn
We shall come back to this point in  Sect. 4.3.

\section{The $SO(10)\times SU(3)_H$ model} 

In the context of $SO(10)$ model all fermion representations 
(\ref{f}) are unified into 16-plets $\psi_i$,    
where $i=1,2,3$ is a family ($SU(3)_H$) index. 
In terms of the $SU(5)$ subgroup, the fermion content 
is represented as $\psi_i=(\bar5+10+1)_i$.  
So, in addition to the superfields $\bar5_i$ and 
$10_i$ (\ref{f}), $\psi_i$ contain also the $SU(5)$ 
singlets $1_i$, usually referred to as RH neutrinos. 
	
The $SO(10)$ symmetry can be spontaneously broken  
to $SU(5)$ by the Higgs supermultiplets 
$C\sim 16$, $\bar{C}\sim \ov{16}$, 
having the VEVs towards the singlet components. The $SU(5)$ 
symmetry can be further reduced down to 
$SU(3)\times SU(2)\times U(1)$  
by Higgses in tensor representations 54 and 45. 
As for the MSSM Higgs doublets $H_{u,d}$  
needed for  breaking the electroweak symmetry and 
generating the fermion masses, they are contained 
in a single superfield $\phi \sim 10$.\footnote{ 
We remind the content of $SO(10)$ multiplets 
with respect to the $SU(5)$ subgroup: 
$16=\bar5+10+1$, $\ov{16}=5+\ov{10}+ 1$, 
$54=24+15+\ov{15}$, $45=1+24+10+\ov{10}$ and 
$10=\bar5+5$. 
}   
It is well known that the $SO(10)$ framework also 
offers the possibility to work out the basic problem of 
the doublet-triplet splitting which has to be unavoidably 
adressed in any realistic grand unified theory. 
In particular, one can implement 
the so called missing VEV mechanism \cite{DW} 
which can be justified by additional symmetry reasons 
(see \cite{BabuBarr,tavzur} and references therein). 
It allows to achieve the mass splitting between the 
components of the Higgs 10-plet so that the doublets 
$H_{u,d}$ remain massless while   
their colour triplet partners get masses of the order 
of the GUT scale,  without unnatural fine tuning 
of the Higgs superpotential parameters. 

Since the  10-plet $\phi$ can couple only to the symmetric 
combination of the fermion 16-plets, 
the higher-order operators inducing the fermion masses 
can only involve the (anti)sextet horizontal Higgs $S$ 
(cfr. the $SU(5)$ case (\ref{hoo}) where also triplet 
Higgses $A$ can contribute).  
Hence, in the context of the $SO(10)\times SU(3)_H$ 
model the operators (\ref{hoo}) are represented as: 
\be{o16} 
\cO\sim \frac{S^{ij}}{M} \psi_i\psi_j \phi , 
\ee
where, as in Sect. 2, the flavour scale $M$ 
can be of the order of the GUT scale $M_G \sim 10^{16}$ GeV, 
and $Y_t\sim 1$ implies that $\cS_3 \sim M$. 
One can still define $\bS = M^{-1}\langle S\rangle$.  
Thus, in the low-energy limit this operator provides equal  
Yukawa matrices for all fermions, including that 
for the neutrino Dirac mass terms:  
$\bY_u,\bY_{d,e},\bY_\nu^D = \bS$.  
In addition, the operator 
\be{o16-M} 
\cO_R \sim \frac{S^{ij}}{M^2} \psi_i\psi_j \bar{C}^2.  
\ee
involving the Higgs 16-plet with a VEV $V_C \sim M$    
induces the Majorana mass terms for the RH neutrinos: 
$\bM_R \propto \bS$. Thus, the effective 
operator for the neutrino masses emerges from the 
familiar see-saw mechanism, and it has a form
(\ref{Yuk-nu}) 
with $\bY_\nu =\eta \bY_u$, where $\eta\sim (M/V_C)^2$. 

Therefore, in such a straightforward  case 
we obtain a highly unrealistic pattern 
for the fermion masses and vanishing mixing 
angles in both the quark and lepton sectors. 
Recalling the discussion in Sect. 2.1, the way 
out from this situation is to invoke  $SO(10)$ and 
$SU(5)$ breaking Higgses (45- and 54-plets) 
in the effective operators for fermion masses, 
as well as the $SU(3)_H$ octet Higgs 
for breaking the horizontal symmetry.

In this section we present a consistent 
$SO(10)\times SU(3)_H$ model to 
motivate the desired Yukawa textures (\ref{Stech}).  
By explicitly specifying the extra symmetries, 
generically denoted by ${\cal G}$ in the previous section, 
we will show, for example, how the Clebsch structure of 
the ansatz {\bf B}  may be obtained.
Interestingly, the same symmetry reasons can be 
utilized for solving the problem of the 
doublet-triplet splitting via the missing VEV mechanism. 

Let us describe the ingredients of the  model. 
For the sake of simplicity,  
we assume that all `vertical'
Higgses are singlets of $SU(3)_H$ and vice versa, 
all `horizontal' Higgses are singlets of $SO(10)$. 
In particular, the `vertical' sector of $SO(10)$ Higgses 
includes chiral superfields in the representations  
54, 45, 16 and $\ov{16}$, needed for the $SO(10)$ symmetry 
breaking down to $SU(3)\times SU(2)\times U(1)$, and 
10-plets for the electroweak symmetry breaking and 
the fermion mass generation. 
The "horizontal" sector of $SU(3)_H$ Higgses includes 
the triplet, sextet and octet representations. 
Namely, as in the previous section, 
we introduce one anti-sextet $S\sim \bar6$,  
three triplets $A_{1,2,3}\sim 3$ 
(plus the conjugated "spectator" superfields 
$\bar{S} \sim 6$ and $\bar{A}_{1,2,3} \sim \bar3$),  
and  the $SU(3)_H$ adjoint $\Sigma \sim 8$. 
In addition, there 
can be some fields which are singlets under  both 
$SO(10)$ and $SU(3)_H$.

Besides the matter superfields $\psi_i \sim (16,3)$ 
we introduce a number of vector-like representations 
containing the fragments like (\ref{T}), (\ref{N}) 
and (\ref{F}) which are needed for the fermion mass 
generation. Therefore, these should be matter  
superfields in $SO(10)$ representations $16 + \ov{16}$, 
10 or singlets, to be specified below. 
These states mediate the see-saw like diagrams and 
hence the  
quark and lepton Yukawa structures emerge after 
integrating them out.  

We invoke the additional symmetry in the form  
$\cG= U(1)_A \times Z_6 \times \cR$,   
where $U(1)_A$ is an anomalous gauge symmetry \cite{Anom} 
implemented in the same spirit as in ref. \cite{tavzur}, 
${\cal R}$ is a discrete symmetry (as in Sect. 2.2) 
under which all superfields in the theory as well as 
the superpotential change the sign, and  
$Z_6$ is a discrete symmetry  acting on the  superfields
as $\varphi_n \to \varphi_n\exp(\frac{\pi}{3}i Q_n)$.   

The superfield content of the model with respect to 
$SO(10)\times SU(3)_H$ is given below, and the superfield
charge assignment  $[\cQ,Q]$ with respect to $U(1)_A$ 
and $Z_6$ symmetries, respectively,  is shown in  subscript. 
Namely, we assume that the Higgs sector contains the 
following superfields: 

(i)  the $SO(10)$ representations   
\be{s-so10}
\begin{array}{llll}
\Theta \sim (54,1)_{[0,0]},\, &   
\Phi \sim (45,1)_{[0,3]}, \,& 
\Omega \sim (45,1)_{[0,2]},\,&  
\Omega' \sim (45,1)_{[0,1]},  \\  
C \sim (16,1)_{[-1/2,1]},\,& 
\bar{C} \sim (\ov{16},1)_{[1/2,3]}, \,&  
\phi \sim (10,1)_{[-1,0]},\, &
\phi' \sim (10,1)_{[1,3]},  \end{array}
\ee 

(ii)  the $SU(3)_H$ representations  
\be{s-su3}
\begin{array}{llll}
S \sim (1,\bar6)_{[0,0]},\, &  
A_1 \sim (1,3)_{[0,1]},\, &  
A_2 \sim (1,3)_{[0,-2]},\,&
A_3 \sim (1,3)_{[0,1]},   \\  
\bar{S} \sim (1,6)_{[0,0]},\, & 
\bar{A}_1 \sim (1,\bar3)_{[0,3]},\, & 
\bar{A}_2 \sim (1,\bar3)_{[0,0]},\, &
\bar{A}_3 \sim (1,\bar3)_{[0,3]},    
\end{array}
\ee 
and $\Sigma \sim (1,8)_{[-2,0]}$,  

(iii) and the gauge singlets 
\be{s-sing}
Y \sim (1,1)_{[0,0]}, ~~~~
Z \sim (1,1)_{[0,2]}, ~~~~
\bar Z \sim (1,1)_{[0,-2]} ~~~~
I \sim (1,1)_{[-2,0]} . 
\ee

It is convenient to split the complete Higgs superpotential 
$W_{\rm Higgs}$ consistent with the 
above symmetries, into the following parts: 
\beqn{sup-so10}
W_1 & =  & \Lambda^2 Y + Y^3 + Y\Theta^2 + 
\Theta^3 + (Y+\Theta)\Phi^2 + YZ\bar{Z} + YS\bar{S} 
\nonumber \\ 
& &  + Z^3 + \bar{Z}^3 + Z\Omega^2 + 
ZC\bar{C} + \Omega C\bar{C} + \bar{Z}\Omega'^2 + 
\Phi\Omega\Omega' 
\nonumber \\ 
&& 
+ S^3 + \bar{S}^3 + Z A_n \bar{A}_n 
+ A_1A_2A_3 + \bar{A}_1\bar{A}_2\bar{A}_3 ,
\nonumber \\ 
W_2 & = & \Phi\phi\phi' + I\phi'^2 + \phi\bar{C}^2 , 
\eeqn 
(all $SO(10)$ and $SU(3)_H$ index contractions 
are left understood). Here  order 1 coupling constants 
are implied in the trilinear terms,
and $\Lambda$ is a mass parameter of the order of 
the GUT scale, $M_G\sim 10^{16}$ GeV.\footnote{
As far as R-parity is concerned, generally it is not 
automatic in $SO(10)$ models involving the Higgs 
16-plets. However, in the context of our model one can 
easily verify that all dangerous R-violating terms 
are suppressed by the $\cG$ symmetry. 
}

An important role is played by the D-term 
of the anomalous $U(1)_A$ symmetry 
\be{anomal}
D_A = \cM^2 + \sum \cQ_n |\varphi_n|^2, ~~~~~ 
\cM^2 = \frac{ {\rm Tr} \cQ}{192\pi^2} M^2_{\rm str} 
\ee 
where the sume runs over all scalar components $\varphi_k$ 
in the theory and $\cQ$  are their $U(1)_A$ charges. 
Therefore, the spontaneous breaking scale of the 
$U(1)_A$ symmetry is naturally related to the string 
scale $M_{\rm str} \sim 10^{17-18}$ GeV with a coefficient 
determined by the charge content of the superfields 
in the game. 

Notice that  $\Lambda$ is the unique mass parameter contained 
in the superpotential (\ref{sup-so10}), while another 
mass scale in the theory comes from the Fayet-Iliopulos 
term $\cM^2$. Therefore, the VEV magnitudes of all fields 
into the game should be determined by these two mass 
scales, which in addition  can be both of the order of 
the GUT scale.   
In particular, the Higgs fields participating 
in the superpotential $W_1$ can get $\sim \Lambda$ VEVs 
violating the $SO(10)$ and $SU(3)_H$ symmetries. 
In fact, the linear term $\Lambda^2Y$ induces the VEVs 
of the singlets $Y,Z,\bar{Z}$ which in turn play the role of mass 
terms for the superfields in non-trivial representations 
like $\Omega,\Omega',\Phi, \Theta$, etc.
Thus, with an accuracy of $\cO(1)$ coupling constants 
in the superpotential terms (\ref{sup-so10}), 
the magnitude of all non-zero VEVs is $\sim M_G$.  
As for the fields $I$ and $\Sigma$, they can get 
 VEV of the order $\cM$ from the anomalous 
D-term (\ref{anomal}). 
We assume that the trace ${\rm Tr}\cQ$ of the $U(1)_A$ 
charges over all superfields is positive: namely, 
the superfields presented in (\ref{s-so10}) and 
(\ref{s-sing}) plus the octet $\Sigma$ 
contribute as  ${\rm Tr}\cQ \sim 10^2$. 
(The theory may include also other superfields 
with charges arranged so that the Green-Schwarz 
cancellation mechanism is at work, but this does not 
affect the orders of magnitude in our consideration).  
Therefore, $\cM$ can naturally be about the GUT scale 
$M_G$ or a bit larger. One has also to remark that 
the anomalous term fixes only the overall VEV combination 
$\langle I\rangle^2 + \langle \Sigma\rangle^2$, 
but not each term separately. The VEV values of 
$I$ and $\Sigma$ will be fixed by the soft supersymmetry 
breaking terms, as well as orientation of 
$\langle \Sigma\rangle$ with respect to the VEVs of 
other horizontal scalars, $S$ and $A_{1,2,3}$, 
in the spirit discussed in sect. 2.2. 
This means that the horizontal fields should contain  
some light fragments, with masses order TeV, which, on the other 
hand,  do not lead to phenomenological problems 
(see e.g. discussion in ref. \cite{AB}). 
The only important point is that, due to the simple 
structure of the $SU(3)_H$ group, the VEV of $\Sigma$ 
in either case would acquire the simplest possible 
configuration 
$\langle \Sigma\rangle \propto {\rm Diag}(1,1,-2)$.   

Remarkably, the terms contained in the superpotential 
$W_1$ provide the  pattern needed to break 
$SO(10)$ to the standard model group. 
In particular, there exists a solution 
for the 16-plet $C,\bar{C}$ VEVs   
which breaks $SO(10)$ down to $SU(5)$:  
$\langle C\rangle = \langle \bar{C}\rangle = 
V_C\cdot |+,+,+,+,+\rangle$ (in terms of the 
corresponding Cartan sub-algebra generators). The condition 
$\langle C\rangle = \langle \bar{C}\rangle$ follows from the 
vanishing of the SO(10) D-terms.  
The 54 and 45 Higgses are needed to subsequently 
break  $SU(5)$ down to $SU(3)\times SU(2)\times U(1)$. 
The necessary VEVs are given as:  
\beqn{Theta}
&&
\langle \Theta\rangle = V_\Theta \cdot 
{\rm Diag}(1,1,1,-3/2,-3/2) \otimes \sigma_0 , 
 \nonumber \\ 
&&
\langle \Phi\rangle = V_\Phi \cdot 
{\rm Diag}(1,1,1,0,0) \otimes \sigma ,  \nonumber \\
&&
\langle \Omega \rangle = V_\Omega \cdot 
{\rm Diag}(1,1,1,1,1) \otimes \sigma ,  \nonumber \\
&&
\langle \Omega'\rangle = 0,
\eeqn
where 
\be{duo} 
\sigma_0 =\mat{1}{0}{0}{1} , ~~~~~~~ \sigma = \mat{0}{1}{-1}{0} .
\ee
One can easily verify that these VEVs are consistent with a 
supersymmetric vacuum and their configurations 
are not affected by some unknown factors. 
In particular, in the 54-plet $\Theta$ only its 24 fragment 
(in terms of $SU(5)$ subgroup) acquires VEV and, noticeably, into  
the only phenomenologically allowed direction 
which induces the $SO(10)$ symmetry breaking down to 
the Pati-Salam subgroup: 
$SO(10)\to SU(4)\times SU(2)\times SU(2)'$.   
The coupling of the 45-plet $\Phi$  to $\Theta$ in the 
superpotential  forces the VEV of the former to lie  
in the $B-L$ direction, 
as a result of a combination of its $SU(5)$ 1 and 24 fragments,  
with {\it exact} zeros on the last two components, 
as it is demanded by 
the missing VEV mechanism \cite{DW,BabuBarr}.\footnote{ 
As we will see below in the analysis of the Yukawa 
sector, such a VEV structure of $\Phi$ is also 
important for obtaining the Clebsch factors needed 
for the ansatz B, namely $k_{1,3}=-3$ and $k_2=0$. }
As for the other 45-plet $\Omega$, it is coupled to the  
16-plets $C,\bar{C}$, but does not couple to the 54-plet 
$\Theta$. Because of this, its VEV is directed entirely 
towards the $SU(5)$-singlet direction, with a vanishing 
VEV towards the 24 fragment.\footnote{This is also  
relevant  for the fermion Clebsch structures,  
because a $SU(5)$-singlet VEV -- namely, the  combination 
$\langle Z+\Omega \rangle$ --  
participating in the fermion mass generation (see below, 
diagrams in Fig. \ref{so10-2})  
does not induce uncontrollable corrections to the 
Clebsch factors.}  
And finally, the third 45-plet $\Omega'$ does not couple 
neither to $\Theta$ nor to $C,\bar{C}$, and thus it should 
have a vanishing VEV. 

It is also important that all non-singlet 
(under the standard model group) fragments  
in the $SO(10)$-breaking superfields   
get  masses of the order of $M_G$.  
In such a case, no light fragment (with mass $\ll M_G$) is left 
behind which could affect the gauge coupling unification 
at the scale $M_G$. The presence of the term 
$\Phi\Omega\Omega'$ is crucial for achieving this result 
\cite{BabuBarr}, since otherwise, in its absence,  
the theory would have  flat directions towards 
non-singlet fragments which are related to extra global 
symmetry in the superpotential.  
Remarkably, this term does not affect the VEV pattern 
of 45-plets while lifting the dangerous flat directions.

Finally, the term $W_1$ in (\ref{sup-so10}) 
is also responsible for the horizontal $SU(3)_H$ 
breaking via the  VEVs of $S$ and $A_{1,2,3}$ fields 
(see (\ref{S}) and (\ref{A}))  
and its features are elaborated in Appendix A. 
Notice, that even if all coupling constants 
are taken real in the superpotential,  non-trivial 
CP-violating phase in these VEVs could appear in case 
of solutions with different phases for the singlet  
VEVs $\langle Y\rangle $ and $\langle Z\rangle$, 
in the spirit discussed in Appendix B. 


Let us address now the doublet-triplet splitting problem. 
As well-known, the VEV of the 54-plet $\Theta$ 
guarantees  the  VEV of the 
45-plet $\Phi$ to be towards the $B-L$ direction, 
$\langle \Phi\rangle \propto {\rm Diag}(1,1,1,0,0)\times \sigma$.  
On the other hand, 
the superpotential $W_2$ contains the necessary terms 
to incorporate the missing VEV mechanism for 
the doublet-triplet splitting.   
In addition, they also induce the mixing between the 
$\bar5$ fragments in $\phi$ and $C$ through the large 
$SU(5)$ conserving VEV  $V_{C}= \langle C\rangle$.  
All this can be more easily represented in terms 
of the following mass matrix:
\be{Higgs5-mix}
\begin{array}{ccc}
 & {\begin{array}{ccc} 5(\phi) ~
& 5(\phi') &   
5(\bar{C}) 
\end{array}}\\ \vspace{2mm}
\begin{array}{c}
\bar{5}(\phi) \\ \bar{5}(\phi') \\ 
\bar{5}({C})   \end{array}\!\!\!\!\!&{\left(\begin{array}{ccc}
0 & \,\,\langle\Phi\rangle & \,\,V_C\, \\ 
 \,\langle\Phi\rangle\,& \,\,M & 0\, \\ 
\,0 & \,\,0 & \,\,M_C   
\end{array}\right)} 
\end{array} \! 
\ee 
where $M \sim \langle I\rangle$ and 
$M_C \sim \langle Z+\Omega \rangle$ is the mass of the 
5-plet fragments contained in $C,\bar{C}$. 
Owing to the form of  the VEV 
$\langle\Phi\rangle$ in (\ref{Theta}),  non-zero  
entries are induced only between the triplet fragments 
in $\phi$ and $\phi'$ while 
the corresponding entries  for the doublet fragments vanish. 
Therefore, all the triplets get  masses of the order 
of $M_G$, while two combinations of the doublet 
components, to be identified with the MSSM Higgs doublets 
$H_{u,d}$, remain light.   
The doublet $H_u$ comes entirely from $H\equiv 5(\phi)$, 
and so no portion of it is contained in $5(\bar{C})$ or  
$5(\phi')$.  
As regards $H_d$, it emerges from the combination 
of $\bar{5}$ fragments in $\phi$ and $C$.  
Namely,  
$\bar{H} = c_\omega \bar5(C) - s_\omega \bar5(\phi)$, 
with $\tan\omega \sim M_C/V_C$  (see also \cite{Yad}).   

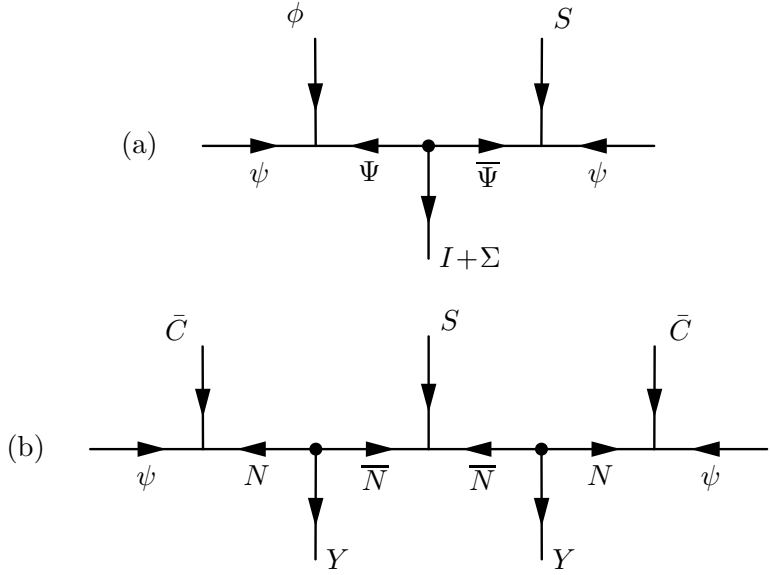
\begin{figure}[t]
\begin{center}
%
\begin{fmfgraph*}(60,30)
\fmfleft{t1}
\fmfright{t2}
\fmftop{x1,H,S,X,x2}
\fmfbottom{b1,b2,a,b4,b5}
\fmf{fermion,label=$\psi$}{t1,v1}
\fmf{fermion,label=$\Psi$,label.side=left}{v2,v1}
\fmf{fermion,label=$\ov{\Psi}$}{v2,v3}
\fmf{fermion,label=$\psi$,label.side=left}{t2,v3}
\fmfv{label=(a),label.dist=17,label.angle=180}{t1}
\fmfv{label=$I\!+\!\Sigma$,label.dist=4,label.angle=0}{a}
\fmffreeze  
\fmf{fermion}{H,v1}
\fmf{fermion}{v2,a}
\fmf{fermion}{X,v3}
\fmfdot{v2}
\fmflabel{$\phi$}{H}
\fmflabel{$S$}{X}
\end{fmfgraph*} \\
\vspace{1.cm}
\begin{fmfgraph*}(90,30)
\fmfleft{t1}
\fmfright{t2}
\fmftop{x1,H,S,X,S1,H1,x2}
\fmfbottom{b,b1,b2,b3,b4,b5,be}
\fmf{fermion,label=$\psi$}{t1,v1}
\fmf{fermion,label=$N$,label.side=left}{v2,v1}
\fmf{fermion,label=$\ov{N}$}{v2,v3}
\fmf{fermion,label=$\ov{N}$,label.side=left}{v4,v3}
\fmf{fermion,label=$N$}{v4,v5}
\fmf{fermion,label=$\psi$,label.side=left}{t2,v5}
\fmfv{label=$Y$,label.dist=4,label.angle=0}{b2}
\fmfv{label=$Y$,label.dist=4,label.angle=0}{b4}
\fmfv{label=(b),label.dist=17,label.angle=180}{t1}
\fmffreeze
\fmf{fermion}{H,v1}
\fmf{fermion}{v2,b2}
\fmf{fermion}{X,v3}
\fmf{fermion}{v4,b4}
\fmf{fermion}{H1,v5}
\fmfdot{v2,v4}  
\fmflabel{$\bar{C}$}{H}
\fmfv{label=$S$,label.dist=5,label.angle=30}{X}
\fmflabel{$\bar{C}$}{H1}
\end{fmfgraph*} 
%
\caption{\small The see-saw diagrams 
giving rise to the effective higher order operators 
for the symmetric contributions to the charged fermion Yukawa matrices
and to the neutrino Dirac mass term  
$\bY^D_\nu$ -- diagram (a), 
and to the Majorana mass matrix $\bM_R$ of 
the RH neutrinos $1$ contained in $\psi$ -- diagram (b). 
In $SU(5)$ language the relevant fragments are 
$\psi=\bar5 + 10 + 1$, $\phi =5+\bar5$, 
and the singlet component of $\bar{C}$ with  VEV $V_C$. 
\label{so10-1}}
\end{center}
\end{figure}

Let us discuss now the fermion mass generation mechanism. 
Apart from the  already familiar 
$\psi \sim (16,3)_{[-1/2,0]}$, let us introduce 
the following vector-like fermions:   

(I) 16-plets 
\be{fermion-16} 
\Psi \sim (16,\bar3)_{[3/2,0]}, ~~~
\ov{\Psi} \sim (\ov{16},3)_{[1/2,0]}; ~~~
\Psi' \sim (16,3)_{[3/2,3]}, ~~~
\ov{\Psi}' \sim (\ov{16},\bar3)_{[1/2,3]}, 
\ee 

(II) 10-plets 
\be{fermion-10} 
\begin{array}{llll}
t \sim (10,3)_{[-1,3]}, \,& 
\bar{t} \sim (10,\bar3)_{[1,-1]}, \,&
t_1 \sim (10,3)_{[1,2]},\,& 
\bar{t}_1 \sim (10,\bar3)_{[-1,2]}, \\ 
{} \,& 
{} \,& 
t_2 \sim (10,3)_{[-1,0]},\,&  
\bar{t}_2 \sim (10,\bar3)_{[1,0]}   \end{array}
\ee
 
(III) and singlets 
\be{fermion-1}
N \sim (1,\bar3)_{[0,3]}, ~~~~
\ov{N} \sim (1,3)_{[0,3]}. 
\ee

Let us start by discussing the role played by the 
additional fermion 10-plets 
$t_i = (\bar5 + 5)_i$, $\bar{t}^i = (\bar5 + 5)^i$. 
All the symmetries of the model allow in particular the following terms:  
\be{reduce}
C\psi\bar{t} + \ov{Z} t\bar{t} .
\ee 
Correspondingly, the mass matrix connecting the matter states 
reads as:
\be{fermion5-mix}
\begin{array}{cc}
 & {\begin{array}{cc} 5(t_i) ~
& 5(\bar{t}^i)    
\end{array}}\\ \vspace{2mm}
\begin{array}{c}
\bar{5}(\psi_i) \\ \bar{5}(t_i) \\ 
\bar{5}(\bar{t}^i)   
\end{array}\!\!\!\!\!&{\left(\begin{array}{cc}
\,0 &\, V_C  \\ 
\,0& \,\langle \ov{Z}\rangle\\ 
\,\langle \ov{Z}\rangle & \,0  \end{array}\right)} 
\end{array} \! ,
\ee
which shows that the light states $\bar5_i$ are  composed by  
a combination of $\bar{5}$ fragments in $\psi_i$ and $t_i$ i.e.  
$\bar5_i = -s_\theta \bar5(\psi)_i + c_\theta \bar5(t)_i$, 
where the angle $\theta$ is defined as 
$\tan\theta = \langle \ov Z \rangle/ V_C$, 
while the orthogonal combinations   
$\bar5'_i = c_\theta \bar5(\psi)_i + s_\theta \bar5(t)_i$ 
and  $\bar5(\bar{t})^i$  are heavy states.

Now we are in position to fix the fermion mass pattern 
which emerges in our theory. 
The diagonal entries in the Yukawa matrices are generated by the 
following superpotential terms:
\beqn{so10-16}
&&
W_\Psi= \psi\phi\Psi + \Psi({I + \Sigma})\ov{\Psi} + \ov{\Psi} S\psi 
\nonumber \\ 
&&
W_N = \psi\bar{C}N + N Y\ov{N} + S\ov{N}^2 , 
\eeqn
which through the see-saw like diagrams shown in 
Fig. \ref{so10-1}, give rise to the effective operators 
like (\ref{o16}) and  (\ref{o16-M}). 
For an easy identification of the contributions to the Yukawa 
matrices, the terms in $W_\Psi$ and $W_N$ can be rewritten 
according to their $SU(5)$ decomposition: 
\beqn{so16-su5}
W_\Psi& = & 10(\psi) 5(\phi) 10(\Psi) + 
\bar5(\psi)\bar{5}(\phi)10(\Psi) + 
\ov{10}(\ov{\Psi}) S 10(\psi) + 
10(\Psi) (I+\Sigma)\ov{10}(\ov{\Psi}) +
\nonumber  \\
&& 10(\psi)\bar{5}(\phi) \bar5(\Psi) + 
5(\ov{\Psi})S \bar5(\psi)  + 
\bar5(\Psi)(I+\Sigma)5(\ov{\Psi}) + 
\nonumber  \\
&& \bar5(\psi)5(\phi) 1(\Psi) + 1(\ov{\Psi}) S 1(\psi) +
1(\Psi) (I+\Sigma) 1(\ov{\Psi}) , \nonumber \\
W_N &= &1(\psi) 1(\bar{C}) N + N Y\bar{N}  + S\bar{N}^2 .
\eeqn
Therefore, as already discussed in Sect. 2.1,  
the VEV combination $\langle I+ \Sigma \rangle$ provides 
the $SU(3)_H$ breaking and leads to the $SU(2)_H$ invariant 
form of the heavy fermion mass matrix 
$\bM_\Psi = M \bb'$, where 
$\bb'= {\rm Diag}(1,1,b')$ and $M \sim \cM$. 
The relevant Yukawa terms for the light fermions emerge 
after integrating out the heavy states.  
In particular, for the Yukawa coupling matrix of up quarks 
(which are entirely contained in $10(\psi)$ fragment) 
to the Higgs doublet $H_u \subset 5(\phi)$ we obtain  
$\bY_u = \bM^{-1}_\Psi \langle S \rangle = 
\bb'^{-1} \bS$, where 
$\bS\equiv M^{-1} \langle S\rangle$   
and it is diagonal once the VEV $\langle S \rangle$ 
is chosen as (\ref{S}).  
On the other hand, recalling that $\bar5(\psi)$ contains 
the light $\bar5$-plet with a weight $s_\theta$ and 
the Higgs $\bar{5}(\phi)$ contains $H_d$ with a weight 
$s_\omega$, the terms (\ref{so16-su5}) provide also 
the diagonal contributions to the down quark and charged
lepton Yukawa matrices 
$\bY^{\rm diag}_{d,e} = s_\theta s_\omega \bb'^{-1}\bS$. 
From the decomposition (\ref{so16-su5}) 
it is apparent that the (heavy fermion) 
exchange contributions to $\bY_u$ are of the 
$T$ -scheme type  (cfr. (\ref{ren-T}) ) 
and those to $\bY^{\rm diag}_{d,e}$ are of both $T$ - and $F$ -scheme 
type (cfr. also (\ref{ren-F})).
    
Finally, for the Dirac Yukawa matrix of neutrinos 
we obtain $\bY_{\nu}^D = s_\theta \bb'^{-1}\bS$. 
On the other hand, by considering the terms in $W_N$, 
we see that the VEV $\langle Y\rangle$ 
induces the $SU(3)_H$ invariant masses between the states 
$N$ and $\bar{N}$, $\bM_N = M_N \cdot {\bf 1}$, 
(${\rm \bf 1}={\rm Diag}(1,1,1)$) and $M_N\sim M_G$,   
and so the RH neutrinos $1(\psi)$ acquire the Majorana 
masses $\bM_R = \bM^{-2}_N V^2_C \langle S \rangle = 
\kappa M \bS$, where $\kappa \sim (V_C/M_N)^2 \sim 1$. 
Thus, as long as both Dirac and Majorana terms are present 
for neutrinos, the operator (\ref{Yuk-nu}) emerges via the  
standard see-saw mechanism: 
\be{so10-Ynu}
\frac{\bY_\nu}{M_L} = \bY_\nu^D \bM_R^{-1} \bY_\nu^D = 
\frac{s^2_\theta}{\kappa M} \bb'^{-2} \bS
\ee  
Therefore, both $\bY_u$ and $\bY_\nu$ are diagonal 
and we have: 
\be{numass} 
\bY_\nu = \eta \bb'^{-1} \bY_u 
\ee
where $\eta = s_\theta^2/\kappa$. 

Let us discuss now the origin of the off-diagonal entries 
in $\bY_{d,e}$, recalling that they have already got the 
diagonal contributions $\propto \bY_u$.  
We also remember that the off-diagonal entries 
should contain non-trivial Clebsches between  quarks and leptons.

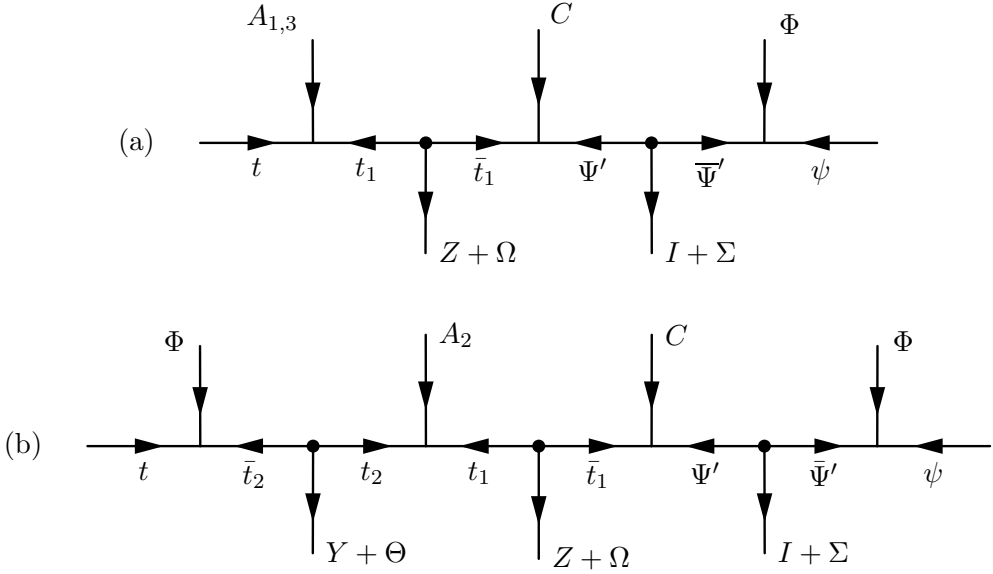
\begin{figure}[t]
\begin{center}
%
%
\begin{fmfgraph*}(90,30)
\fmfleft{t1}
\fmfright{t2}
\fmftop{x1,H,S,X,S1,H1,x2}
\fmfbottom{b,b1,b2,b3,b4,b5,be}
\fmf{fermion,label=$t$}{t1,v1}
\fmf{fermion,label=$t_1$,label.side=left}{v2,v1}
\fmf{fermion,label=$\bar{t}_1$}{v2,v3}
\fmf{fermion,label=$\Psi'$,label.side=left}{v4,v3}
\fmf{fermion,label=$\ov{\Psi}'$}{v4,v5}
\fmf{fermion,label=$\psi$,label.side=left}{t2,v5}
\fmfv{label=$Z+\Omega$,label.dist=5,label.angle=0}{b2}
\fmfv{label=$I+\Sigma $,label.dist=5,label.angle=0}{b4}
\fmfv{label=(a),label.dist=17,label.angle=180}{t1}
\fmffreeze
\fmf{fermion}{H,v1}
\fmf{fermion}{v2,b2}
\fmf{fermion}{X,v3}
\fmf{fermion}{v4,b4}
\fmf{fermion}{H1,v5}
\fmfdot{v2,v4}  
\fmflabel{$A_{1,3}$}{H}
\fmfv{label=$C$,label.dist=5,label.angle=30}{X}
\fmflabel{$\Phi$}{H1}
\end{fmfgraph*} \\
\vspace{1cm}
\begin{fmfgraph*}(120,30)
\fmfleft{l}
\fmfright{r}
\fmftop{tl,t1,t2,t3,t4,t5,t6,t7,tr}
\fmfbottom{bl,b1,b2,b3,b4,b5,b6,b7,br}
\fmf{fermion,label=$t$,label.side=right}{l,v1}
\fmf{fermion,label=$\bar{t}_2$,label.side=left}{v2,v1}
\fmf{fermion,label=$t_2$,label.side=right}{v2,v3}
\fmf{fermion,label=$t_1$,label.side=left}{v4,v3}
\fmf{fermion,label=$\bar{t}_1$}{v4,v5}
\fmf{fermion,label=$\Psi'$,label.side=left}{v6,v5}
\fmf{fermion,label=$\bar{\Psi}'$}{v6,v7}
\fmf{fermion,label=$\psi$,label.side=left}{r,v7}
\fmfv{label=$Y+\Theta$,label.dist=5,label.angle=0}{b2}
\fmfv{label=$Z+\Omega$,label.dist=5,label.angle=0}{b4}
\fmfv{label=$I+\Sigma$,label.dist=5,label.angle=0}{b6}
\fmfv{label=(b),label.dist=17,label.angle=180}{l}
\fmffreeze
\fmf{fermion}{t1,v1}
\fmf{fermion}{v2,b2}
\fmf{fermion}{t3,v3}
\fmf{fermion}{v4,b4}
\fmf{fermion}{t5,v5}
\fmf{fermion}{v6,b6}
\fmf{fermion}{t7,v7}
\fmfdot{v2,v4,v6}  
\fmflabel{$\Phi$}{t1}
\fmfv{label=$A_2$,label.dist=5,label.angle=0}{t3}
\fmfv{label=$C$,label.dist=5,label.angle=0}{t5}
\fmflabel{$\Phi$}{t7}
\end{fmfgraph*}
\caption{\small  Diagrams
giving rise to the effective higher-order operators  
for the non-diagonal contributions to $\bY_{e,d}$. 
The relevant 
contributions of these diagrams involve  
$H_d \subset \bar5 (C)$  and 
$\bar5_i \subset \bar5 (t_i)$ states.  
\label{so10-2}}
\end{center}
\end{figure}
%
To this purpose, we recall that the light fermion 
$\bar5$-plet is also contained in $\bar5(t)$ with a weight 
$c_\theta$, while the doublet $H_d$ is 
contained in 
the fragment  $\bar5(C)$ with a weight $c_\omega$.
Then,  the symmetry of the theory prescribes other  
superpotential couplings for   
the superfields $t$ and $C$,  
involving other (than $\Psi, \ov{\Psi}$)  
heavy fermions, 
so that the exchange of the latter 
induce the off-diagonal contributions 
to $\bY_{d,e}$ via the VEVs of the triplet 
horizontal Higgses $A_{1,2,3}$.  
The relevant diagrams\footnote{For the 
sake of brevity we do not write down the relevant 
superpotential terms responsible for the generation of the 
off-diagonal Yukawa couplings: they can 
be read out directly from the diagrams in Fig. \ref{so10-2}.} 
are shown in  Fig. \ref{so10-2}.
The diagram (a) provides the entries (2,3) and (1,2) 
as only the $SU(3)_H$ triplets $A_1$ and $A_3$ are allowed to 
have a direct coupling to the matter 10-plets $t_i$. 
In this case, the 
coupling of the 45-plet $\Phi$ (with its VEV into the $B-L$ direction) 
to the 16-plets ($\psi$ and $\ov{\Psi}'$) 
gives rise to the  Clebsch coefficients $k_{1,3} = -3$.
We should also observe that these off-diagonal contributions 
are effectively induced by the  exchange of the $SU(5)$ 10+ $\ov{10}$ 
fragments contained in 
the vector-like states $\Psi'+ \ov{\Psi}'$,   
according to the $T'$ -scheme (cfr. (\ref{ren-T1})).   
 
The triplet $A_2$ is instead involved in the diagram (b) which 
therefore contributes to the (1,3) entries. 
Because of the requirement of $\cG$ invariance,   
this triplet cannot  
couple to the states $t_i$ but it couples to the heavy states $t_2$ 
which are mixed to the former through the VEV of the 45-plet $\Phi$.  
This occurs through  the couplings  on the left 
of the $A_2$ incoming-line 
in the diagram (b), while the part on the right side contains exactly the 
same couplings as that on the right of the $A_{1,3}$ line 
in diagram (a).
However,  the coupling of the  45-plet $\Phi$ 
with the fermion 10-plets,  
$t \Phi \bar{t}_2$, 
gives vanishing contribution to the lepton doublets 
contained in $\bar5(t)$
(similarly to what happens in the missing VEV mechanism 
for the doublet-triplet splitting) and therefore  we obtain $k_2=0$.
In either case of diagram (a) or (b), 
the $SU(3)_H$ breaking Clebsch $b$ emerges from the same 
combination $I+\Sigma$ (i.e. from the matrix 
$\bM'_\psi \sim M {\bf b}_{\Psi'}= M\cdot {\rm diag} (1,1,b_{\Psi'})$)
so that the 13 and 23 entries have the same asymmetry. 

Thus, the final form of the down quark and 
charged lepton Yukawa matrices reads as:
\be{YD}
\bY_{d} = \rho \bY_u^D + \bb^{-1} \bA_d, ~~~~~~~
\bY_{e}^T = \rho \bY_u^D + \bb^{-1} \bA_e, 
\ee 
where the anti-symmetric matrices $\bA_{d,e}$ have the form 
(\ref{Ade}) with $k_{1,3}= -3$ and $k_2=0$,  $\bb = {\rm diag}(1,1,b)$ where  
$b = b^{-1}_{\Psi'}$. 
In this way we have reproduced the Clebsch factors  
demanded by the ansatz ${\bf B}$ (\ref{CL-B}). 
The coefficient  $\rho$ is identified with  $s_\omega s_\theta$.  
It may be  smaller than 1 and thus  
moderate or even small values of $\tan\beta$ can  
naturally be accommodated.

In conclusion, 
we have presented a complete $SO(10)\times SU(3)_H$ 
model in which the symmetry breaking VEV pattern as well 
as the form of the superpotential couplings needed to 
obtain the desired 
predictive mass texture for the fermions  
(specifically, the ansatz {\bf B} presented in Sect. 2) 
are fully motivated by the additional 
$\cG=U(1)_A\times Z_6 \times \cR$ symmetry.  
Interestingly, the missing VEV
mechanism for the doublet-triplet splitting can be motivated on 
the basis of the same symmetry $\cG$. 
The other Clebsch structures demanded by the ans\"atze 
{\bf A} or {\bf C} can be reproduced  
along similar lines 
by rearranging the charge assignments of 
the superfields in the game 
with respect to the $Z_6$ symmetry, however 
we will not show the corresponding models here.

\section{Phenomenological analysis}

This part is devoted to the phenomenological study 
of the  ans\"atze suggested. First, we provide 
the corresponding 
expressions of the physical quantities  -- quark masses as well 
as mixing angles. Second, the criteria of the fit are presented. 
Finally the results are shown and discussed.

\vspace{0.5cm}

\subsection{Physical observables from 
the theoretical model}

We now deal with the explicit form for the 
Yukawa matrices $\bY_{e,d}$ in (\ref{Stech-F}). 
In general the coupling constants 
and the VEVs present in the theory are complex.  
However, all the phases except three can be pulled out 
by field redefinitions and the Yukawa 
matrices   can be presented in this   
form:\footnote{As far as  $\bY_u$ and 
$\bY_\nu$ are diagonal, these phase transformations 
are not relevant for the form of the mixing matrices 
$V_{q,l}$ and we do not show them. 
}
\be{Y-diago}
\bY_u = \matr{Y_u}{0}{0} {0}{Y_c}{0} {0}{0}{Y_t}, ~~~~~~~
\bY_\nu = \eta \matr{Y_u}{0}{0} {0}{ Y_c}{0} 
{0}{0}{ \frac{Y_t}{b}}, 
\ee
\be{Y-gen2}
\bY_d = \matr{Y_{ut} D\mbox{e}^{{\rm i}\zeta} } 
{ A\mbox{e}^{{\rm i}\sigma} }{C}
{-A\mbox{e}^{{\rm i}\sigma}}
{ Y_{ct} D\mbox{e}^{{\rm i}\xi} }{B}
{\frac{1}{b}C}{\frac{1}{b}B}{ D}~, ~~~~~~~
\bY_e = \matr{ Y_{ut} D\mbox{e}^{{\rm i}\zeta } } 
{- k_{3} A\mbox{e}^{{\rm i}\sigma}} {\frac{1}{b} k_{2} C }
{ k_{3} A \mbox{e}^{{\rm i}\sigma} } 
{Y_{ct} D \mbox{e}^{{\rm i}\xi}} {\frac{1}{b}k_{1} B}
{k_2 C}{k_1 B} {D}~, 
\ee
Here $D= \rho Y_t$, $Y_{ut}= Y_u/Y_t$ and 
$Y_{ct}= Y_c/Y_t$. 
Notice that in $\bY_\nu$ $b'=b$ has been taken 
for concreteness.
As compared to the expressions in (\ref{Stech-F}) 
the off-diagonal terms  have been redefined as 
$A^d_1 = B$, $A^d_2 = C$ and $A^d_3 = A$ 
and understood to be real.\footnote{In general 
the relative "lepton versus quark" 
Clebsches $k_n = A^e_n/A^d_n$ can be complex, 
however in the following they are taken  real.}   

The following observations can be made. 
The 11 entry in the matrices $\bY_{e,d}$ is negligible 
and in the following we set it to zero.  
Indeed, the ratio $Y_u/Y_t$ is two orders of magnitude 
less than $Y_e/Y_\tau$ and $Y_d/Y_b$ (cfr. Table \ref{tab1}) 
and thus its contribution in $Y_{d,e}$ is irrelevant.       

As regards the 22 entry, this also leads to small 
corrections -- the ratio $Y_c/Y_t$ is one order of magnitude 
less than $Y_s/Y_b$ (cfr. Table \ref{tab1}).  
However, we find that these corrections are relevant 
on comparing with the present precision in the CKM angles. 
So this contribution in $Y_{d,e}$ cannot be neglected.       

In general case the phases in (\ref{Y-gen2}) 
are arbitrary. However, they can have specific 
values in the context of the spontaneous CP 
violation. Namely, one can assume that the original 
theory has an exact CP invariance, i.e. all 
couplings in the Yukawa and Higgs sectors can 
be simultanoulsy made real by  phase redefinitions 
of the superfields. However, CP invariance can be 
spontaneously broken by  non-vanishing  
phases of some VEVs in the theory. 
In particular,  as we show in Appendix B, the 
triplets $A_n$ can have complex phases.  
 
For example, one can choose a basis where the 
VEVs of the sextet $S$, $\cS_{1,2,3}$ are real. 
Thus, in this basis all three diagonal entries 
of $\bY_d$ in (\ref{Stech-F}) are real. 
For definiteness, let us take them also to 
be positive. 
On the other hand, the off-diagonal 
entries keep on the phases originated from the VEVs of 
$A_n$. As we show in Appendix B, these VEVs can 
exhibit a spontaneous CP violation 
when the product $A_1 A_2 A_3$ is imaginary.   
In particular, if all entries $A_{1,2,3}$ are imaginary, 
the  matrix $\bY_d$ in (\ref{Stech-F}), once recasted in  
the form (\ref{Y-gen2}),  implies the phase assignment 
$\xi =\pi$, $\sigma=\pi/2$. In another possible 
case, when $A_{1,2}$ are real but $A_3$ is imaginary, 
one has instead $\xi=0$ and $\sigma=\pi/2$.    
In principle, the spontaneous CP breaking could 
provide some other interesting phase assignments 
such as  $\xi$ and $\sigma$ have some 
"inverse $\pi$-fold" values like $2\pi/3$, 
$\pi/4$, etc. 
In the following, to be more general, we keep 
the phases $\xi$ and $\sigma$ arbitrary, however 
we shall pay special attention to those specific phase 
values which might be obtained in the context 
of the spontaneous CP violation. 

Let us now proceed our analysis. 
The  (1,3) and (3,1) entries in $\bY_{d,e}$ 
can be simultaneously rotated away   
by orthogonal and phase transformations 
in the 12-flavour space:
\beqn{P-O12} 
&& 
P_- \tilde{O}^{dT}_{12} \bY_d \tilde{O}^d_{12} P_+ 
= \tilde{\bY}_d , ~~~~~~
P_+ \tilde{O}^{eT}_{12} \bY_e \tilde{O}^e_{12} P_- 
= \tilde{\bY}_e , 
\nonumber \\
&&
P _\pm = \matr{ \pm {\rm e}^{ {\rm i}\sigma} }{0}{0} 
{0}{1}{0} {0}{0}{1} , ~~~~~~ 
\tilde{O}^{d(e)}_{12}= 
\matr{\tilde{c}^{d(e)}_{12}}{\tilde{s}^{d(e)}_{12}}{0} 
{-\tilde{s}^{d(e)}_{12}}{\tilde{c}^{d(e)}_{12}}{0} { 0}{0}{ 1}~,
\eeqn 
where 
$\tilde{c}^{d(e)}_{12}=\cos\tilde{\theta}^{d(e)}_{12}$ etc. 
and it is taken  $\tan\tilde{\theta}^d_{12} = C/B$
and  $\tan\tilde{\theta}^e_{12}= 
\frac{k_{2}}{k_{1}} \tan\tilde{\theta}^d_{12}$. 
Notice that these rotations do 
not affect the (1,2) and (2,1) elements 
as the Yukawa matrices are antisymmetric in the 12 block. 
On the contrary the 23 block gets modified and 
by further re-defining everywhere 
the parameter $B$ as $\tilde{B}\equiv \frac{B}{\tilde{c}^d_{12}}$, 
the matrices 
become: 
\be{pha1}
\tilde{\bY}_d = \matr{0}{-A}{0} 
{-A }{Y_{ct}D \mbox{e}^{{\rm i}\xi}}{ \tilde{B} } 
{0}{ \frac{1}{b} \tilde{B} }{D} ,  ~~~~~~
\tilde{\bY}_e = 
\matr{0}{- k_3 A }{0} {- k_3 A }
{Y_{ct} D \mbox{e}^{{\rm i}\xi}}
{\frac{1}{b} \tilde{k}_1 \tilde{B} } 
{ 0}{ \tilde{k}_1 \tilde{B} }{ D}~ . 
\ee
where 
$\tilde{k}_1= k_1 \sqrt{(\tilde{c}^{d}_{12})^2 + 
(\tilde{s}^{d}_{12} k_2/k_1)^2 }$  
Finally, these matrices 
$\tilde{\bY}_e$ and $\tilde{\bY}_d$ can be diagonalized 
by bi-unitary transformations as in eq. (\ref{unit}) with 
the matrices $U_{e,d}$ and $ U'_{e,d}$ understood as:
\beqn{unitary}
&&U  = U_{32}U_{13}U_{12} 
\equiv \\ 
&&\matr{1}{0}{0} {0}{c_{23}}{s_{23} \mbox{e}^{i\phi_{23}}} {0}{-s_{23}
\mbox{e}^{-i\phi_{23}}}{c_{23}} 
\matr{c_{13}}{0}{-s_{13}\mbox{e}^{-i\phi_{13}}} {0}{1}{0} 
{s_{13}\mbox{e}^{i\phi_{13}}}{0}{c_{13}}  
\matr{c_{12}}{s_{12}\mbox{e}^{i\phi_{12}}}{0} 
{-s_{12}\mbox{e}^{-i\phi_{12}}}{c_{12}}{0} {0}{0}{1}. \nonumber 
\eeqn 
Therefore, the CKM matrix is 
$V_q= \tilde{O}^d_{12} P_+ U_d$ and  is 
parameterised as 
\be{CKM2}
V_q=\matr{\tilde{c}^d_{12} e^{-i\sigma}v_{ud} + \tilde{s}^d_{12}v_{cd}}
{\tilde{c}^d_{12}e^{-i\sigma}v_{us}+\tilde{s}^d_{12}v_{cs}}
{\tilde{c}^d_{12}e^{-i\sigma}v_{ub}+\tilde{s}^d_{12}v_{cb}} 
{\tilde{c}^d_{12}v_{cd} -\tilde{s}^d_{12}e^{-i\sigma}v_{ud}}
{\tilde{c}^d_{12} v_{cs}- \tilde{s}^d_{12}e^{-i\sigma}v_{us}}
{\tilde{c}^d_{12}v_{cb}- \tilde{s}^d_{12}e^{-i\sigma}v_{ub}} 
{v_{td}}{v_{ts}}{v_{tb}} 
\ee
where  by $v_{ud}, v_{us}$ {\it etc.} 
with {\it lower} $v$ we 
 mean the matrix elements of the quark 
mixing $U_d$ as defined in eq. 
(\ref{CKM}) and parameterised as in (\ref{MNS}).
Analogously, the effective  leptonic 
mixing matrix is $V_l = U^\dagger_e P^\dagger_{-} \tilde{O}^{e T}_{12}$: 
\be{lept-mix}
V_l=\matr{-\tilde{c}^e_{12} e^{i\sigma}v_{e1} + \tilde{s}^e_{12}v_{e2}}
{\tilde{c}^e_{12}v_{e2}+ \tilde{s}^e_{12}e^{i\sigma}v_{e1}}
{v_{e3}} {-\tilde{c}^e_{12}e^{i\sigma}v_{\mu 1} + \tilde{s}^e_{12}v_{\mu2}}
{\tilde{c}^e_{12}v_{\mu2}+\tilde{s}^e_{12}e^{i\sigma}v_{\mu1}}{v_{\mu3}} 
{-\tilde{c}^e_{12}e^{i\sigma}v_{\tau1} +\tilde{s}^e_{12}v_{\tau2}}
{\tilde{c}^e_{12}v_{\tau2} +\tilde{s}^e_{12}e^{i\sigma}v_{\tau1}}
{v_{\tau3}} 
\ee
where with {\it lower} $v_{e1}, v_{e2}$ {\it etc.} are denoted  
the matrix elements of the 
leptonic mixing matrix  $U^\dagger_e$. 
These forms  of 
quark and  leptonic mixing matrices 
are convenient since 
the contributions $v_{us}, v_{ud}$ and  
$v_{e1}, v_{e2}$ {\it etc.} 
from the   
Yukawa matrices $\tilde{\bY}_{e,d}$ of 
(\ref{pha1}), 
are separated 
from that arising from 
the initial 12-rotation $\tilde{O}^{d,e}_{12}$.
Had the  matrices $\tilde{\bY}_{e,d}$ the form 
in (\ref{Fred})  (i.e. 
vanishing 22 entry) the above 
transformations $U_{e,d}$
would be pure rotations. 
Then the only source of CP violation would 
reside in the residual phase 
$\sigma$ in $P_{\pm}$ driven by the (1,3), (3,1) entries. 
It is worth  noticing that 
the CP violating phase appearing in the
quark and leptonic mixing matrices 
differs by $\pi$ (cfr. 
(\ref{P-O12})).
This feature comes from the antisymmetry of the 12 block, 
while it does not depend 
on the initial antisymmetry 
in the 23 and 13 blocks.

The GUT scale Yukawa eigenvalues 
$Y_{u,c,t},Y_{d,s,b}, Y_{e,\mu,\tau},
Y_{1,2,3}$,  are 
linked to the physical 
fermion masses  through the  
renormalization group equations (RGE). 
For moderate values of $\tanb=v_u/v_d$, one obtains at 
one-loop\footnote
{The notation is the standard one: for the heavy quarks $t,b,c$,
$m_{t,b,c}$ are  their running masses respectively at $\mu=m_{t,b,c}$,
while for the light quarks $m_{u,d,s}$ are given  at $\mu=1$ GeV. 
} (see e.g. \cite{RG}):
\be{RG}
\begin{array}{lll}
m_u=Y_uR_u R_u^{SM}\eta_u B_t^3 v_u\,,~~~~& 
m_d=Y_dR_d R_d^{SM}\eta_d v_d\,,~~~~& 
 m_e=Y_e R_e R_e^{SM} v_d  \nonumber  \\
m_c=Y_cR_u R_u^{SM}\eta_c B_t^3 v_u\,,~~~~&
m_s=Y_sR_d R_d^{SM}\eta_s  v_d\,,~~~~& 
m_{\mu}=Y_{\mu} R_e R_e^{SM} v_d  \nonumber\\
m_t=Y_t R_u R_u^{SM} B_t^6 v_u\,,~~~~ &
m_b=Y_bR_d R_d^{SM}\eta_b B_t v_d\,,~~~~ &
m_{\tau}=Y_{\tau} R_e R_e^{SM} v_d  , \end{array}
\ee
and for  the neutrino masses
\be{nu-RGE}
m_{1,2,3} = \frac{Y_{1,2,3}}{M} R_\nu R_\nu^{SM} B^6_{\tilde{t}} v^2_2
\ee
where the factors $R_{u,d,e,\nu}$ and $R_{u,d,e,\nu}^{SM}$  
account for the gauge-coupling 
induced running from the GUT scale $M_G\simeq 10^{16}$ GeV 
to the SUSY breaking scale $M_S\simeq M_t$ and from $M_S$ to the 
electroweak scale $M_Z$, respectively.   
The factors $\eta_f$ encapsulate the QCD+QED running from 
$M_S$ down to $m_f$ for $f=b,c$ 
(or to $\mu=1$ GeV for the light quarks $f=u,d,s$). 
Namely, for $\al_s(M_Z)=0.119\pm 0.004$ we have 
\be{etas}
\begin{array}{llll}
R_u R^{SM}_u  =3.53^{+0.06}_{-0.07} , ~~~~~& 
R_d R^{SM}_d =3.43^{+0.07}_{-0.06}  
~~~~~ & R_e R^{SM}_e =1.50 , ~~~~~& R_\nu R^{SM}_\nu =1.15 ,
   \nonumber \\ 
\eta_b=1.53^ {+0.03}_{-0.04} , ~~~~~& \eta_c=2.05^{+0.13}_{-0.11} , 
~~~~~ & \eta_{u,d,s}=2.38^{+0.24}_{-0.19} . ~~~~~ &  \end{array}   
\ee 
The factor $B_t$ includes the running induced by the 
large top quark Yukawa constant ($Y_t\sim 1$).\footnote{
In the RG running of neutrino masses (\ref{nu-RGE}), the factor 
$B_{\tilde{t}}$ is function of $Y_t/\sin\beta$.} 
For $Y_t$ varying from the lower limit $Y_t=0.5$, imposed by the 
top pole-mass,  to the perturbativity limit $Y_t\approx 3$, the function 
$B_t$ decreases from 0.9 to 0.7.
Regarding the CKM elements, their physical values are related to 
the corresponding GUT-scale quantity 
(labelled by the superscript $G$) as follows
\footnote{ 
We shall not take into account 
the analogous RGEs \cite{RG-nu} 
for the neutrino mixings,  
since the experimental data   
still  contain big error bars 
and such an improvement is not justified. Moreover, 
renormalization 
effects are mostly expected in 
case of strong mass degeneracy 
\cite{RG-rec} which is not our case.}:
\beqn{low-ckm}
&&V_{us (d)} = V^G_{us(d)}~, ~~~~~~~  
V_{cs (d)} = V^G_{cs(d)}~, ~~~~~~~
V_{tb} = V^G_{tb}~, \\ \nonumber
&&
V_{c (u)b} = V^G_{c(u)b} B^{-1}_t~,   ~~~~~
 V_{ td(s)} = V^G_{td(s)} B^{-1}_t~.
\eeqn
Using 
the procedure outlined 
in detail in \cite{BR}, 
the following  relations (valid at the GUT scale) 
are derived from the matrices (\ref{pha1}):
\beqn{b-tau1}
&& D=
Y_\tau\left[\frac{1-(b+b^{-1})\frac{Y_\mu-Y_e}{Y_\tau}}
{F_e}
\right]^{1/2} 
=Y_b\left[\frac{1-(b+b^{-1})\frac{Y_s-Y_d}{Y_b} }
{F_d}
\right]^{1/2}~,\\
&& F_{e (d)}= 1 +\tilde{c}^{e (d) 2}_{12}Y_{ct} (b+b^{-1}) \cos\xi
\eeqn 
and 
\be{dsb1}
A^2D = Y_eY_\mu Y_\tau = k_{3}^2 Y_dY_sY_b~,
\ee
\be{sb1}
\tilde{B}^2=\frac{b}{\tilde{k}_{1}^2}
(Y_\mu-Y_e)Y_\tau \frac{I_e }{F_e} = 
b (Y_s-Y_d)Y_b\frac{I_d}
{F_d}~, 
\ee
where we have defined 
\be{Ied}
I_e = 1 +\tilde{c}^{e 2}_{12}\frac{Y_\tau}{Y_\mu-Y_e}Y_{ct}\cos\xi~, ~~~~~~~
I_d= 1 +\tilde{c}^{d 2}_{12}\frac{Y_b}{Y_s-Y_d}Y_{ct}\cos\xi~.
\ee
In the following we directly substitute the Yukawa constant 
ratios with the corresponding mass ratios 
whenever the latter are RGE invariant, 
e.g. $Y_\mu/Y_\tau=m_\mu/m_\tau$, $Y_d/Y_s=m_d/m_s$, etc. 
Now, by dividing the squared of 
the l.h.s and r.h.s in eq. (\ref{b-tau1}) 
by the corresponding sides of  eq. (\ref{sb1})  and inverting, 
we obtain the following implicit relation:
\be{s/b1}
\frac{Y_s-Y_d}{Y_b} =  
\frac{I_e F_d \tilde{k}_{1}^{-2}}{I_d F_e Z^2}
\frac{m_\mu-m_e}{m_\tau}
\approx 0.059 \frac{I_e F_d \tilde{k}_{1}^{-2}}{I_d F_e Z^2} ,
\ee
where 
\be{zeta}
Z =
\left(\frac{F_d}{F_e}\right)^{1/2}
\left[{  
1 - \left( 1 - \frac{I_e}{I_d \tilde{k}^2_1}
\right) 
(b+b^{-1})\frac{m_\mu-m_e}{m_\tau} }\right]^{1/2} .
\ee
Substituting the expression (\ref{s/b1})  
back into eq. (\ref{b-tau1}) we get: 
\be{b/tau}
\frac{Y_b}{Y_\tau} =  Z  ,
\ee
which shows the modification induced on the  expression of the 
$b$--$\tau$ 
Yukawa unification at the GUT scale by  the asymmetry 
in the 23 block.
Analogously, by utilizing the relation (\ref{b/tau}) 
in  eq. (\ref{sb1}), the latter can be rewritten as 
\be{s/d1}
\frac{Y_s-Y_d}{Y_\mu-Y_e} = 
\frac{\tilde{k}_{1}^{-2}}{Z}\dfrac{ I_e F_d}{ I_d F_e}~,
\ee
while 
by dividing the squared of the r.h.s. and l.h.s. of 
(\ref{sb1}) by the corresponding sides of (\ref{dsb1}) we have:
\be{s/d2}
 \frac{m_s}{m_d} + \frac{m_d}{m_s} -2  = 
\frac{k_{3}^2} {\tilde{k}_{1}^4 Z}
\left(\frac{  I_e F_d}{I_d F_e  }\right)^2 
\left(\frac{m_\mu}{m_e}+\frac{m_e}{m_\mu} -2 \right) 
\approx 204.7 
\frac{k_{3}^2} {\tilde{k}_{1}^4 Z}
\left(\frac{I_e F_d}{I_d F_e}\right)^2 .  
\ee
Compare the present expressions (\ref{s/b1},\ref{s/d1},\ref{s/d2}) and 
 (\ref{zeta}) with the corresponding ones in eqs. 
(\ref{bsd},\ref{s/d}) and (\ref{btau}) obtained in the 
Fritzsch-like case ($C=0, Y_{ct}, Y_{ut}\to 0$, i.e. $F_{e,d}, I_{e,d} 
\to 1$).  
From eqs. (\ref{b/tau}) and (\ref{s/d1}) we can extract the following 
physical masses:  
\beqn{bottom}
&&
m_b = \frac{R_d R^{SM}_d\eta_b}{R_e R^{SM}_e} B_t Z m_\tau = 
B_tZ\cdot (6.22^{+0.25}_{-0.27}) ~ {\rm GeV}~, 
\nonumber \\
&& 
m_s-m_d = \frac{R_d R^{SM}_d\eta_d}{R_e R^{SM}_e} 
\frac{ I_e F_d}{ \tilde{k}_{1}^{2}Z I_d F_e}  
(m_\mu-m_e) 
= \frac{ 4 I_e F_d}{ \tilde{k}_{1}^{2}Z I_d F_e}  
 \cdot (143^{+18}_{-14})~{\rm MeV} . \label{stran}
\eeqn

We now turn  to the quark and lepton mixing. For the angles 
$\theta^{e,d}_{23}$ and the corresponding phases appearing in the 
unitary transformation (\ref{unitary}) we find:
\beqn{ang-23} 
&&\tant^{e}_{23} = 
 2\sqrt{b}\sqrt{(\frac{m_\mu-m_e}{m_\tau})
(1 + \frac{2}{b} Y_{ct} \cos\xi)}
\frac{ \left[(1-I_e(b+b^{-1})\frac{m_\mu-m_e}{m_\tau}) \right]^{1/2} }
{1 - 2b \frac{m_\mu-m_e}{m_\tau}+( b-b^{-1})  Y_{ct} 
\cos\xi}  ,   \\  
&&\tant^{d}_{23} =\!  
 \frac{2}{\sqrt{b}}\sqrt{(\frac{Y_s-Y_d}{Y_b})
(1 + {2}{b} Y_{ct} \cos\xi)}
\frac{ \left[ (1-I_d(b+b^{-1})\frac{Y_s-Y_d}{Y_b}) \right]^{1/2} }
{1 - 2b^{-1} \frac{Y_s-Y_d}{Y_b}+ ( b-b^{-1})Y_{ct}\cos\xi}~,   \\
&& 
\tan\phi^{e}_{23}= \frac{ \frac{Y_{ct}}{b}\sin \xi}{1 +
\frac{Y_{ct}}{b}\cos\xi} \approx \frac{Y_{ct}}{b}\sin\xi~,\\
&&
\tan\phi^{d}_{23}= \frac{ b Y_{ct}\sin\xi}{1 +
bY_{ct}\cos\xi} \approx bY_{ct}\sin\xi~,
\eeqn 
while the right rotation angles $\theta'^{e,d}_{23}$ and $\phi'^{e,d}_{23}$ 
are 
obtained from these expressions substituting respectively 
$b \to b^{-1}$.
The phases  appear strongly suppressed by the tiny 
$Y_{ct}$  parameter: for $Y_{ct}\sim 3\cdot 10^{-3}$ and $b\sim 10$,  
$~\phi^d_{23}\lsim 0.01$ .
The 13 mixing angles are 
\beqn{ang-13}
&&
\sin\theta^e_{13} = \sqrt{a_e} \frac{s'^e_{23}}{c'^e_{23}}
\frac{\sqrt{m_em_\mu}}{m_\tau} = 
\left(\frac{m_em_\mu^2}{bc^e_{23} m_\tau^3}\right)^{1/2} \leq 10^{-3}, 
~~~~~~~~ a_e = \frac{c'^e_{23}}{c^e_{23}}~,\\
&&
\frac{s^d_{13}}{s^d_{23}} = \frac{1}{\sqrt{a_d}} 
\frac{s'^d_{23}}{c'^d_{23} s^d_{23}} 
\sqrt{\frac{m_d}{m_s}}\frac{Y_s}{Y_b}~,
~~~~~~~~~a_d = \frac{c^d_{23}}{c'^d_{23}}~.
\eeqn 
One can easily verify that for the corresponding phases   $\phi^{e,d}_{13}=
\phi^{'e,d}_{23}$.
Finally the $U_{12}$ transformations are expressed by the following angles:
\beqn{ang-12}
&&
\tant^{e}_{12} = 2\sqrt{a_e}\sqrt{\frac{m_e}{m_\mu}}
\frac{ \left[1-(a_e+a_e^{-1})\frac{m_e}{m_\mu} \right]^{1/2} }
{1 - 2a_e \frac{m_e}{m_\mu} }~  
\approx \frac{2}{\sqrt{c^e_{23}} }\sqrt{\frac{m_e}{m_\mu-m_e}}
\approx \frac{0.140}{\sqrt{c^e_{23}} }~,   \\ 
&& 
\tant^{d}_{12} = \frac{2}{\sqrt{a_d}}\sqrt{\frac{m_d}{m_s}}
\frac{ \left[1-(a_d+a_d^{-1})\frac{m_d}{m_s} \right]^{1/2} }
{1 - 2a_d^{-1}\frac{m_d}{m_s} } \approx
2\sqrt{c'^d_{23}}\sqrt{\frac{m_d}{m_s-m_d} }~ .  \\ 
\eeqn
while the  corresponding phases are
\beqn{pha-12}
&&
\tan\phi^e_{12}\approx 
\frac{\sin\xi}{Y_{ct}^{-1} \frac{m_\mu}{m_\tau} -\cos\xi}\lsim
\left[ (Y_{ct}^{-1} \frac{m_\mu}{m_\tau})^2 -1 \right]^{-1/2} \approx 
4.6\times10^{-2} \nonumber \\
&&
\tan\phi^d_{12}\approx  
\frac{\sin\xi}{Y_{ct}^{-1} \frac{Y_s}{Y_b} -\cos\xi}\approx 
\frac{\sin\xi}{ \frac{Y_{ct}^{-1}}{(k_{1} Z)^2} 
\frac{m_\mu}{m_\tau} -\cos\xi}\lsim 
\left[ ( Y_{ct}^{-1} 
\frac{m_\mu}{m_\tau} )^2 (\frac{1}{k_1 Z})^4 -1 \right]^{-1/2} 
\eeqn
We can easily estimate that $\phi^d_{12}$ can be at most $11^{\circ}$ 
and  $25^{\circ}$  with $b\gsim 1$, $~Y_{ct} = 3 \times 10^{-3}$  and 
for $k_{1}= 2$ and $k_{1}= -3$ , respectively. 
Moreover,  the larger
is the asymmetry parameter $b$, 
the  smaller the function $Z$ becomes and therefore  
the smaller the phase $\phi^d_{12}$ appears.
In conclusion,  the phase $\xi$ 
cannot provide by itself 
a sizeable source of CP violation since it is driven by the tiny ratio 
$Y_{ct}$.  That is the reason motivating us to introduce 
non-vanishing (1,3), (3,1) entries  driving the extra phase $\sigma$. 
Hence, as a matter of fact 
we may think of the matrices $U_{e,d}$ as pure rotations, parameterised 
in the standard form (\ref{CKM}) with the phase $\delta =\pi$ \cite{BR}.

\begin{table}[ht]
 \begin{center}
 \begin{tabular}{||c|c||}
	\hline \hline
	Observables & Values  \\
	\hline
	$\bullet~ m_{e}\left[\mbox{MeV}\right]$ & $0.511$  \\
	\hline
	$\bullet~ m_{\mu} \left[\mbox{MeV}\right]$ & $105.7$   \\
	\hline
	$ \bullet~ m_{\tau}\left[\mbox{GeV}\right]$ & $1.777$  \\
	\hline
	$\bullet~ m_u/m_t $ & $(0.9 \div 3)\times10^{-5}$ \\
	\hline
	\hline
	$\bullet~ m_{c}/m_t $ & $(8.0\pm 1.3)\times10^{-3} $   \\
	\hline
	$\bullet~Q$ & $22.7\pm0.8$\\ 
	\hline
        $\bullet~ |\eps_K|$ & $(2.280\pm0.013)\times10^{-3}$   \\
	\hline
	\hline
	$m_{s}\left[\mbox{MeV}\right] $ & $155\pm75$  \\
	\hline
	$m_b\left[\mbox{GeV}\right] $ & $4.25\pm0.15$   \\
	\hline
	$m_s/m_d$ & $21\pm4$   \\
	\hline
	$m_u/m_d$ & $0.2\div 0.7$  \\
        \hline
	$M_t \left[\mbox{GeV}\right]$ & $173.8\pm5.2$ \\
	\hline
	$|V_{cb}|$ & $0.0395\pm0.0017$  \\
	\hline
	$|V_{us}|$ & $0.2196\pm0.0023$  \\
	\hline
	$\left|{V_{ub}}/{V_{cb}}\right|$ & $0.093\pm0.014$   \\
	\hline
	$|V_{td}|$ & $0.0088\pm 0.0018$  \\
	\hline
	$\left|{V_{td}}/{V_{ts}}\right|$ & $< 0.22$   \\
	\hline
	$B_K$& $0.8\pm0.2$   \\
	\hline
	\hline 
 \end{tabular}
\caption[]{\small Physical quantities used in the phenomenological 
analysis. Those marked by $\bullet$ are used as input for the fits.
The bounds on $|V_{td}|$ and $|V_{td}/V_{ts}|$ are inferred from 
the evaluation of $B^0_d-\bar{B}^0_d$ and $B^0_s-\bar{B}^0_s$ 
transitions in the standard model \cite{maiani} and 
thus they are not literally valid in the context 
of supersymmetric model; even in the MSSM framework, 
with flavour-aligned soft terms,  
one can expect substantial supersymmetric contributions 
to these transitions.  
}
 \label{tab4}
 \end{center}
 \end{table}

In the analysis, we test the amount of  CP-violation through
the  parameter  $\eps_K$ describing 
the CP violation in $K\rightarrow \pi\pi$. More precisely,  
we use the more uncertain parameter $B_K$  parameterising 
the  deviation from the vacuum saturation limit in 
$K^0-\bar{K}^0$ transition which we `extract' from the 
expression of $\eps_K$  as given 
in the standard model \cite{buras}:
\be{epsk}
B_K  = |\eps_K| \left[\frac{f_K^2 M_K}{\Delta M_K} 
\frac{G_F^2 m_W^2}{12\sqrt{2}\pi^2} 
\frac{{\rm Im}({\cal F}^\star t_u^2) }
{|t_u|^2} \right]^{-1} .
\ee
and compare with its theoretical value which 
ranges from 0.6 to 1. 
Here $f_K=161$ MeV is the kaon decay constant, 
$M_K= 497.7$ MeV is the $K^0$ mass, 
and  $\Delta M_K=0.53 \times 10^{10}$ s$^{-1}$ 
is the experimental value of $K_L-K_S$ mass difference. 
The mixing angles enter through the combinations 
$t_{\alpha} = V_{\alpha s}^\star V_{\alpha d}$: 
$t_u$ appears explicitly,\footnote{
Note that in most of $V_{CKM}$ parameterizations, 
the factor $t_u$, coming from the amplitude of the
decay $K^0\to 2 \pi$, is made real and thereby
the factor $t_u^2/|t_u^2|$ becomes 1. In our 
final parametrization of the CKM matrix this is not the case --  
see (\ref{CKM2}).} 
whereas $t_c, t_t$ are contained in the function 
\be{sm}
{\cal F} = \eta_1 t_c^2 S_0(x_c) +  \eta_2 t_t^2 S_0(x_t)+
 2\eta_3 t_c t_t S_0(x_c, x_t)
\ee
where $\eta_1 \sim 1.38$, 
$\eta_2 \sim 0.57$,  $\eta_3 \sim 0.47$ are QCD correction factors and 
$S_0$ is one of  the Inami-Lim functions 
($x_{c,t}= \frac{m_{c,t}^2}{m_W^2}$).
In this way, we are testing whether the experimental value 
of $\eps_K$ can be reproduced entirely by the standard 
model contribution associated with the box diagrams 
involving the charm, top quark and the W boson. 
It is known, that in the MSSM or more generally, 
in models with flavour-aligned soft terms like in 
 \cite{ZB}, the supersymmetric contributions are 
negligible.  
However, in general SUSY GUT context 
$\eps_K$ can receive substantial contributions from the 
super-partners at the weak scale \cite{FCNC,CP-susy}. 
In view of this, we shall not use the information 
from $\eps_K$ to constrain the parameter space. 

\subsection{Model parameters versus physical observables}

We can now proceed by confronting the  parameters 
with the physical observables described by the model.
The number of the former  amount 
to 14, namely $Y_{u,c,t},  A, B, D, 
\tilde{t}^{d}_{12}~, b, \sigma, \xi,
k_1, k_2,\\ k_3, M_L $ while 
 the number of  observables is 20: 
the up-quark masses 
$m_{u,c,t}$, 
the down-quark masses $m_{d,s,b}$, the charged-lepton masses 
$m_{e,\mu,\tau}$, 
the neutrino masses $m_{1,2,3}$, 
four independent CKM parameters and other four parameters of the 
leptonic mixing matrix. 
As it will be discussed 
more extensively below, 
the
three Clebsch coefficients $k_1, k_2, k_3$ 
are treated as external parameters, 
fixed at some specific GUT-inspired  value.

The various physical observables deserve 
a different treatment in the analysis for at least two reasons. 
First, they are known with 
different degree of accuracy. 
Second, some of them have 
a minor impact on the determination of the 
`flavour' parameters themselves. 
Therefore, the observables are classified and used as follows:
\begin{enumerate}
\item
$m_e, m_{\mu}, m_\tau$:  these masses are determined 
with high 
accuracy so they   are used as inputs  to extract 
the parameters $A, B, D$ via eqs. (\ref{b-tau1}-\ref{sb1}).
Notice that in this way, also the explicit dependence 
on $\tan\beta$ of the fermion masses 
is automatically absorbed in the parameters  $A, B, D$. 
\item
$\frac{m_u}{m_t}, \frac{m_c}{m_t}$:  
these mass ratios enter only into the 
determination of $Y_{ut}, Y_{ct}$ and 
are used as inputs\footnote{Now for the sake of 
discussion we consider also $Y_{ut}$ but in practice, as already 
mentioned, we neglect it at all.}. 
Notice, however, that $Y_{ut}, Y_{ct}$ quite depend on $Y_t$ 
through the renormalization factor $B_t$: 
e.g. $Y_{ct} = m_{c}/m_t B^3_t \eta^{-1}_{c}$. 
 As 
this ratio is tiny, 
the dependence on $m_c/m_t$ is weak and hence our results 
are not sensitive to the corresponding errors on $m_c$ and $m_t$. This 
justifies the fact that we use  the central values of 
the ratio $m_c/m_t$ 
shown in Table \ref{tab4}.  
On the other hand, the Yukawa constant  $Y_t$ is 
left as free parameters. 
Indeed, it is not sensible  to fix $Y_t$ by the top mass 
due to its infrared behaviour.  
It will turn out that (in almost every case) 
the  preferred value for  
$Y_t$ is   0.5 which enhances the effect of $Y_{ct}$.  
This value is indeed 
the smallest one compatible with the top mass 
for which $m_t = (160.9\pm 2.2)\sin\beta$ GeV  
for $\alpha_s(M_z) =0.119\pm 0.004$.
Correspondingly, we have for  the  ratio
$Y_{ct}= (2.7^{-0.1}_{+0.2})\cdot 10^{-3}$.  

\item
$m_d, m_s, m_b, 
|V_{us}|, |V_{cb}|, |V_{ub}/V_{cb}|$: 
these observables are fitted 
by a $\chi^2-$like procedure (see Sect. 4.4) 
to constrain 
the remaining parameters  they depend on, 
namely $Y_t, b, 
\tilde{t}^d_{12}, \sigma, \xi$ once the Clebsch factors 
$k_1, k_3$ are prescribed (see Sect. 4.3).

\item
$M_t, m_u/m_d$: the top pole mass $M_t = m_t \cdot\left
(1 + \frac{4}{3\pi}\alpha_s(M_t) + \frac{11.4}{\pi^2}\alpha^2_s(M_t)
\right)$
is determined as output 
once $Y_t$ is fixed by the fit in point 3., 
modulo $\sin\beta$. Though in our analysis we cannot 
determine univocally   (and in a sense is not necessary) 
$\tan\beta$, the model we presented in Sect. 3  naturally 
yields  $\tan\beta$ in the small or moderate regime. 
Using the determination of $m_s/m_d$ from the fit, we extract 
the ratio $m_u/m_d$ from the  combination \cite{Q-el}
\be{Q-leu}
Q =   \dfrac{m_s/m_d}{\sqrt{1 - (m_u/m_d)^2}} .
\ee

\item
$B_K$: the CP violation parameter is given as a prediction from the 
result of the fit performed in point 3.
\item
$|V_{\mu 3}|, |V_{e 2}|, |V_{e3}|$ $\&$  lepton CP-phase: 
these are also given as predictions. 
More precisely, for the sake 
of comparison with the experimental data, 
we trade these three elements for the    
corresponding `oscillation' parameters:  
$\sin^2 2\theta^l_{23} = 4 |V_{\mu 3}|^2 (1 - |V_{\mu 3}|^2)$,      
$\sin^2 2\theta^l_{12} = 4 |V_{e2}|^2 (1 - |V_{e2}|^2)$ and  
$\sin^2 2\theta^l_{13} = 4 |V_{e3}|^2 (1 - |V_{e3}|^2)$. 
Notice, however, that the predictions for both 
CP violation phase and  the element $|V_{e 2}|$  depend 
also on $\tilde{t}^e_{12} = \frac{k_2}{k_1}\tilde{t}^d_{12}$ i.e. on 
the Clebsch $k_2$. 
\item
$m_1, m_2, m_3$: the neutrino masses follow immediately as predictions 
$m_2/m_3= b Y_{ct}$, $m_1/m_2= Y_{uc}$. However, the determination 
of any single mass does require  the last parameter i.e. the 
mass scale $M_L = \frac{M}{\eta}$.  
\end{enumerate}
In summary,  we 
have that  from the sets 1) and 2) five parameters are 
fixed -- $A, B, D$, $Y_{ut},Y_{ct}$. From the set 3) other 
five parameters are constrained by the fit --
$b, \tilde{t}^d_{12},  \sigma, \xi, Y_t$ once  the 
two Clebsches $k_1, k_3$ have been assigned. 
The determination 
of these twelve parameters allows us to make five {\it clean} 
predictions: $m_1/m_2, m_2/m_3$,   
$B_K$, $|V_{\mu 3}|, |V_{e 3}|$. 
Also $M_t$ is given as a prediction within the uncertainty 
of $\tan\beta$ expected to be in the range $1 -10$. 
From the result of the fit, we also gain the ratio 
$m_u/m_d$ by using the information on 
the parameter $Q$ (fixed at its central value).
 
Finally, 
the neutrino masses depend on  $M_L$ and  
both  $|V_{e 2}|$ and the amount of CP violation 
in the lepton mixing matrix depend on 
the Clebsch coefficient $k_2$.  For definiteness, 
the flavour mass scale $M$ and the lepton-violation mass scale $M_L$ 
are  set at the GUT scale $M_G$, i.e. $M=M_L = M_G$  which  
as already discussed 
means that $\dfrac{b}{\eta Y_t} \sim 10^{-1}$ or 
$\eta \sim 10^2$ for $Y_t \sim 1, b \sim 10$.

In the next Section we shall discuss 
more extensively how we handle the Clebsch factors.

Finally, the phenomenological analysis 
requires also the parameters 
associated with the standard model gauge sector: $\alpha_s(M_Z),  
\alpha_{em}, \sin^2\theta_w$. For $\alpha_{em}, \sin^2\theta_w$ 
we use the central values quoted in \cite{maiani}. 
As regards $\alpha_s(M_Z)$, 
the present world average is   
$\alpha_s(M_Z)=0.119\pm 0.002$~ \cite{maiani}. 
On the 
other hand the $SU(5)$ unification of 
gauge couplings implies   
a larger value, 
$\alpha_s(M_Z)\approx0.125$. Therefore, 
as a compromise, we have taken $\alpha_s(M_Z)=0.123$. 
We remind that only $m_b$ and $m_s$ are 
sensitive to variations of 
$\alpha_s(M_Z)$ within the quoted error. 
The corresponding variations 
can be inferred from the 
gauge-renormalization factors given in (\ref{etas}). 
On the other hand, the $\alpha_s(M_Z)$ 
dependence of the running factor $B_t$ 
is less than 1\% and 
therefore  all other observables 
are not sensitive to $\alpha_s(M_Z)$.

\subsection{Clebsch prescription: three different ans\"atze}
First let us recall  how 
the Clebsch coefficients
enter into the determination of the physical 
quantities.   
The Yukawa eigenvalues $Y_{e,\mu,\tau}$ 
and $Y_{d,s,b}$ 
depend only on $k_1$ and $k_3$ 
(see the analytical expressions in 
Sect. 4.1).\footnote{The dependence on $k_2$ 
contained in $\tilde{k}_1$ in eq. 
(\ref{pha1}) is mild for $\tilde{t}^d_{12}<0.1$.} 
Similarly, the mixing angles $v_{us}, v_{ud}, \cdots $ 
also depend  
only on $k_1, k_3$. 
The coefficient $k_2$ affects only the lepton mixing matrix 
(\ref{lept-mix}) via the 
initial  
12 rotations $\tilde{t}^{e}_{12}= 
\frac{k_2}{k_1}\tilde{t}^d_{12}$.  
Thereby 
$k_2$ can only be fixed for example by the 12 lepton mixing angle 
$V_{e2}$ through the SN parameter 
$\sin^2 2\theta^l_{12} = 4 |V_{e2}|^2 (1 - |V_{e2}|^2)$. 
Notice, indeed, that  both 
$\sin^2 2\theta^l_{13}$ and 
 $\sin^2 2\theta^l_{23}$
do not depend on $k_2$. 

We can guess very easily the needed 
range for such $k_{1,2,3}$. 
Indeed $k_{1,3}$  are mainly fixed 
by the strange mass in (\ref{bottom}) and 
the ratio $m_s/m_d$ in (\ref{s/d2}). 
From $m_s$  
we infer that $k_1$ is to lie within $\frac53 \div 3$ -- for 
$b\gsim 8$. Correspondingly, from $m_s/m_d$, $k_3$ is forced 
to be in the range $ 1\div \frac83$. 
Hence, we have performed  
a preliminary scanning of 
the parameter space, by fitting  
the quark observables in point 3 (Sect. 4.2)   
with    $k_1, k_3$  free to float  
in the range $0\div 3$. 
In such a way we have specified   
three   Clebsch prescriptions  
for $k_1$ and $k_3$ and they are  given 
in  eqs. (\ref{CL-A}, \ref{CL-B}) and (\ref{CL-C}), 
featuring three different 
ans\"atze {\bf A, B, C}, recalled here for 
convenience:

\vspace{0.2cm}
\noindent
ansatz {\bf A}:$~~~~~~~~~~~~k_1=2~, ~~~~~~~~k_3=1$

\vspace{0.3cm}
\noindent
ansatz {\bf B}:$~~~~~~~~~~~~k_1=k_3= -3~,$

\vspace{0.3cm}
\noindent
ansatz {\bf C}:$~~~~~~~~~~~~k_1=-3~, ~~~~~~~~k_3=2$

\vspace{0.2cm}

On the other hand {\it a priori}  
we cannot stick to any specific value of $k_2$: 
only the result of the  
fit will allow us to envisage 
the value that can give the better 
prediction for $\sin^22\theta^l_{12}$. 
For that it is 
useful to write down explicitly\footnote{For the sake 
of brevity  we are assuming that  
the elements of the mixing matrices  
$U_{e,d}$ are real as we have seen that the phases 
$\phi^{e,d}_{ij} $ are typically tiny.} $V_{e2}$:
\be{Ve2_ex} 
|V_{e2}|^2= \tilde{c}^{e2}_{12}v^2_{e2}+\tilde{s}^{e2}_{12}v^2_{e1} 
+ 2\tilde{c}^e_{12}\tilde{s}^e_{12}v_{e2} v_{e1}\cos\sigma
\ee   
The present MSW range in eq. (\ref{SN}) translates into 
$|V_{e2}|= (1.6 - 6)\cdot 10^{-2}$.
The {\it lower} elements of $U^{\dagger}_e$ read as 
\be{low_ve}
v_{e2} \approx - s^e_{12} c^e_{23}~, ~~~~~~
v_{e1} \approx  c^e_{12} c^e_{13}\sim 1~,
\ee
then from eqs. 
(\ref{ang-12}) we have $|v_{e2}| \approx 
\sqrt{\frac{ c^e_{23} m_e}{m_\mu}}\sim 
0.07 \sqrt{c^e_{23}}$ 
which slowly decreases 
with the asymmetry parameter $b$ -- for $b\gsim 8$ it becomes 
around $(5 -6)\cdot 10^{-2}$  saturating by itself the present 
experimental  upper limit. 
Therefore, the r.h.s. of eq.(\ref{Ve2_ex}) tells us that 
the third term  can reduce $|V_{e2}|$ 
if 
$\tilde{t}^e_{12}\lsim 0.1$ and $\sigma < 90^\circ$. 
In this way $\sin^2 2\theta^l_{12}$ may become smaller than $10^{-2}$. 
Interestingly, it will turn out that 
the quark-sector fit demands  the corresponding 
initial 12 rotation $\tilde{t}^d_{12}$ to lie in a similar range,    
$\tilde{t}^d_{12}\lsim 0.1$,  in all the three ans\"atze. 
We also write the 
CKM elements in (\ref{CKM2}) to explicitly show the 
interplay between 
the 12 rotation $\tilde{t}^d_{12}$ and the CP-phase $
\sigma$:
\beqn{vud_ex}
&& |V_{us}|^2= \tilde{c}^{d2}_{12}v^2_{us}+\tilde{s}^{ d2}_{12}v^2_{cs} 
+ 2\tilde{c}^d_{12}\tilde{s}^d_{12}v_{us} v_{cs}\cos\sigma~, \\ \nonumber
&& |V_{cb }|^2= \tilde{c}^{d2}_{12}v^2_{cb}+\tilde{s}^{ d2}_{12}v^2_{ub} 
- 2\tilde{c}^d_{12}\tilde{s}^d_{12}v_{cb} v_{ub}\cos\sigma~, \\ \nonumber
&& |V_{ub }|^2= \tilde{c}^{d2}_{12}v^2_{ub}+\tilde{s}^{ d2}_{12}v^2_{cb} 
+ 2\tilde{c}^d_{12}\tilde{s}^d_{12}v_{cb} v_{ub}\cos\sigma~,   \\      
\eeqn
where 
\be{vud_low}
 v_{us} = c^d_{13} s^d_{12}\approx s^d_{12}~, ~~~~~
v_{cs} = c^d_{12} c^d_{23}\sim 1 ~~~~~~
 v_{cb} \approx s^d_{23}~,~~~~~~ v_{ub} = - s^d_{13}~. 
\ee
Therefore, 
if the fit of the quark quantities   selects $\sigma < 90^\circ$
and $\tilde{t}^d_{12} \lsim 0.1$, 
then the choice 
$k_2=k_1$ is  favoured 
as it implies  $\tilde{t}^{e}_{12}=\tilde{t}^{d}_{12} $.
On the other hand, if it comes out that 
$\sigma \geq 90^\circ$ (and irrespectively 
of $\tilde{t}^d_{12}$) then $k_2=0$ (i.e. $\tilde{t}^e_{12} =0$) 
is preferable in order not to 
raise $\sin^2 2\theta^l_{12}$ above $1.4\cdot 10^{-2}$.    
As we shall see in the next section 
the three ans\"atze will prefer 
different values of the phase $\sigma$ such that 
in the ansatz {\bf A} $k_2=k_1 = 2$ can be chosen and as a result  
$\sin^2 2 \theta^l_{12} < 0.01$ can be obtained. On the other hand, 
in both the other scenarios we have to select $k_2=0$ and 
content ourselves with $\sin^2 2 \theta^l_{12} > 0.01$. 
Needless to say that this picture in which all the 
Clebsch $k_{1,2,3}$ are 
taken real is ``minimal''. 
For example, by allowing 
for complex $k_n$, the lepton and 
quark mixing matrices would have 
different CP violating phases and therefore, 
for example, 
also in the ans\"atze {\bf B, C} we could achieve 
$\sin^2 2\theta^l_{12}< 0.01 $  
with non-vanishing $\tilde{t}^e _{12}$.
Finally, one may wonder about 
the alternative  possibility 
to account  for  
the large-mixing angle MSW solution of the   SN anomaly requiring   
$\sin^2 2\theta^l_{12} \sim 0.5 - 0.98$. That parameter range 
translates into $|V_{e2}|> 0.4$ 
which could be achieved within our
ans\"atze
by large $\tilde{t}^e _{12}$ -- 
$\tilde{t}^e_{12}\gsim 0.7-0.8$, or $\frac{k_2}{k_1} \gsim 7 -8$. 
Such $O(10)$ Clebsch coefficients
seem to come less naturally and 
probably  
at the price of a less economical Higgs content.

\subsection{Strategy for the fit}
To study the constraints on the parameters of the model from 
the down-quark masses and the CKM mixing 
angles  
we have performed an 
alternative $\chi^2-$ analysis.\footnote{Those 
observables are computed by numerically  diagonalizing  the 
Yukawa matrices  (\ref{pha1}) 
once the parameters $A, B, D$ are determined 
from the charged-lepton masses 
(for given $k_{1,2,3}$) through 
eqs. (\ref{b-tau1}-\ref{sb1}). 
The agreement between the numerical 
and the analitycal out-comings -- 
presented in Sect. 4.1 -- 
is at the level of one per mil.} 

The  CKM elements $|V_{us}|, |V_{cb}|, 
|V_{ub}/V_{cb}|$  are assumed to be mainly affected by Gaussian 
errors and therefore the standard $\chi^2$-function can be assumed for them:
\be{chi_s}
\chi^2 = \sum_a \left(\dfrac{\Delta_a}{\sigma_a}\right)^2~, ~~~~~~~~
\Delta_a={x_a - x_a^{exp}} 
\ee
where   the theoretical outcome for a certain  
observable is denoted by $x_a$ and 
the corresponding experimental  value and statistical error 
by   $x^{exp}_a$ and $\sigma_a$, respectively (see Table \ref{tab4}).
On the contrary, the uncertainties affecting   
the determination  of the down-quark masses are due to 
 the  dependence on the theoretical-model used to extract the masses from 
the measurements. For this reason we think that 
it is not correct   
to assign them a Gaussian distribution and  hence  to 
define the canonical $\chi^2$- function.  
Nevertheless, we interpret the 1-$\sigma$ range 
reported in the Table \ref{tab4} as a `reasonable' interval accounting  
for all possible (different) determinations. Therefore, we prescribe 
 the masses a flat distribution in that range. More precisely,  
we define the following   ${\chi}^2$-like function for the masses: 
\beqn{chi2_ns}
&& \tilde{\chi}^2_{b} =   \left\{ \begin{array}{ll}
0 & \mbox{if} ~ \abs{\Delta_b} < 0.15  \\
\left(\frac{\abs{\Delta_b}-0.15}{0.075}\right)^2 & 
\mbox{if}~ \abs{\Delta_b} \geq 0.15 ~ 
 \end{array}\right.  ~, \nonumber \\
&&\tilde{\chi}^2_{s} =   \left\{ \begin{array}{lll}
0 & \mbox{if} ~ \abs{\Delta_s} < 75 \\
\left(\frac{m_s-80}{10}\right)^4 & \mbox{if} ~ \Delta_s \leq  -75~ \\
\left(\frac{m_s-230}{30}\right)^4 & \mbox{if} ~\Delta_s \geq 75 ~ 
 \end{array}\right.  ~, \nonumber \\
&&\tilde{\chi}^2_{s/d} =   \left\{ \begin{array}{ll}
0 & \mbox{if} ~ |\Delta_{s/d}| < 3 \\
(|\Delta_{s/d}|- 3)^4 & \mbox{if}~ |\Delta_{s/d}| \geq 3 \end{array}
\right. ~, 
\eeqn
where $\tilde{\chi}^2_{b}$, $\tilde{\chi}^2_{s}$, $\tilde{\chi}^2_{s/d}$ 
refer to $m_b, ~ m_s, ~m_s/m_d$, respectively and 
$\Delta_b,~ \Delta_s, ~\Delta_{s/d}$ are the corresponding deviations 
defined as  in eq. (\ref{chi_s}).
We can note from (\ref{chi2_ns}) that 
mass values outside the 1-$\sigma$ range are strongly penalized by the 
higher-power dependence of the (arbitrary) assigned distribution.
The sum of the above functions will be  denoted as  $\tilde{\chi}^2 = 
\tilde{\chi}^2_{b} +\tilde{\chi}^2_{s} +\tilde{\chi}^2_{s/d}$. 

{\small
\begin{table}
\begin{center}
\begin{tabular}{||c||c|c|c|c||}  \hline \hline
Ansatz ${\bf A}$: & $I.~C=0$ & $II. ~Y_{ct}=0$ 
& $III.~{\rm Complete}$ & $IV.~ {\rm Spont. CP}$ \\
$k_{1,2}=2, k_3=1$ & $(Y_t,b,\xi)$ & 
$(Y_t,b,\sigma,\tilde{t}_{12}^d)$ & 
$(Y_t,b,\xi,\sigma,\tilde{t}_{12}^d)$ 
& $(Y_t,b,\tilde{t}_{12}^d)$ \\ 
\hline\hline 
$ m_b\left[\mbox{GeV}\right]$ & $4.35$ & $4.00$ & $4.08$& 
${4.04}$ \\
\hline
$ m_s \left[\mbox{MeV}\right] $ & ${219}$ & $
{249}$ & ${253}$ & 
${254}$
\\ \hline
$ m_s/m_d $ & $17.5$ & $20.8$  & 
$21.9$  & $22.0$ 
 \\\hline
$\left|V_{us}\right| $ & ${0.2138}$ & 
${0.2197}$ & ${0.2193}$ & ${0.2198}$ 
  \\\hline
$\left|V_{cb}\right| $ & ${0.0507}$ & $
{0.0447}$ &  ${0.0437}$ &
${0.0433}$  \\\hline
$\left|V_{ub}/V_{cb}\right|$ & ${0.080}$ & 
${0.097} $ & ${0.094} $ & 
${0.098}$ 
 \\\hline
$\star\left|V_{td}\right| $ & $0.0148$ & 
$0.0140$ & $0.0133$ & 
$0.0133$\\
\hline
$\star\left|V_{td}/V_{ts}\right| $ & ${0.30}$ & 
${0.33}$ & ${0.32}$ & 
${0.32}$\\
\hline
$\star \, \mbox{B}_k$ & $\infty$ & ${1.9}$& ${0.84}$& 
${0.77}$
\\\hline
$\star\sin^2 2\theta^{l}_{23}$ & $0.96 $ & ${0.83}$  &${0.87}$  
& ${0.85}$  \\ \hline
$\star\sin^2 2\theta^{l}_{12}$ & $1.2 \cdot10^{-2}$ & 
${2.0}\cdot10^{-3}$  & ${5.2}\cdot10^{-3}$  & 
${5.7}\cdot10^{-3}$ \\ \hline 
$\star\sin^2 2\theta^{l}_{13}$ & $1.8\cdot10^{-2}$ & ${2.6}\cdot10^{-2}$  & 
${2.4}\cdot10^{-2}$  & 
${2.4}\cdot10^{-2}$ \\ \hline
$\star m_u/m_d $ & $0.6$ & $0.4$  & 
$0.3$  & $0.3$ 
 \\\hline 
$\star M_t/\sin\beta$ & $172.7$ & $172.7$  & 
$172.7$  & 
$172.7$ \\ \hline\hline 
$Y_t$ & $0.5$ & $0.5$ 
& $0.5$ & $0.5$ 
\\\hline  
$b$ & ${10.0}$ & ${12.1}$ 
& ${11.8}$ & ${12.0}$ 
\\\hline
$\xi $ & $0$ & $-$ &  $180^\circ$ & $\pi$ \\ \hline
$\sigma $ & $-$ & $19.7^\circ$ & $43.4^\circ$ & $\pi/4$ \\ 
\hline
$\tilde{t}^d_{12}$ & $-$ & $3.4 \cdot 10^{-2}$& $4.4\cdot 10^{-2}$ 
& $4.7\cdot 10^{-2}$\\
\hline \hline 
$\tilde{\chi}^2_{\rm min}~\& ~\chi^2_{\rm min}$ & 
$0.06~\&~50.7$ & $ 2.1~\&~ 9.5$ & 
${0.4}~ \&~ 6$  & ${1}~\&~ 5.1$ 
\\ \hline \hline
 \end{tabular}
\caption[]{\small The analysis of the ansatz {\bf A}. 
Taking as reference eq. (\ref{Y-gen2}) for ${\bf Y}_{e,d}$,  
the columns refer to: $ I.~$ (1,3) and (3,1) entries are set 
to zero ($C=0$), i.e. three-parameter fit $Y_t,b,\xi$; 
$II.~$  (2,2) entry is set to zero ($Y_{ct}\to 0$), i.e. 
four-parameter fit $Y_t,b, \sigma, \tilde{t}^d_{12}$; 
$III.~$ complete pattern 
-- five-parameter fit $Y_t,b, \xi, \sigma, \tilde{t}^d_{12}$. 
In all cases, from $Y_t=0.5$ 
it comes out  $Y_{ct}= 2.6\cdot 10^{-3}$.
The quantities marked by the symbol $\star$, are not        
included in the fits but are predictions of the fit itself 
in correspondence of 
best-fit parameters (also shown). 
The values at the minima of the (non-Gaussian) $\chi^2-$like function 
  $\tilde{\chi}^2_{min}$ (see eq. (\ref{chi2_ns})) and of the Gaussian 
function, $\chi^2_{min}$ are also given. 
}
\label{pattA}
\end{center}
\end{table}}


By  minimizing 
the sum  $\tilde{\chi}^2 +{\chi}^2$, we rather test  
the  consistency between the model and the 
Gaussian-distributed data {\em within} a reasonable range 
for the down-quark masses.   
In the next we shall report both the value 
of $\tilde{\chi}^2_{min} $ and ${\chi}^2_{min}$ separately,  in  Tables 
\ref{pattA}, \ref{pattB} and \ref{pattC}
containing all the results of the fits. 
We have to remark that  descriptions of the 
quark masses outside the `reasonable' range may be accounted, for example, 
by uncertainties in $\alpha_s(M_Z)$.

\subsection{Testing theoretical models: fit and predictions}

We have found  it to be instructive to consider first  the case with 
(1,3), (3,1) entries set to zero  ($C=0$), 
to investigate 
the effect of a non-zero 22 entry on the pattern 
(\ref{Fred}).
This pattern (denoted by $I$) 
has three parameters, $b,\xi$ and $Y_t$. 
As a second step, we have restored  
the (1,3) (3,1) entries and set to zero the 22-entry 
(formally, $Y_{ct}=0$) 
leaving as free parameters $b, \tilde{t}^d_{12}, 
\sigma, Y_t$ (case denoted by $II$).

Finally, the complete pattern is analysed with five free parameters,   
$b, \xi, \tilde{t}^d_{12}, \sigma, Y_t$ (case denoted by $III$). 
We can  first make some general considerations. 
We can note that the  dependence on  $Y_{ct}$ of the 
physical quantities is mainly encoded 
in the factors $I_e, ~I_d$.  
The former can be easily estimated, $I_e \approx  1+ 0.044 \cdot \cos\xi~$ and 
thus $Y_{ct}$ can induce  quite a small correction 
on the leptonic mixing angles. 
Therefore, from  the expressions 
(\ref{ang-23}), we can infer 
that  $|V_{\mu 3}| \equiv |v_{\mu 3}| \approx s_{23}^e$ 
(see (\ref{lept-mix})) 
depends mainly  on the asymmetry parameter $b$,  increasing  
roughly as $\sqrt{b}$. 
As a result, the AN bound $\sin^2 2\theta^l_{23} = 4 |V_{\mu 3}|^2
(1 -|V_{\mu 3}|^2)  >0.8$ requires 
$6 <b < 12$ as it was found      
in the case (\ref{Fred}) \cite{BR}. 
Analogously, in the case $I$ it is $|V_{e 2}| \equiv
|v_{e 2}| \approx s_{12}^e c_{23}^e$ 
which weakly decreases with $b$. 
Then the MSW range, in the case $I$  
can be recovered for $b\gsim 7$ where  
$\sin^2 2\theta^l_{12} = 4 |v_{e 2}|^2
(1 -|v_{e 2}|^2) $ gets smaller than $1.5\cdot 10^{-2}$. 
As we already know, this mixing can be affected by 
the initial 12 rotation $\tilde{t}^e_{12}$ present 
in the full model (case $III$).

The discussion about $I_d$ affecting the quark observables is more 
involved as it depends,
through the ratio $Y_b/(Y_s-Y_d)\propto k^2_1$,  
on the specific ansatz considered.
However, in the case {\bf A} a rough estimate  gives   
$~I_d \sim 1 + 0.12\cos\xi~$, while  larger correction can be achieved 
in the ans\"atze {\bf B}, {\bf C},  
$~I_d \sim 1 + 0.25\cos\xi$.  
In the following we discuss separately our results for the three ans\"atze, 
reported in  Tables \ref{pattA}, \ref{pattB} and \ref{pattC}.

\begin{figure}[t]
\vskip -3.7cm
\centerline{\protect\hbox{
\epsfig{file=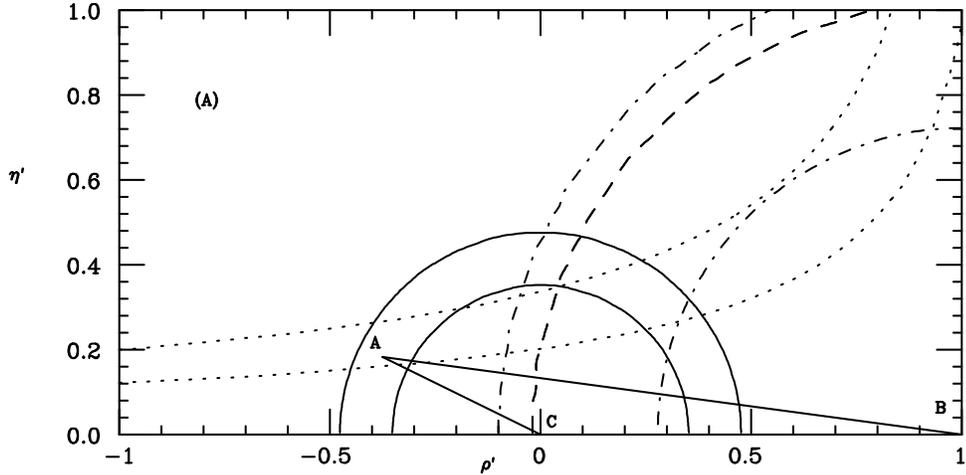,height=12.1cm,width= 16.0cm,angle=90}}}
\vskip -3.3cm
\caption{\small The unitarity triangle as emerging from the fit of the 
ansatz {\bf A} (case $III$) in the $(\rho',\eta')$ plane.
The angles defined in eqs. (\ref{UT}) are to be identified as 
$\al \equiv C\hat{A}B~,\beta\equiv A\hat{B}C~, \ga \equiv B\hat{C}A~$.   
The parameters  $\la, A$ are accordingly fixed as 
$\lambda = 0.2193$ and $A  = 0.909$. 
We have drawn the 1-$\sigma$ contours of 
$R_b= 0.414\pm 0.062$ (solid lines)  
$R_t= 0.913 \pm 0.191$ -- as inferred from $\Delta M_d$ -- (dot-dashed lines) 
and $R_t < 1.02$ -- from the lower bound on $\Delta M_s$ (dashed-line).
The band enclosed by the hyperbola (dotted) refers to 
$B_K= 0.8\pm0.2$. 
}
\vskip -0.5cm
\label{f1}
\end{figure}

\vspace{0.5cm} 
\noindent 
{\bf Ansatz A}~
In the  case $I$ (Table \ref{pattA} -- second column), 
the result of the fit is  acceptable though, due to the  
high accuracy achieved in the determination of the CKM angles, 
we can observe some discrepancy in 
the Cabibbo angle $|V_{us}|\equiv |v_{us}| \approx s^d_{12} 
\propto I_d/I_e$  and $|V_{cb}|\approx s^d_{23}$. 
The former appears to be quite 
small while the latter is somehow too large. 
These two quantities require in turn  $\xi =0$ (to maximize 
$I_d$) and quite a  large 
23 asymmetry,  $b=10$ (to minimize $s^d_{23}$)  as best fit points. 
Notice that $b>10$, though preferable to further reduce $|V_{cb}|$, 
 is not tolerated by $|V_{us}|$ which would further decrease.
 
On the other hand, the quark masses fall in their 
`reasonable' range. 
Noticeably, $Y_t=0.5$ is preferred in all the fits and as a result 
the prediction 
$M_t/\sin\beta \leq 
172.7$ GeV turns out to be consistent.   
Interestingly, the predictions (marked by $\star$) 
for the leptonic mixing angle are quite good. 
We have to recall that whenever the quark mixing are `reasonably' 
accommodated, the leptonic mixing angles fall   {\em automatically}  
into the presently most favoured range thanks to the  remarkable 
 product rule   (\ref{rule23}) inherent in the 
theoretic structure of the model.  
For example  the AN 
oscillatory mixing $\sin^2 2\theta^l_{23}$ 
is almost maximal, as the large $b= 10$ implies. The 13 neutrino mixing 
remain well below the upper bound shown in Table 1 
and $\sin^2 2\theta^l_{13}$ 
lies in the range $ (1\div 3)\cdot 10^{-2}$ in all the  ans\"atze, 
as we will see 
(cfr. Tables \ref{pattB}, \ref{pattC}).  
This range in the case of $\dem_{23}$ close 
to the upper bound in (\ref{AN}), can be of interest 
for the experimental search of $\nu_e\to \nu_\tau$ 
oscillation in the future CERN Neutrino Factory \cite{CERN}.

We do not show in the tables 
the prediction for the neutrino mass ratios. 
However, as regards $m_1/m_2$ 
the prediction  is just 
$m_1/m_2=Y_u/Y_c$ (in any of the 
ans\"atze) and so it is typically about 
$\sim 3\cdot 10^{-3}$.
On the other hand, 
the prediction for the ratio  $m_2/m_3$ 
depends on the value of the asymmetry parameter $b'$. 
In the simplest case $b'=b$, it can 
be easily computed as 
$m_2/m_3 = b Y_{ct}$ and so it is in the    
range  $\sim (2 - 3)\cdot  10^{-2}$ 
in agreement with the present experimental hint. 
This out-coming will be similar in all 
the ans\"atze and it is a byproduct 
of the link between the neutrino  and 
the up-quark masses and the large 23-asymmetry $b$ 
demanded by the quark phenomenology. 

The other predictions,  $V_{td},~V_{td}/V_{ts}$   
are marginally compatible with the experimental values. 
We have also reported the expected mass ratio $m_u/m_d$ 
extracted from the `ellipse' parameter $Q$ (\ref{Q-leu}) 
which appears to be consistent with 
the determination of $m_s/m_d$ obtained from the fit. 

Clearly there is no CP violation in the CKM matrix as 
the infinite value of $B_K$ does reflect. 
The situation gets improved in the case $II$. 
The presence of the (1,3), (3,1) entries introduces the initial 12 rotation 
$\tilde{t}^d_{12}$ and the phase $\sigma$ 
which  strongly modifies the CKM elements (\ref{CKM2}). 
From the expressions of  the  Cabibbo angle in (\ref{vud_ex}) 
and from the fact that the element $v_{us}$ itself is around 0.2, 
we can   deduce  
that  $\tilde{O}^d_{12}$ 
is to be  a small rotation, $\tilde{t}^d_{12} \lsim 0.1$, to prevent 
too large a correction 
from the element $v_{cs} \sim 1$. So 
$|V_{us}|$ is increased up to its experimental value  
by  $\tilde{t}^d_{12}\sim 0.03$ and $\sigma \sim 20^\circ$.
At the same time $|V_{cb}|$ is reduced, 
thanks to the larger value of the asymmetry parameter,  
$b\sim 12$  which is now  not prevented by $V_{us}$.   
We can notice that also $|V_{ub}/V_{cb}|$ is successfully reproduced. 
The effect of $b$ larger, i.e. of the decreasing of the ratio 
$Y_b/Y_\tau$ 
has in turn induced a bigger $m_s$ and a smaller $m_b$, 
both slightly outside their `reasonable' range.
The amount of CP violation from the CKM matrix, 
though increased thanks 
to the non-vanishing phase $\sigma$, is still not enough, $B_K \sim 2$.  
The predicted values of the leptonic mixings are  good. 
Accordingly to the approach elucidated in the Sect. 4.4, 
we have fixed the initial 12 lepton rotation as 
$\tilde{t}^e_{12}= \tilde{t}^d_{12}$, i.e. $k_2=k_1$.   
Then,  as expected,    $\sin^2 2\theta^l_{12}$ is 
strongly reduced below $10^{-2}$ with respect to the case $I$.
Notice, that  
the amount of CP violation 
in both the quark and lepton mixing matrix is the same as 
it is controlled by the same phase 
$\sigma$.

Finally, by comparing the results obtained in the case $III$ -- where 
the 22-entry is restored --   we conclude that 
the presence of the 22-entry does not play a significative role 
 -- the quality of the fit is quite 
stable in the two cases.
{\small
\begin{table}[t]
\begin{center}
\begin{tabular}{||c||c|c|c|c||}  \hline \hline
Ansatz ${\bf B}$ & $I.~C=0$ & $II. ~Y_{ct}=0$ 
& $III.~{\rm Complete}$ & $IV.~ {\rm Spont. CP}$ \\
$k_{1,3}=-3, k_2=0$ & $(Y_t,b,\xi)$ & 
$(Y_t,b,\sigma,\tilde{t}_{12}^d)$ & 
$(Y_t,b,\xi,\sigma,\tilde{t}_{12}^d)$ & 
$(Y_t,b,\tilde{t}_{12}^d)$ \\ 
\hline \hline
$m_b\left[\mbox{GeV}\right]$ & $4.65$ & $4.82$ & $4.24$ & 
${4.37}$ \\
\hline
$ m_s \left[\mbox{MeV}\right] $ & ${76}$ & ${83}$ & ${87}$ & 
${83}$
\\ \hline
$ m_s/m_d $ & $20.3$ & $27.3$ & $24.4$ & 
$23.0$ 
 \\\hline
$\left|V_{us}\right| $ & ${0.2127}$ & 
${0.2194}$ & ${0.2196}$ & 
${0.2194}$ 
  \\\hline
$\left|V_{cb}\right| $ & ${0.0430}$ & $
{0.0575}$ & $
{0.0403}$ & 
${0.0409}$  \\\hline
$\left|V_{ub}/V_{cb}\right|$ & ${0.014}$ & 
${0.102} $ & ${0.094} $ & 
${0.099}$ 
 \\\hline
$\star\left|V_{td}\right| $ & $0.0097$ & 
$0.011$ & $0.0086$ & 
$0.0089$\\
\hline
$\star\left|V_{td}/V_{ts}\right| $ & ${0.23}$ & 
${0.20}$ & ${0.22}$ & 
${0.22}$\\
\hline
$\star \, \mbox{B}_k$ & $\infty$ & ${0.22}$& ${0.69}$& 
${0.62}$
\\\hline
$\star\sin^2 2\theta^{l}_{23}$ & $0.97 $ & ${0.64}$  &${1}$  
& ${1}$  \\ \hline
$\star\sin^2 2\theta^{l}_{12}$ & $1.5 \cdot10^{-2}$ & 
${1.7}\cdot10^{-2}$  & ${1.3}\cdot10^{-2}$  & 
${1.4}\cdot10^{-2}$ \\ \hline 
$\star\sin^2 2\theta^{l}_{13}$ & $1.0\cdot10^{-2}$ & $
{4.5}\cdot10^{-3}$  & ${1.5}\cdot10^{-2}$  & 
${1.4}\cdot10^{-2}$ \\ \hline
$\star m_u/m_d $ & $0.4$ & $-$  & 
$-$  & $0$ 
 \\\hline 
%
$\star M_t/\sin\beta$ & $172.7$ & $195.6$  & 
$172.7$  & 
$172.7$ \\ \hline\hline
$Y_t$ & $0.5$ & $0.97$ 
& $0.5$ & $0.5$ 
\\\hline    
$b$ & ${6.8}$ & ${3.5}$ & ${8.8}$  
& ${8.2}$ 
\\\hline
$\xi$ & $0$ & $-$ & $9.7^\circ$& ${0}$\\ \hline
$\sigma $ & $-$ & $95.7^\circ $ & $93.5^\circ$ & $\pi/2$
\\\hline
$\tilde{t}^d_{12}$ & $-$ & $0.1$ &$9.2\cdot 10^{-2}$ & $9.7\cdot 10^{-2}$\\
\hline \hline 
$\tilde{\chi}^2_{\rm min}~\&~\chi^2_{\rm min}$ & $11.4~\&~ 45.2 $ 
& $ 150.5~\& ~ 112$ & 
$0.02 ~\&~ 0.24$ &  $0 ~\& ~0.82$ 
\\ \hline \hline
 \end{tabular}
\caption[]{\small The analysis of the ansatz {\bf B}. 
In the fit $II$ from $Y_t=0.97$ 
it follows  $Y_{ct}= 1.9\cdot 10^{-3}$.
See also the caption of Table \ref{pattA}.
} 
\label{pattB}
\end{center}
\end{table}}
However, the predictions are better. Notice that the right amount of
 CP violation is achieved -- $B_K \sim 0.7$ -- 
as the phase $\sigma$ is now bigger. For the same reason, also 
$\sin^2 2 \theta^l_{12}$ 
lies exactly in the range required by the MSW solution.
    
The fact that the best-fit points of the phases,  $\xi = \pi, 
~\sigma \approx \pi/4$,  are just integers (or half-integers) of 
$\pi$  may suggest that the CP violating phase is originated by 
a  spontaneous-breaking mechanism. 
For this reason 
in the column $IV$ we have performed again the  fit 
fixing the phases $\xi$ and $\sigma$ at the  
nearest "$\pi$-fold" values  
of the kind 
$q\pi $ ($q=1,~ 1/2,~ 1/4, ~ 
\cdots $) as respect to those given by  the fit $III$.
In this case we do not observe any variation. 
Therefore we can conclude that the ansatz {\bf A} 
provides quite a good fit of the quark observables for large 23 asymmetry 
$b\sim 12$ and $\tilde{t}^d_{12} \sim 0.05$ and for $\xi =180^\circ~, 
\sigma = 45^\circ$. 
The 
corresponding amount of CP violation appears to be in the right range  as    
indicated by $B_K \sim 0.8$. We shall come back to the  CP issue  below.


\begin{figure}[t]
\vskip -3.7cm
\centerline{\protect\hbox{\epsfig{file=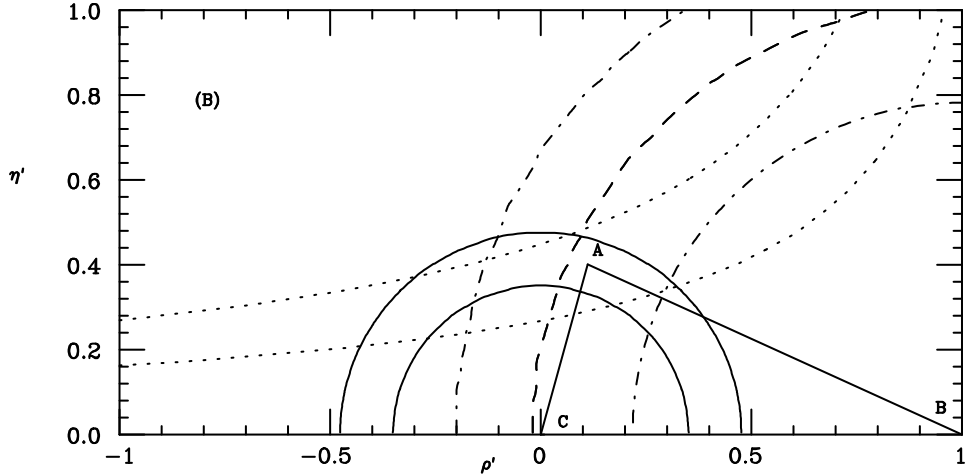,height=12.1cm,width= 16.0cm
,angle=90}}}
\vskip -3.3cm
\caption{\small As in Fig. (\ref{f1}) for the \ans {\bf B} (case $III$). 
Here $\lambda = 0.2196$ and $A  = 0.836$,   
$R_b= 0.413 \pm 0.062$, $R_t= 0.989\pm 0.207$ (from $\Delta M_d$)  
and $R_t< 1.02$ (from $\Delta M_s$).       
}
\vskip -0.5cm
\label{f2}
\end{figure}

\vspace{0.5cm} 
\noindent 
{\bf Ansatz B}~
We have first to remind that this Clebsch-pattern in its Fritzsch version  
 (vanishing 22 entry, $C=0$ and $b=1$) 
would give $m_s/m_d \approx 25$ and 
hence too a small $|V_{us}| \approx 0.20$. 
Let us  consider the effect of the non-vanishing 22 entry.  
From eq. (\ref{s/d2}) we see that 
a proper trend can be achieved 
for $\xi \sim 0$ and $b \geq 6$ so that  to maximise $I_d$. 
On the other hand, the moderate asymmetry $b\sim 6-7$ required by the 
fit implies 
quite a  large $m_b$,   small $m_s$ and consequently also 
a very tiny $|V_{ub}/V_{cb}|$. The fit is indeed rather poor (see Table 
\ref{pattB}, column $I$). 
However,  
setting to zero the  22-entry and taking $C\neq 0$ 
spoils completely the fit (Table 
\ref{pattB}, column $II$). 
Among the CKM elements only $|V_{us}|$ and the ratio 
$|V_{ub}/V_{cb}|$ get  
improved thanks to  the $\tilde{t}^d_{12}\sim 0.1$ rotation. 
The same rotation is instead less important for $|V_{cb}|$ which remains 
too large due to the small asymmetry, $b \sim 3-4$. For the same reason
 also  $m_s/m_d$ is quite large. 
In particular as $m_s/m_d$ is larger 
than the parameter $Q$,  the light-quark mass ratio $m_u/m_d$ 
cannot consistently be evaluated.      
As a whole this case is definitely disfavoured. Remarkably, the leptonic 
mixing angles as well as $B_K$ are not well predicted, too.       

Nevertheless, the interplay of the 22-entry with the (1,3), (3,1) entries 
can offer a satisfactory description as the results in the case $III$ show. 
All the quark quantities are perfectly fitted (notice that both 
$\tilde{\chi}^2_{min}$ and ${\chi}^2_{min}$  are $< 1$). 
The fit requires  
$b\sim 9$, $\tilde{t}^d_{12}\sim 0.1$, 
small $\xi$ and large CP phase,  $~\sigma \sim 94^\circ$.  
All the predicted 
quantities are within their experimental ranges. 
The 23 leptonic mixing is predicted to be 
maximal as required by the AN anomaly. 
The large phase $\sigma$ 
enforces $\tilde{t}^e_{12}=0$ or $k_2=0$ and  
the SN oscillation mixing 
comes out to be close to the upper limit, $\sin^2 2 \theta^l_{12} =1.3 
\cdot 10^{-2}$. 
Indeed,  were $\tilde{t}^e_{12} =\tilde{t}^d_{12}$ ($k_2=k_1$)  
and $\sigma \sim 90^\circ$ -- as preferred by the fit -- 
the 12 leptonic mixing 
$|V_{e2}|$  would get further increased 
(see eq. (\ref{Ve2_ex}) ), 
contrary to what happens in the ansatz {\bf A}.
As a further consequence of $\tilde{t}^e_{12}=0$ 
the CP phase is vanishing in the lepton mixing matrix except for the 
small contribution driven by the phase $\xi$.
Finally in the last column, we have considered the case with 
$\xi=0$ and $\sigma =\pi/2$ fixed. 
The quality of the fit remains very good.  

{\small
\begin{table}[ht]
\begin{center}
\begin{tabular}{||c||c|c|c|c||}  \hline \hline
Ansatz ${\bf C}$: & $I.~C=0$ & $II. ~Y_{ct}=0$ 
& $III.~{\rm Complete}$ & $IV.~ {\rm Spont. CP}$ \\
$k_1=-3, k_3=2, k_2=0$ & 
$(Y_t,b,\xi)$ &  $(Y_t,b,\sigma,\tilde{t}_{12}^d)$ & 
$(Y_t,b,\xi,\sigma,\tilde{t}_{12}^d)$ & 
$(Y_t,b,\tilde{t}_{12}^d)$ \\ 
\hline \hline
$ m_b\left[\mbox{GeV}\right]$ & $4.04$ & $4.02$ & $4.17$ & 
${4.09}$ \\
\hline
$ m_s \left[\mbox{MeV}\right] $ & ${115}$ & ${111}$ & ${114}$ & 
${120}$
\\ \hline
$ m_s/m_d $ & $18.2$ & $16.7$ & $18.9$ & 
$20.4$ 
 \\\hline
$\left|V_{us}\right| $ & ${0.2197}$ & 
${0.2196}$ & ${0.2195}$ & 
${0.2205}$ 
  \\\hline
$\left|V_{cb}\right| $ & ${0.0378}$ & ${0.0381}$ & ${0.0395}$ & 
${0.0381}$  \\\hline
$\left|V_{ub}/V_{cb}\right|$ & ${0.041}$ & 
${0.093} $ & ${0.093} $ & 
${0.082}$ 
 \\\hline
$\star\left|V_{td}\right| $ & $0.0098$ & 
$0.010$ & $0.0098$ & 
$0.0093$\\
\hline
$\star\left|V_{td}/V_{ts}\right| $ & ${0.27}$ & 
${0.28}$ & ${0.26}$ & 
${0.25}$\\
\hline
$\star \, \mbox{B}_k$ & $9.3$ & ${0.74}$& ${0.65}$& 
${0.83}$
\\\hline
$\star\sin^2 2\theta^{l}_{23}$ & $0.97 $ & ${0.97}$  &${1}$  
& ${0.99}$  \\ \hline
$\star\sin^2 2\theta^{l}_{12}$ & $1.2 \cdot10^{-2}$ & 
${1.2}\cdot10^{-2}$  & ${1.3}\cdot10^{-2}$  & 
${1.3}\cdot10^{-2}$ \\ \hline 
$\star\sin^2 2\theta^{l}_{13}$ & $1.8\cdot10^{-2}$ & ${1.8}\cdot10^{-2}$  & 
${1.4}\cdot10^{-2}$  & 
${1.6}\cdot10^{-2}$ \\ \hline 
$\star m_u/m_d $ & $0.6$ & $0.7$  & 
$0.5$  & $0.4$ 
 \\\hline 
$\star M_t/\sin\beta$ & $172.7$ & $172.7$  & 
$187.0$  & 
$181.2$ \\ \hline\hline 
$Y_t$ & $0.5$ & $0.5$ 
& $0.7$ & $0.6$ 
\\\hline     
$b$ & ${10.0}$ & ${10.1}$ &${8.7}$ 
& ${9.4}$ 
\\\hline
$\xi$ & $107^\circ$ & $-$ &  $135^\circ$ & $\pi$\\ \hline
$\sigma$ & $-$ & $106^\circ$ & $93.3^\circ$ & $\pi/2$
\\ \hline
$\tilde{t}^d_{12}$ & $-$ & $7.3\cdot 10^{-2}$ &
$8.6\cdot 10^{-2}$ & $7.3\cdot 10^{-2}$\\
\hline \hline 
$\tilde{\chi}^2_{\rm min} ~\&~\chi^2_{\rm min}$ & $0.5~\& ~ 15 $ & 
$ 4.0~\&~ 0.7$ & $ 0~\&~ <0.01$ & 
$0.02 ~\& ~1.5$ 
\\ \hline \hline
 \end{tabular}
\caption[]{\small The analysis of the ansatz {\bf C}. 
In the fit $III$ ($Y_t=0.7$) and $IV$ ($Y_t=0.6$)  
it is $Y_{ct}= 2.2\cdot 10^{-3}$ and $ 2.4\cdot 10^{-3}$, respectively. 
See also the caption of Table (\ref{pattA}).
} 
\label{pattC}
\end{center}
\end{table}}

\vspace{0.5cm} 
\noindent 
{\bf Ansatz C}~
The Fritzsch case for this pattern shows an opposite behaviour 
as compared to the previous ansatz. Indeed, it is $m_s/m_d \approx 12$ 
implying too a big $|V_{us}| \approx 0.29$. 
Therefore, in the variant $I$, 
this can be cured with $\xi > 90^\circ$ and a large $b$ (see Table 
\ref{pattC}, column $I$). 
In this case only $|V_{ub}/V_{cb}|$ is strongly incompatible   with its 
experimental range. Notice that, for the same asymmetry $b = 10$, 
$m_b$ is smaller 
as  respect to the value obtained in the ansatz {\bf A} ($I$), 
since as $k_1$ is larger as  $Z$ becomes smaller. 
All the predictions are reasonable (except for $B_K$). 
In the second case ($II$) all the CKM mixings are well reproduced 
thanks to the initial 12 rotation, but $Y_{ct}=0$ makes $m_s/m_d$ 
below the `reasonable' range. Finally, the complete model provides quite 
an excellent fit ($III$). 
Interestingly, also this \ans points to maximal CP violation -- 
$\sigma \sim 90^\circ$.
In  correspondence of the 
best-fit parameters
all the predicted quantities are in agreement 
with the present experimental status.  
We should remark that like in the previous \ans, 
the presence of the 22-entry 
is crucial. 
Moreover, also in this case the choice $k_2=0$ does not induce 
CP violation in the lepton mixing matrix.  
In the last column, we have performed 
a three-parameter fit, setting $\sigma =90^\circ$ and $\xi = 180^\circ$. 
One could  conclude  that, by endowing    
 the same pattern  with spontaneous breaking 
of CP-symmetry,  the  description remains very good.

\vspace{0.5cm}
Finally, we can try to give an overlook at all the three ans\"atze 
analysed.  
The case {\bf A} requires the CP-phase $\sigma < 90^\circ$, whereas 
both the ans\"atze {\bf B} and {\bf C} prefer  maximal CP-violation,  
$\sigma \gsim 90^\circ$. This should imply in general 
a different relations between the corresponding CKM elements. 
For example  the three ans\"atze will  show up  
a different shape of the ``unitarity'' triangle characterized by the 
following  angles    
\be{UT}
\al \equiv \mbox{arg}\left(-\frac{V^\star_{tb}V_{td}}{V^\star_{ub}V_{ud}}
\right)~,~~~~~~
\beta \equiv \mbox{arg}\left(-\frac{V^\star_{cb}V_{cd}}{V^\star_{tb}V_{td}}
\right)~,~~~~~~
\ga \equiv \mbox{arg}\left(-\frac{V^\star_{ub}V_{ud}}{V^\star_{cb}V_{cd}}
\right)~.
\ee

\begin{figure}[t]
\vskip -3.7cm
\centerline{\protect\hbox{\epsfig{file=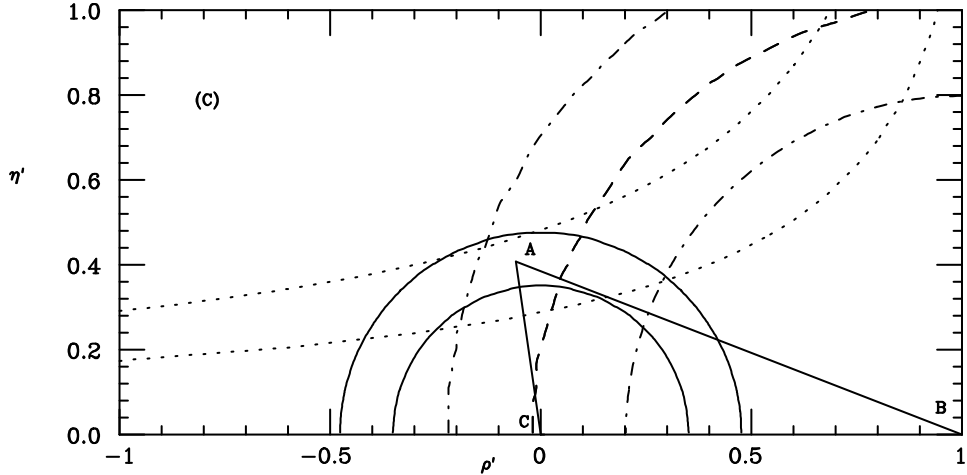,height=12.1cm,width= 16.0cm
,angle=90
}}}
\vskip -3.3cm
\caption{\small 
As in Fig. (\ref{f1}) for the \ans {\bf C} (case $III$). 
Here $\lambda = 0.2195$ and $A  = 0.820$,   
$R_b= 0.413 \pm 0.062$, $R_t= 1.01\pm 0.21$ (from $\Delta M_d$)  
and $R_t< 1.02$ (from $\Delta M_s$).       
}
\vskip -0.5cm
\label{f3}
\end{figure}

Therefore in correspondence of the results of the fits obtained in each
ansatz for the complete case $III$,  
we have plotted the  unitarity triangle in the $\rho', \eta'$ 
plane  (Fig. 6, 7, 8). 
We recall that $\rho, \eta, A,  \lambda$ are the Wolfenstein parameters 
and $\rho'= \rho (1-\lambda^2/2)$, $\eta'= \eta(1-\lambda^2/2)$.   
More precisely, the side CB corresponding to 
$V_{cd} V^\star_{cb}$ has been chosen real and rescaled to 
unit length. Hence the length CA and CB in the rescaled triangle,  
usually denoted by $R_b$ and $R_t$, respectively, are:
\beqn{sides}
&& R_b \equiv \left|\dfrac{V_{ud}V^\star_{ub}}{V_{cd}V^\star_{cb}}\right|=
\sqrt{\rho'^2 +\eta'^2} = (1 - \frac{\la^2}{2}) \frac{1}{\la} 
 \left|\dfrac{V_{ub}}{V_{cb}}\right|~, \\
&& R_t \equiv \left|\dfrac{V_{td}V^\star_{tb}}{V_{cd}V^\star_{cb}}\right|=
\sqrt{(1-\rho')^2 +\eta'^2} = \frac{1}{\la} 
 \left|\dfrac{V_{td}}{V_{cb}}\right|~. 
\eeqn
Once  the parameters $\la, A$ have been fixed 
according to the results of the $III$ fit, 
$\la = |V_{us}|~, ~ A = |V_{us}|/\la^2~$,    
we have depicted the $1$-$\sigma$ circles of $R_b$  from the measured value 
of $\left|\frac{V_{ub}}{V_{cb}}\right|$ (solid lines), and  
the circles of $R_t$ from the the mass difference 
$\Delta M_{d,s}$ decribing the strength of the $B^0_{d,s} - \bar{B}^0_{d,s}$ 
mixings (dot-dashed and dashed lines, respectively). 
The values used are $\Delta M_{d}= 0.471\pm 0.016 {\rm ps}^{-1}$ and 
$\Delta M_{s}> 12.4 {\rm ps}^{-1}$ \cite{NP}. 
In the same plane 
the hyperbola represent the iso-contours for $B_K$ (dotted lines).   
This overall picture of the CKM elements shows that 
 the \ans {\bf A} (see Fig. \ref{f1}) 
is characterized 
by a large angle $\ga$, $~\ga > 100^\circ$,   
and by a small angle $\beta$, $~\beta \approx 8^\circ$. 
They both appear to be in disagreement with 
the  result reported by the CDF collaboration $\sin2\beta 
=\cvb{0.79}{+0.41}{-0.44}$ and with the indirect bounds on $\gamma$ which 
disfavour   $\gamma\gsim  100^\circ$ \cite{NP}. 
Moreover, the vertex $A$ lies above the region delimited by 
the circles $R_t$ due to the large value 
of $|V_{td}| \sim 0.013$ predicted. 
On the contrary, the \ans  {\bf B} 
provides a consistent scenario  where the observed amount of 
CP-violation can be accounted by the CKM picture (Fig. \ref{f2}). 
The same holds for the \ans  {\bf C} (Fig. \ref{f3}).      
Ultimately, a better discrimination of the specific ansatz will be possible 
when the program of the measurements of the present and future 
$B$ factories is completed.


\section{Conclusions} 

We have considered a particular set of textures
for the Yukawa matrices of  quarks and leptons which share
some features with both the Stech ansatz proposed long time ago
\cite{Stech, su3H} and
the  Fritzsch-like textures suggested in \cite{BR}.
Our main motivation was to show that they are not only 
successful from the phenomenological point of view but 
also grounded and compelling on the theoretical side.
We have discussed how these textures could emerge 
in the context of grand unified theories, in terms of 
the prototype $SU(5)$ model complemented 
by the horizontal symmetry $SU(3)_H$ (Sect. 2).

The $SU(3)_H$ group may seem too `large' as compared to 
$U(2)_H$. Indeed the latter, providing {\it ab initio} the
$2 + 1$ representation structure for  the fermion fields,
directly singles out the heavy top quark from the lighter 
fermions of the first and second  generation. 
Namely, the models based on $U(2)_H$ \cite{su2H} 
invoke the  familiar paradigm according to which  
the third generation 
has {\it a priori} order 1 (tree-level) 
Yukawa couplings 
while the small Yukawa constants of  
the lighter generations 
emerge from  higher-order terms containing the 
$U(2)_H$ symmetry breaking Higgses, 
with VEVs smaller than the cutoff scale.

However, there are many good points in favour of $SU(3)_H$.
First, it  accounts by itself for three fermion families, 
and thus can be more predictive than $U(2)_H$.   
Second, the spontaneous breaking features of $SU(3)_H$ 
may turn  
the Yukawa constants of the 
low energy theory (MSSM) into  dynamical degrees of freedom 
and fix the inter-family hierarchy in a pretty natural 
manner. 
Namely, the third generation becomes heavy 
($Y_t\sim 1$), while the second and first ones become 
lighter by successively increasing powers of small 
parameters. 

In the general discussion of theoretical issues, we 
have put forward the following argument: 
just like the $SU(5)$ theory contains 
the  adjoint Higgs, 24-plet, which,  by 
breaking down $SU(5)$ 
to $SU(3)\times SU(2)\times U(1)$, does hide 
the existence of the large GUT symmetry, 
also $SU(3)_H$ may be first broken down to $U(2)_H$ 
by its adjoint Higgs, an octet, which makes 
the existence of a larger 
symmetry among all three families less visible.

In particular, the horizontal octet which breaks $SU(3)_H$ 
down to $SU(2)_H$, not only diversifies 
the third family from the lighter ones, 
but also induces a non-trivial Clebsch structure 
between the 1-2 and 3 generations 
in the Yukawa matrices of quarks and leptons. 
The effect of the latter is twofold. 
First, it induces the complementary "large -- small" 
(see-saw like) pattern between the neutrino and quark 
mixing angles,  
which is indeed exhibited by the observed small value of 
$V_{cb}$ and the nearly maximal $\nu_\mu-\nu_\tau$ mixing.
Second, it may link  the neutrino  mass hierarchy to that
of the up-quarks in realistic way.

On the other hand, 
the $SU(5)$ adjoint has proved necessary to break the 
`standard' down-quark and lepton degeneracy, providing
different Clebsch coefficients in the Yukawa matrices. 
Though its role is not univocal, 
we have featured three Yukawa patterns characterized
by what we consider as the most `natural' Clebsch factors.
In Sect. 3 we have elaborated a consistent $SO(10)\times SU(3)_H$ model 
which, supplemented by the symmetry $\cG = U(1)_A \times Z_6 \times \cR$,  
represents an existence proof of the Yukawa texture suggested. In 
particular, the specific model outlined provides an understanding 
of the Clebsch pattern of the ansatz {\bf B}. 
Remarkably, the same symmetry content of $\cG$ 
also motivates the missing VEV 
mechanism necessary to achieve the doublet-triplet splitting.        

A careful fit of the quark observables  demonstrates that 
all the ans\"atze {\bf A, B} and {\bf C} reproduce quite well 
the down-quark masses and CKM mixing angles (Sect. 4).
What appears very interesting is that for the best-fit ranges 
of the parameters the predictions for the neutrino mixing angles
are in good agreement with the present experimental hints.
About the lepton mixing angles, we have to stress that
while the outcome for the 23 neutrino mixing is a `genuine'   
prediction, in strict connection {\it only}
with the down-quark sector through
the product rule (\ref{rule23}), the 12 mixing angle 
becomes effectively a free parameter fixed by the
rotation $\tilde{t}^e_{12}$. 
Should the small angle MSW solution
be confirmed in the future, then all the ans\"atze can account
for that with different values of $\tilde{t}^e_{12}$.
On the contrary, should the 12 neutrino mixing angle be much
larger, as indicated by the large-mixing angle regime
of the MSW solution,
then  only the lepton sector would need some refinement.

Interestingly, also the prediction for CP violation in the 
quark sector may lie in the observed range.
In particular, all the ans\"atze generate a CKM  matrix 
with the correct amount of CP violation. 
Though the experimental test of CP violation 
in the SM is not yet accomplished, 
this result may imply  that 
superpartners contributions to $\eps_K$ must be 
adequately suppressed \cite{CP-susy} and it would be interesting 
to study in the theoretical models presented here 
the implications of the horizontal symmetry on 
the SUSY spectrum.

\vspace{1cm}
\noindent 
{\Large \bf Acknowledgements} 
\vspace{0.5cm}

We thank Andrea Brignole and Denis Comelli for useful 
discussions. This work is partially supported by the MURST 
research grant of national interest "Astroparticle Physics".

\vspace{1.0cm}
\noindent
{\Large \bf Appendix A} 
\vspace{0.5cm}

Consider the following superpotential terms including  
antisextet superfield $S=S^{ij}$ and its sextet 
partner $\bar{S}=\bar{S}_{ij}$:   
\be{H-super}
W_S= -\mu S\bar{S} + S^3 + \bar{S}^3 ,  
\ee
where,  to  be explicit,  
$S^3 = \frac13 \eps_{ijk}\eps_{lmn} S^{il} S^{jm} S^{kn}=\det S$ 
(similarly for $\bar{S}^3$), 
and order 1 coupling constants are absorbed. 
Observe that this superpotential is manifestly invariant 
under $Z_3$ symmetry:  
$S \to \exp(i\frac{2\pi}{3}) S$ and  
$\bar{S} \to \exp(-i\frac{2\pi}{3}) \bar{S}$. 

Without loss of generality, the VEV of $S$ can be chosen 
in the diagonal form, 
$\langle S\rangle ={\rm Diag}(\cS_1,\cS_2,\cS_3)$. 
Then the condition of vanishing $F$-terms 
$F_S,F_{\bar{S}}=0$ implies that 
$\langle\bar{S}\rangle$ is also diagonal, 
$\langle\bar{S}\rangle=
{\rm Diag}(\bar{\cS}_1,\bar{\cS}_2,\bar{\cS}_3)$, 
and 
\beqn{FS}
& \cS_1\cS_2 = \mu \bar{\cS}_3 , ~~~ 
&\bar{\cS}_1\bar{\cS}_2 = \mu \cS_3 , 
\nonumber \\   
&  \cS_1\cS_3 = \mu \bar{\cS}_2 , ~~~ 
&  \bar{\cS}_1\bar{\cS}_3 = \mu \cS_2 , 
\nonumber \\   
&  \cS_2\cS_3 = \mu \bar{\cS}_1 , ~~~ 
&  \bar{\cS}_2\bar{\cS}_3 = \mu \cS_1 . 
\eeqn 
In the exact supersymmetric limit the VEV pattern of $S$ and 
$\bar{S}$ is not fixed unambiguously and there remain 
flat directions which represent a two-parameter vacuum valley.   
In other words, 
the six equations (\ref{FS}) reduce to four conditions:
\be{FS1}
\cS_1\bar{\cS}_1 = \cS_2\bar{\cS}_2 = 
\cS_3\bar{\cS}_3 = \mu^2, ~~~~~ 
\cS_1\cS_2\cS_3 = \mu^3 ;  ~~~~~
\bar{\cS}_1\bar{\cS}_2\bar{\cS}_3 = \mu^3, 
\ee 
while the others are trivially fulfilled 
(the last in eq. (\ref{FS1}) is not independent, 
it just follows from the others). 
Thus, in principle the eigenvalues $\cS_{1,2,3}$ can be 
different from each other, say $\cS_3 > \cS_2 >\cS_1$. 
Then eqs. (\ref{FS}) imply that $\bar{\cS}_{1,2,3}$ should
have an inverse hierarchy, 
$\bar{\cS}_3 < \bar{\cS}_2 < \bar{\cS}_1$. 
More precisely, we have  
$\bar{\cS}_1:\bar{\cS}_2:\bar{\cS}_3=
\cS_1^{-1}:\cS_2^{-1}:\cS_3^{-1}$. 

The flat directions of the VEVs can be fixed from the 
soft supersymmetry breaking terms. The relevant ones 
are D-terms:\footnote{The F-terms $\sim \int d^2\theta zW$ 
are not relevant for the VEV orientation as far as they 
just repeat the holomorphic invariants like 
$S\bar{S}$ and $\det S$ whose values are already fixed 
by the conditions (\ref{FS1}). }  
\be{SSB-S}
\cL = - \int d^4\theta z\bar{z} 
\left[\alpha {\rm Tr}S^\dagger S + 
\frac{\beta}{M^2} ({\rm Tr}S^\dagger S)^2 + 
\frac{\gamma}{M^2} {\rm Tr}(S^\dagger S S^\dagger S) + ...\right] 
+ [ S \to \bar{S}] 
\ee 
having a similar form for $S$ and $\bar{S}$.    
Here $z=m\theta^2$ and $\bar{z}=m\bar{\theta}^2$ are 
supersymmetry breaking spurions, with $m\sim 1$ TeV.  
The cutoff scale $M$ is taken as the flavour scale, i.e. 
the same as the one in superpotential terms (\ref{hoo}),  
and we assume that $M\gg \mu$. 
Therefore, for $S_{1,2,3}$ these 
terms translate explicitly into the following scalar 
potential: 
\be{SSB-exp}
\alpha m^2 (|\cS_1|^2 + |\cS_2|^2 + |\cS_3|^2) +   
\beta \frac{m^2}{M^2} (|\cS_1|^2 + |\cS_2|^2 + |\cS_3|^2)^2 +   
\gamma \frac{m^2}{M^2} (|\cS_1|^4 + |\cS_2|^4 + |\cS_3|^4) + 
[|\cS_k| \to |\bar{\cS}_k|]   
\ee 
The stability of the potential implies that 
$\beta >0$ and $\gamma > -\beta$, whereas $\alpha$ can be 
positive or negative. In the former case the minimization of 
the potential (\ref{SSB-exp}), under the conditions (\ref{FS1}), 
would imply that $\cS_1=\cS_2=\cS_3=\mu$, i.e. no hierarchy 
between the fermion families.  
In the latter case, however, the largest eigenvalue of 
$S$ and $\bar{S}$, respectively $\cS_3$ and $\bar{\cS}_1$,  
grow up above the typical VEV size $\mu$ and reach  
values of the order of the cutoff scale $M$: 
\be{VEV3}
\cS_3,\bar{\cS}_1 =  
\left(\frac{\alpha}{2(\beta+\gamma)}\right)^{1/2} M
\ee
that is $\cS_3,\bar{\cS}_1 \sim M$. 
Then it follows from (\ref{FS1}) that 
\be{VEV12}
\cS_2,\bar{\cS}_2 = \mu = \eps \cS_3 , ~~~~  
\cS_1,\bar{\cS}_3 = \frac{\mu^2}{\cS_3} = \eps^2 \cS_3 ,   
\ee 
where $\eps = \mu/\cS_3 \sim \mu/M$. 

Let us assume now that the theory contains three triplets 
superfields $A_n=A_{ni}$, 
and their partners $\bar{A}_n=\bar{A}_{n}^i$,  
$n=1,2,3$.  
One can incorporate them by the following terms 
in the superpotential:   
\be{A-super}
W_A= -\sum_n \mu_n A_n\bar{A}_n  + A_1 A_2 A_3 + 
\bar{A}_1 \bar{A}_2 \bar{A}_3  
\ee
where order 1 coupling constants are understood and 
$A_1A_2 A_3 \equiv \eps^{ijk} A_{1i}A_{2j} A_{3k}$.  
For simplicity, we shall take all masses $\mu_n$ equal, 
$\mu_{1,2,3}=\mu' < M$. 
In addition, by assuming that the triplets 
have $Z_3$ charges different from that of $S$, 
we do not include terms like $SA^2\equiv S^{ij} A_iA_j$. 
In this way, in the exact supersymmetric limit the ground 
state has a continuous degeneracy (flat direction) 
related to unitary transformations $A_n \to UA_n$ 
with $U\subset SU(3)_A$. 
In other terms, the superpotential $W=W_S + W_A$ 
has an accidental global symmetry $SU(3)_S\times SU(3)_A$, 
with the  two $SU(3)$ factors independently transforming the 
two sets of horizontal superfields, $S$ and $A$.   
 
Similarly to the case of the Higgs fields $S, \bar{S}$, 
now the conditions 
$F_A,F_{\bar A}=0$ can only fix 
the values of holomorphic invariants $A_n\bar{A}_n$ 
and $A_1A_2A_3$. Namely, by unitary transformation 
$A_n \to UA_n$ ($U\subset SU(3)_A$), one can choose a basis 
where the VEV of $A_1$ points towards the first component. 
Then we see that in this basis the fields $A_2$ and $A_3$ 
should have the VEVs towards the second and third components. 
In other words, in this basis the triplets have the VEVs  
$\langle A_{ni} \rangle = \delta_n^i \cA_i$, 
satisfying the following equations:  
\be{FA1}
\cA_1\bar{\cA}_1 =\cA_2\bar{\cA}_2 = 
\cA_3\bar{\cA}_3 = \mu'^2, ~~~~~
\cA_1\cA_2\cA_3 = \mu'^3 , ~~~~~ 
\bar{\cA}_1\bar{\cA}_2\bar{\cA}_3 = \mu'^3 .  
\ee
The three VEVs can be different and one can take,  say  
$\cA_1 > \cA_2 > \cA_3$. The hierarchy between the latter 
can be fixed by soft D-terms 
\be{SSB-A}
\cL = - \int d^4\theta z\bar{z} 
\left[\alpha' \sum_n A_n^\dagger A_n + 
\frac{\beta'}{M^2} \sum_{mn} (A_m^\dagger A_m)(A_n^\dagger A_n) + 
\frac{\gamma'}{M^2} \sum_{m\neq n} 
(A_m^\dagger A_n)(A_n^\dagger A_m) \right] 
+ [ A_n \to \bar{A}_n] 
\ee 
with $\alpha'<0$. 
Then,  similarly to what happened to the sextets, 
the largest VEV rises  up to the cutoff scale $M$: 
$\cA_1,\bar{\cA}_3\sim M$, while    
$\cA_2,\bar{\cA}_2 \sim \eps' M$ and 
$\cA_3,\bar{\cA}_1 \sim \eps'^2 M$, with 
$\eps'\sim \mu'/M \sim (\mu'/\mu) \eps$.    

As for the relative orientation of the sextet and triplet 
VEVs, these  will also be fixed from the soft D-terms:  
\be{SSB-U}
\cL = - \int d^4\theta z\bar{z} 
\frac{\delta}{M^2}\sum_{n}(A_n^\dagger SS^\dagger A_n)
+ [ S \to \bar{S}, ~A_n \to \bar{A}_n] 
\ee 
Indeed, for positive $\delta$ the favoured orientation 
between the $S$ and $A$ bases, which minimizes the 
energy of the ground state, corresponds to $U=1$. 

The above can be interpreted in the following manner. 
In the view of operators like (\ref{hoo}), 
the MSSM Yukawa constants become dynamical degrees of freedom. 
In particular, the operator $\cO_u$ in (\ref{hoo}) implies 
\be{S-ap} 
\bY_u = \matr{Y_u}{0}{0} {0}{Y_c}{0} {0}{0}{Y_t} 
\sim \frac{\langle S^{ij} \rangle}{M} = 
\frac{1}{M}
\matr{{\cal S}_1}{0}{0} {0}{{\cal S}_2}{0} {0}{0}{{\cal S}_3} . 
\ee
In the exact supersymmetric limit the values of the  
constants $Y_{u,c,t}$ are not fixed -- they have flat 
directions and  eq. (\ref{FS1}) translates  into the 
constraint $Y_u Y_c Y_t = \eps^3$.
 However, we have shown that 
 soft supersymmetry breaking terms could naturally 
split the Yukawa constants so that $Y_t\sim 1$, 
while $Y_c \sim \eps$ and $Y_u\sim \eps^2$, 
which perfectly reflects the observed pattern 
if $\eps \sim 5 \cdot 10^{-3}$.  
(For discussions on  dynamical $Y_t$ in supersymmetry 
see also ref. \cite{Fabio}.) 
 
As for the triplet VEVs, in view of the operator $\cO$ 
(\ref{hoo}) they provide off-diagonal Yukawa entries 
\be{A-ap}
\bA= \matr{0}{A_3}{A_2} {-A_3}{0}{A_1} {-A_2}{-A_1}{0} \sim 
\sum_n \frac{\langle A_n^{ij} \rangle}{M} = \frac{1}{M} 
\matr{0}{\cA_3}{\cA_2} {-\cA_3}{0}{\cA_1} {-\cA_2}{-\cA_1}{0} . 
\ee
with the constraint $A_1 A_2 A_3 = \eps'^3$ 
in the 
supersymmetric limit, while the soft breaking terms 
can dynamically fix their values as $A_1\sim 1$, 
$A_2 \sim \eps'$ and $A_3\sim \eps'^2$. 
In the context of our work it is  clear that such a  
hierarchy of the  off-diagonal entries in $\bY_{d,e}$ 
is also well suited for the observed fermion mass and 
mixing pattern if $\eps' \sim 0.1$.  
 
One has to remark that the above considerations are 
literally valid if the horizontal symmetry is global. 
For the case of local $SU(3)_H$, apart from  F-flatness 
conditions, there are eight additional conditions:  
the gauge $D$ terms $D_a = \sum_n X^\dagger T^a X$, 
should vanish on the vacuum configuration  
($T^a$ are $SU(3)_H$ generators, $a=1,..8$).  
Although this does not occur for the obtained VEV pattern 
of $S$ and $A_n$, one can easily 
imagine the theory 
to contain additional "spectator" superfields in different 
representations of $SU(3)_H$ and their VEVs are oriented 
so that to 
cancel the contributions of $\langle S\rangle$ 
and $\langle A\rangle$ in gauge D-terms of $SU(3)_H$.  
In order not to affect the obtained solutions for 
$S$ and $A_n$, these extra superfields should 
not couple to the latter in the superpotential. 

\vspace{1.0cm}
{\Large \bf Appendix B} 
\vspace{0.5cm}

Let us assume now that the considered theory  
has an exact CP invariance. 
In other words, we can  choose a superfield basis where 
all coupling constants in the theory are real, 
including 
the coupling constants with the matter superfields, 
i.e. those in the terms (\ref{hoo}).
Applied to the superpotential 
(\ref{H-super}) this means that the dimensional parameter 
$\mu$ as well as the understood constants at $S^3$ and $\bar{S}^3$ 
are real. 
Without losse of generality, by means of a $SU(3)_H$ 
transformation, one can chose a basis where 
$\cS_3$ and $\cS_2$ are real and positive. 
Then, since eq. (\ref{FS1}) tells  us 
that the VEV products $\cS_1\cS_2\cS_3 = \mu^3$ 
is real, also the component $\cS_1$ 
should be real. The same consideration is true 
for the VEVs of triplets $A_n$. Therefore, no 
CP-violation can occur in this case.  

Let us now consider an alternative superpotential 
for triplets (\ref{A-super}) not containing the mass 
terms but invoking instead an auxiliary singlet $I$: 
\be{Hsuper}
W_A= \kappa_n I A_n\bar{A}_n - A_1 A_2 A_3 - 
\bar{A}_1 \bar{A}_2 \bar{A}_3 + 3\Lambda^2 I + I^3 , 
\ee
where $\Lambda$ is a dimensional parameter, 
$\kappa_n$ are coupling constants and order one 
constants are also understood  in other terms. 
Applied to this superpotential, CP invariance means that 
$\kappa_n$ as well as $\Lambda^2$ are real. 
For  simplicity, let us assume again 
that all $\kappa_n$ are equal, $\kappa_n=\kappa$,   
and $\kappa \ll 1$.

It is clear, that the $F$-term conditions 
$F_A,F_{\bar{A}}=0$ are the same as in (\ref{FA1}) 
apart from the fact that the 
mass scale $\mu'$ should be 
substituted by $\kappa\cI$, where $\cI= \langle I\rangle$.    
The latter is then fixed by the condition $F_I=0$ given as  
\be{FI}
\Lambda^2 + \cI^2 + \frac13 \kappa 
\sum_{k=1}^3 \cA_k\bar{\cA}_k = 
\Lambda^2 + (1+ \kappa^3)\cI^2 = 0 
\ee
Hence, $\cI^2 \simeq - \Lambda^2$, and if $\Lambda^2$ 
is positive, then $\cI$ should be imaginary and  
so we  obtain  spontaneous CP violation. 
In other words, imaginary $\cI$ means that  
the CP-odd component of the superfield $I$  
acquires a VEV.    

As a result, the `induced' mass $\mu' = \kappa\cI$ 
and so the VEV product $\cA_1\cA_2\cA_3 = \mu'^3$ 
are also imaginary. Therefore, at least one of 
the VEVs $\cA_{1,2,3}$ should be complex 
and thus  would induce the corresponding  
phase in the fermion Yukawa matrices. 

To fix the point, we have to find the 
relative phase orientation between the VEVs of 
$S$ and $A$. These should be determined by soft terms 
like 
\be{SSB-CP}
\cL = - \int d^4\theta z\bar{z} 
\frac{\zeta}{M^2}(S^\dagger S^\dagger A_n A_n + {\rm h.c.}) , 
\ee 
etc., with $\zeta$ being a real constant. 
In particular, the above term induces 
the following couplings in the Higgs potential: 
\be{CP}
\zeta \frac{m^2}{M^2}(\cS_3 \cS_2 \cA_1^2 + 
\cS_3 \cS_1 \cA_2^2 + \cS_2 \cS_1 \cA_3^2) ~ + ~{\rm h.c.}
\ee
We see that for $\zeta >0$ the minimization of 
the ground state energy requires that in the basis where  
$\cS_k$ are real and positive, $\cA_{1,2,3}^2 <0$,  
i.e. all triplets have imaginary VEVs. 
For $\zeta < 0$ we see instead that $\cA_{1,2}$ prefer 
to be real. From the couplings (\ref{CP}) also 
the smallest VEV $\cA_3$ would prefer to be real;  
however it is forced to have a complex phase 
$\pi/2$ by the constraint that the product 
$\cA_1\cA_2\cA_3$ is imaginary. 
Both these cases can be of interest  
for the 
CP-breaking phases of  the fermion Yukawa textures 
discussed in this work.

\end{fmffile}
\end{document}